\documentclass[fleqn,usenatbib]{mnras}

\usepackage{newtxtext,newtxmath}

\usepackage[T1]{fontenc}

\DeclareRobustCommand{\VAN}[3]{#2}
\let\VANthebibliography\thebibliography
\def\thebibliography{\DeclareRobustCommand{\VAN}[3]{##3}\VANthebibliography}

\usepackage{graphicx}	
\usepackage{amsmath}	
\usepackage{lscape}
\usepackage{tabularx}
\usepackage[dvipsnames]{xcolor}
\usepackage{soul}
\usepackage{longtable}
\usepackage{supertabular}

\usepackage{hyperref}
\hypersetup{
  pdfborder={0 0 0},
  linkbordercolor=[rgb]{0.5,0.57,0.72},
  linkcolor=[rgb]{0.0,0.0,0.5},
  urlcolor=[rgb]{0.0,0.0,0.5},
  colorlinks=true,
}

\newcommand{\Teff}{\mbox{$T_{\mathrm{eff}}$}}
\newcommand{\logg}{\mbox{$\log g$}}
\newcommand{\htohe}{\mbox{$\log(\mathrm{H/He})$}}
\newcommand{\hel}[2] {He\,{\sc #1}~$\lambda$#2}
\newcommand{\Line}[3]{#1\,{\sc #2}~$\lambda$#3}

\newcommand{\Ion}[2]{#1\,{\sc #2}}

\newcommand{\Rwd}{\mbox{$R_{\mathrm{WD}}$}}
\newcommand{\Mwd}{\mbox{$M_{\mathrm{WD}}$}}
\newcommand{\Msun}{\mbox{$\mathrm{M}_{\odot}$}}

\newcommand{\ztohe}{\mbox{$\log(\mathrm{Z/He})$}}

\title[Uncertainties in He-dominated white dwarf analyses]{Systematic uncertainties in the characterisation of helium-dominated metal-polluted white dwarf atmospheres}

\author[P. Izquierdo et al.]{Paula Izquierdo$^{1,2,3}$\thanks{E-mail: Paula.Izquierdo-Sanchez@warwick.ac.uk},
Boris T. G\"ansicke$^{1,4}$,
Pablo Rodr\'iguez-Gil$^{2,3}$,
Detlev Koester$^{5}$,
Odette Toloza$^{6}$,
 \newauthor
Nicola P. Gentile Fusillo$^{7,8}$, 
Anna F. Pala$^{7}$ and
Pier-Emmanuel Tremblay$^{1}$\\
$^{1}$Department of Physics, University of Warwick, Coventry CV4 7AL, UK\\
$^{2}$Instituto de Astrof\'isica de Canarias, 38205 La Laguna, Tenerife, Spain\\
$^{3}$Departamento de Astrof\'isica, Universidad de La Laguna, 38206 La Laguna, Tenerife, Spain\\
$^{4}$Center for Exoplanets and Habitability, University of Warwick, Coventry CV4 7AL, UK\\
$^{5}$Institut f\"ur Theoretische Physik und Astrophysik, University of Kiel, 24098 Kiel, Germany\\
$^{6}$ Departamento de F\'isica, Universidad T\'ecnica Federico Santa Mar\'ia, Av. Espa\~na 1680, Valpara\'iso, Chile\\
$^{7}$European Space Agency, European Space Astronomy Centre, Camino Bajo del Castillo s/n, 28692 Villanueva de la Ca\~nada, Madrid, Spain\\
$^{8}$ European Southern Observatory, Karl Schwarzschild Strasse 2, Garching, 85748, Germany}

\date{Accepted XXX. Received YYY; in original form ZZZ}

\pubyear{2015}

\begin{document}
\label{firstpage}
\pagerange{\pageref{firstpage}--\pageref{lastpage}}
\maketitle

\begin{abstract}
White dwarf photospheric parameters are usually obtained by means of spectroscopic or photometric analysis. These results are not always consistent with each other, with the published values often including just the statistical uncertainties. The differences are more dramatic for white dwarfs with helium-dominated photospheres, so to obtain realistic uncertainties we have analysed a sample of 13 of these white dwarfs, applying both techniques to up to three different spectroscopic and photometric data sets for each star. We found mean standard deviations of $ \left < \sigma \Teff \right > = 524$\,K, $\left < \sigma \logg \right > = 0.27$\,dex and $\left < \sigma \htohe \right > = 0.31$\,dex for the effective temperature, surface gravity and relative hydrogen abundance, respectively, when modelling diverse spectroscopic data. The photometric fits provided mean standard deviations up to $\left < \sigma \Teff \right > = 1210$\,K and $\left < \sigma \logg \right > = 0.13$\,dex. We suggest these values to be adopted as realistic lower limits to the published uncertainties in parameters derived from spectroscopic and photometric fits for white dwarfs with similar characteristics. In addition, we investigate the effect of fitting the observational data adopting three different photospheric chemical compositions. In general, pure helium model spectra result in larger \Teff\ compared to those derived from models with traces of hydrogen. The \logg\ shows opposite trends: smaller spectroscopic values and larger photometric ones when compared to models with hydrogen. The addition of metals to the models also affects the derived atmospheric parameters, but a clear trend is not found.

\end{abstract}




\section{Introduction}
\label{sec:intro}

About 20~per~cent of all white dwarfs in the Galaxy are known to have helium-dominated atmospheres \citep{bergeronetal11-1}. These are thought to form either after a late shell flash, if the white dwarf progenitor burns all its residual hydrogen in the envelope \citep{herwig99,althaus05-1,werner06}, or via convective dilution or mixing scenarios, where a thin hydrogen layer is diluted by the deeper convective helium one \citep{fontaine87, cunningham20}. The helium-dominated white dwarfs with effective temperatures, \Teff, between 10\,000 and 40\,000\,K\footnote{The \Ion{He}{i} optical transitions  originate from states with principal quantum number $n=2$. For $\Teff \leq 10\,000$\,K, helium is mostly in its ground state, and hence, the optical spectra of cool white dwarfs with helium atmospheres are featureless and classified DC. For $\Teff \geq 40\,000$\,K, helium is mostly ionised, and the spectra of these hot white dwarfs show \Ion{He}{ii} transitions and are classified DO.} are called DBs and are characterised by \Ion{He}{i} absorption lines dominating their optical spectra.



The first fully characterised DB white dwarf \citep[GD\,40;][]{shipman77} paved the way for numerous studies in the following 25 years \citep[see e.g.][]{wickramasinghe83, koester85, liebert86, wolff02}, establishing the techniques currently used to derive the photospheric parameters of these degenerates. Their \Teff, surface gravity, \logg, and chemical abundances are obtained by means of (1) grids of synthetic spectra to fit the helium (plus hydrogen, if present) absorption lines identified in their observed spectra \citep[see e.g.][]{koester15}, (2) reproducing their photometric spectral energy distribution \citep[SED;][]{bergeron97}, or (3) a hybrid approach that simultaneously fits the spectroscopy and photometry to deliver a more consistent set of parameters \citep[see e.g.][]{izquierdo20}. Even though no major issues have been reported, these techniques do not always lead to consistent parameters \citep[e.g.][]{bergeronetal11-1, koester15, tremblayetal19-2,cukanovaite21}.

The discrepancies are likely a consequence of the several hurdles that determining the atmospheric parameters of DBs has to face. It is hard to obtain accurate \Teff\ values in the $\simeq 21\,000-31\,000$\,K range\footnote{This range coincides with the instability strip of DBs, where most white dwarfs \citep{nitta09} undergo non-radial oscillations which complicate their characterisation \citep[e.g.][]{winget82,vanderbosch22}.}, where a plateau in the strength of the \Ion{He}{i} absorption lines gives rise to similar $\chi^{2}$ values on each side of this temperature range (usually referred to as the ``hot'' and ``cool'' solutions). Likewise, there appears to exist a problem related to the implementation of van der Waals and resonance broadening mechanisms for neutral helium, the two dominant interactions in white dwarfs with $\Teff \leq 15\,000$~K \citep{koester15}. On top of that, as white dwarfs cool, they develop superficial convection zones that grow bigger and deeper with decreasing \Teff\ \citep{tassoul90}. The treatment of convective energy transport is neither fully understood nor implemented, even though \citet{cukanovaite21} presented a complete implementation for DBs with no free parameters, in contrast to the canonical and simplistic mixing-length (ML) theory\footnote{Convection in white dwarfs is thought to be highly turbulent, and currently the most common treatment relies on the ML approximation \citep{prandtl1925,bohm-vitense58}. For white dwarf model atmospheres, this approximation has four free parameters to describe the convective energy flux, among which we find the ratio of the mixing length, $l$, to the pressure scale height, $H_\mathrm{P}$, known as the convective efficiency, $\alpha = l/H_\mathrm{P}$. These four free parameters change from version ML1 to ML2 \citep[see][for further details]{koester10-1}.}. Nevertheless, the actual DB convective efficiency is still under debate, which likely gives rise to uncertainties in the model spectra.

There are other possible sources of systematic uncertainties in the characterisation of helium-dominated white dwarfs. The same analysis of an individual star using independent data sets, even if obtained with the same telescope/instrument, can yield to significantly discrepant results \citep[see e.g.][for spectroscopic and photometric comparisons, respectively]{vossetal07-1,izquierdo20}. This may be partially due to the different instrument setups, which ultimately differ in their spectral ranges and resolutions, the accuracy of the flux calibrations, the atmospheric conditions, and/or the signal-to-noise ratio (SNR) of the data.




An appropriate choice of the grids of synthetic spectra is essential too, since the structure of the photosphere depends on its chemical composition. This is a difficult task when analysing large samples of white dwarfs by means of parallaxes and archival photometry \citep[see e.g.][]{gentile19,gentile21}, where the use of canonical model spectra (pure H or He photospheres) may neglect possible traces of hydrogen, helium or metals. In fact, about 75 per cent of DB white dwarfs do show traces of hydrogen \citep[thus becoming DBAs since the A accounts for the presence on hydrogen;][]{koester15}, whose origin is attributed to the convective dilution and convective mixing mechanisms \citep{strittmatter71, cunningham20}, or to accretion from external sources \citep{macdonald91,gentile17}. Even a relatively small hydrogen abundance, that may go unnoticed depending on the spectral resolution, the SNR and the wavelength range of the observed spectra, may have an effect on the measurements, leading to an incorrect determination of the white dwarf photospheric parameters.



\begin{figure}
\centering
\includegraphics[width=0.46\textwidth]{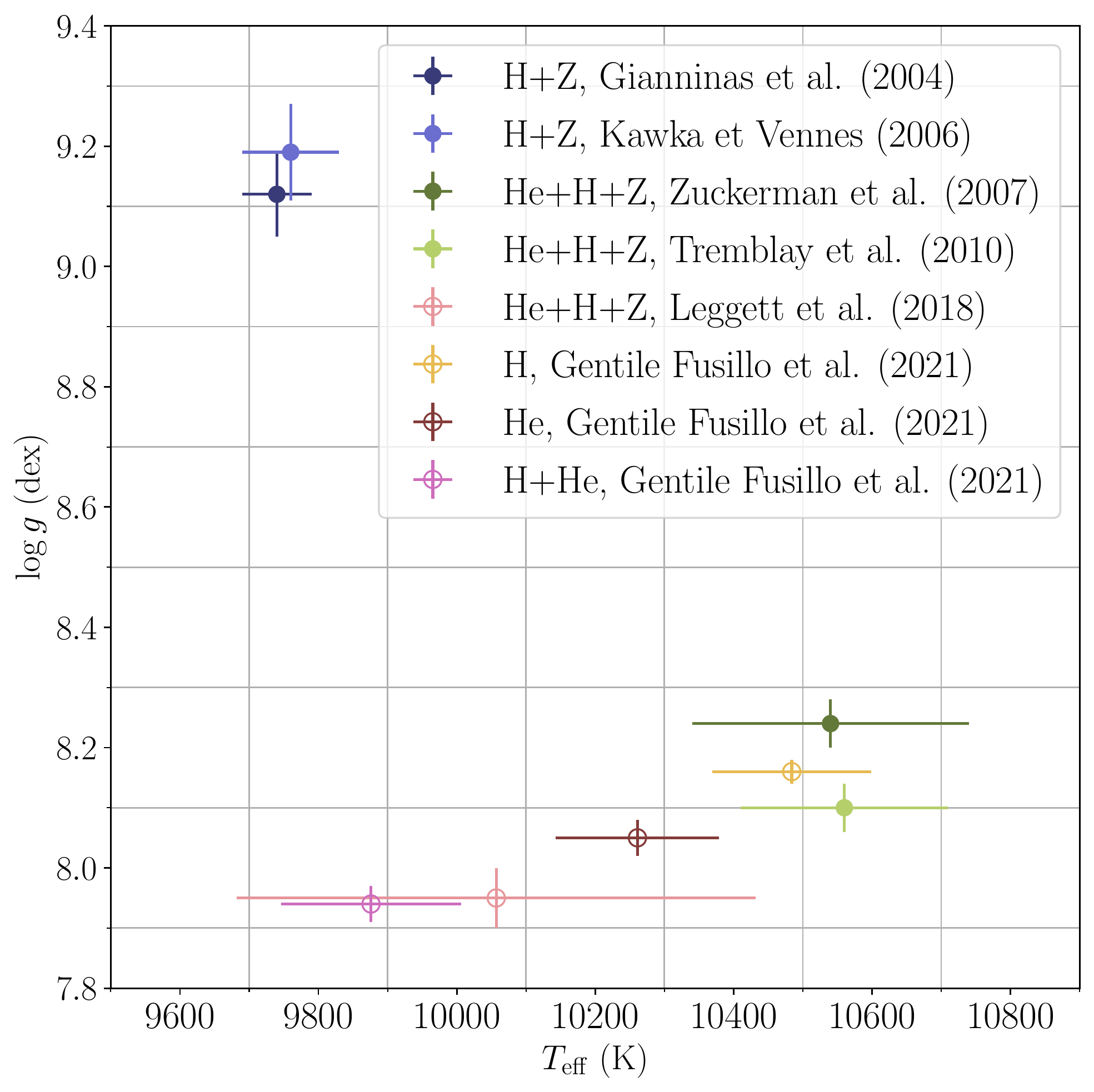}
\caption{Atmospheric parameters of the helium-dominated white dwarf GD\,362 as derived from spectroscopic (filled markers) and photometric (void markers) modellings by different authors, employing models with the chemical compositions displayed in the legend. \citet{gianninas04} and \citet{kawka06} fit spectroscopic data with H+Z model spectra (no He), while \citet{zuckerman07} and \citet{giammichele12} used a He+H+Z model grid. \citet{leggett18} performed a photometric modelling using He+H+Z models, while \citet{gentile21} fit the \textit{Gaia} DR3 photometry with H, He and H+He models. This is an extreme example of the very first studies misinterpreting the strong Balmer absorption lines in GD\,362 as characteristic of a hydrogen-dominated atmosphere. As such, it illustrates the strong dependence of the atmospheric parameters determined from either spectroscopy or photometry on the detailed assumptions about the atmospheric chemical composition.}
\label{fig:gd362}
\end{figure}

Besides some amount of hydrogen, about $10$ per cent of DB white dwarfs also contain traces of metals \citep{koester15}, which furthers the complexity of their atmospheric structure. An iconic example is the metal-polluted GD\,362, which was initially classified as a DAZ white dwarf \citep[the Z denotes the presence of metals;][]{gianninas04,kawka05}, and only later was it found to have a helium-dominated atmosphere \citep{zuckerman07}. Correspondingly, the atmospheric parameters derived using the different chemical compositions diverge dramatically (Fig.~\ref{fig:gd362}).

Whereas GD\,362 is certainly an extreme example, the presence of metals in the photospheres of white dwarfs has often been neglected, maybe due to low spectral resolution and/or SNR observing data, that make the identification of metal lines, and thus the estimate of their abundances, harder. Metals change the atmospheric structure: they contribute to both the opacity and the ionisation balance, as the ionisation of metals occurs at relatively low temperatures, which injects free electrons into the atmosphere. Metal blanketing has a considerable effect on the slope of the continuum due to the numerous strong metal lines in the ultraviolet (UV), which block the outgoing flux in that spectral range. This results in an energy redistribution towards more transparent regions that causes a back-warming effect. As a consequence, the structure of the photosphere is altered, and so is the emitted SED. Hence, to obtain reliable estimates of the \Teff\ and \logg\ of a metal-polluted white dwarf, a realistic treatment of the full chemical composition of its photosphere is needed.


Given the challenges that characterising helium-dominated white dwarfs pose, and the discrepancies encountered in the literature for the same objects \citep[see e.g.][]{tremblayetal19-2}, it is clear that systematic uncertainties intrinsic to each modelling approach must be explored and assessed. In this paper, we present spectroscopic and photometric modellings of a sample of 13 helium-dominated white dwarfs with traces of hydrogen and metals, which allow us to estimate the systematic uncertainties inherent to each technique. 

In what follows, we provide an overview on the most important analyses of DB and DBA white dwarfs to date, where attempts to measure the systematic uncertainties were reported. The details of the model atmospheres, such as the use of different broadening mechanisms, the convective efficiency and the addition of different blanketing sources, fitting procedures, and discrepancies between different studies are presented.

\section{Past studies of DB and DBA white dwarfs}
\label{sec:past_studies}

The first analysis of a large sample of DB white dwarfs was reported in \cite{beauchamp96}, who reviewed previous studies of about 80 DBs and DBAs, and secured high quality spectra of the objects. They compared the \Teff\ derived from UV and optical spectra for 25 of them and found an average standard deviation around the 1:1 correspondence of $1600$\,K (random scatter). They adopted the ML2 version, which has also been employed in all the remaining studies cited in the present paper, but they did not supply any further details of the model atmospheres.

The work by \citet{vossetal07-1} was a milestone in the understanding of the nature and evolution of DBs and DBAs. They used the spectra of 71 white dwarfs with helium-dominated photospheres, observed by the ESO Supernova Ia Progenitor Survey \citep[SPY;][]{napiwotzki03}, to estimate their \Teff, \logg\ and \htohe\ by fitting the absorption-line profiles with helium-dominated model atmospheres with different amounts of hydrogen. These authors adopted the ML2 with a convective efficiency of $\alpha=0.6$, included blanketing effects due to the presence of hydrogen and helium when appropriate, and implemented the treatment of the van der Waals line broadening mechanism \citep[see][for further detail]{finley97,koesteretal05-1}. A comparison of their derived atmospheric parameters with those reported in \cite{beauchamp99}, \cite{friedrich00} and \cite{castanheira06} revealed $\simeq \pm 10$\,per cent differences in \Teff\ and an average of $\pm 0.15$\,dex in \logg. \citeauthor{vossetal07-1} attributed these discrepancies to the different atmospheric models used, the fitting procedures and the SNR of the spectra. In addition, they did the same analysis with independent sets of 22 SPY spectra and found $\left< \frac{\Delta \Teff}{\Teff}\right> = 0.0203$, $\left< \Delta \logg \right> = 0.06$\,dex and  $\left< \Delta \htohe \right> = 0.02$\,dex\footnote{Throughout this paper, the angle brackets denote the mean.}. These revealed that the statistical uncertainties quoted for the derived atmospheric parameters of white dwarfs were unrealistically small (the formal uncertainties from the $\chi^2$ routine they used amounted to a few times 10\,K), and that the true uncertainties are likely dominated by systematic effects. 


A statistical analysis of 108 spectra of helium-atmosphere white dwarfs, of which 44~per cent are DBAs, was published by \cite{bergeronetal11-1}. They computed the model atmospheres with the code described in \cite{tremblay09} and tested various convective efficiencies, accounting for the different element opacities and including the van der Waals line-broadening treatment. \citet{bergeronetal11-1} derived \Teff, \logg\ and \htohe\ by fitting the absorption-line profiles and demonstrated that the smoothest and most uniform distribution of their sample in terms of \Teff\ and \logg\ (as predicted by the white dwarf luminosity function) is obtained for a convective efficiency of $\alpha=1.25$, a value that has been adopted as the canonical choice in many published DB analyses. They assessed the systematic uncertainties due to flux calibration by comparing the atmospheric parameters of 28 DBs with multiple spectra, finding $\left< \frac{\Delta \Teff}{\Teff}\right> = 0.023$ and $\left< \Delta \logg \right> = 0.052\, \rm dex$. A comparison of their atmospheric parameters with those of \citet{vossetal07-1} revealed that \citeauthor{bergeronetal11-1}'s \logg\ values are larger by 0.15\,dex and that a random scatter of $\simeq 3900$\,K in the \Teff\ between the two data sets exists for $\Teff \leq 19\,000$\,K \citep[see fig. 19 in][]{bergeronetal11-1}. 


Using Sloan Digital Sky Survey (SDSS) spectroscopy and photometry of 1107 DBs, \cite{koester15} increased the number of characterised DBs by a factor of 10. They found a DBA fraction of 32~per cent, which increases to 75~per cent when restricting the analysis to spectra with $\mathrm{SNR}>40$. The synthetic spectra used in this study were computed with the code of \cite{koester10-1} and to determine the \Teff, \logg\ and \htohe\ they applied an iterative technique: the photometric data are initially used to estimate the \Teff\ with \logg\ fixed at 8.0\,dex (note that no prior information about the distances was available), which serves to distinguish between the spectroscopic \Teff\ hot and cool solutions. Then, the absorption-line profiles are fitted with pure helium model spectra to derive the \Teff\ and \logg, which are subsequently fixed to measure the \htohe. This procedure is repeated until convergence is obtained. In their study, \citeauthor{koester15} carried out an assessment of their parameter uncertainties using 149 stars with multiple spectra, which resulted in random average differences of 3.1~per cent, 0.12\,dex and 0.18\,dex for \Teff, \logg\ and \htohe, respectively. A comparison of the stars in common with the ones in \cite{bergeronetal11-1} yields average systematic differences of $+1.3$~per cent and $+0.095$\,dex in \Teff\ and \logg, respectively (both parameters being larger in average for the \citeauthor{koester15}'s sample), with mean dispersions of 4.6~per cent and 0.073\,dex.

\cite{tremblayetal19-2} modelled the \textit{Gaia} DR2 photometric data of 521 DBs that had already been spectroscopically characterised \citep{koester15,rolland18}, and compared the resulting atmospheric parameters with the published spectroscopic results. \citeauthor{tremblayetal19-2} used an updated version of the code described in \cite{tremblay09} to compute 1D pure helium model atmospheres. They fit the photometric points, previously unreddened using the 2D dust reddening maps of \cite{schlafly+finkbeiner11-1}, with \Teff\ and the white dwarf radius, \Rwd, as free parameters. To compare the results produced by both fitting techniques, they first derived the \textit{spectroscopic} parallaxes from the atmospheric parameters provided by the spectroscopic technique, the \textit{Gaia} $G$-band apparent magnitude and the theoretical mass-radius relation of \citet{fontaineetal01-1}. They observed reasonable agreement (within 2-$\sigma$) with the \textit{Gaia} parallaxes for $\Teff \ge 14\,000$\,K in the \citet{rolland18} and \citet{koester15} DB sample. However, for cooler white dwarfs larger differences became apparent, again likely caused by problems with the neutral helium line broadening. They also compared the spectroscopic and photometric \Teff\ and \logg\ and found that the fits to the \textit{Gaia} photometry systematically provide lower \Teff\ and randomly scattered differences in the \logg. This points once more to an inadequate treatment of the van der Waals broadening. They concluded that the photometric technique, and in particular the use of \textit{Gaia} photometry and parallaxes, can give solid atmospheric parameters and is, in particular, more reliable in constraining the \logg\ for the cooler DBs ($\Teff \leq 14000$\,K) as compared to the spectroscopic method.

A similar study was presented by \cite{genest-beaulieu19-2}, who also used the \textit{Gaia} DR2 parallaxes and compared the photometric and spectroscopic \Teff, \logg, \htohe, $\log{\mathrm{(Ca/He)}}$, the white dwarf mass, \Mwd, and \Rwd\ of more than 1600 DBs from the SDSS. They adopted the grid of synthetic models of \cite{bergeronetal11-1}, but used an improved version of the van der Waals broadening. The photometric and spectroscopic techniques were carried out as follows: (1) the \Teff\ and the solid angle, $\pi(\Rwd/D)^2$, were obtained from fitting the observed SDSS photometry points (unreddened with the parametrisation described in \citealt{harris06}) and the distance $D$ derived from \textit{Gaia} DR2; (2) the \Teff, \logg\ and \htohe\ were derived by fitting the continuum-normalised absorption lines with synthetic profiles. The results show statistical errors of 10~per cent in the photometric \Teff\ and $\left< \sigma \Mwd \right> = 0.341\,\Msun$, while the uncertainties in the spectroscopic parameters are of $ 4.4$~per cent for \Teff, $\left< \sigma \logg \right> = 0.263$\,dex, $\left< \sigma \htohe \right> = 0.486$\,dex and $\left< \sigma \Mwd \right> = 0.156\,\Msun$. The authors also estimated the uncertainties in the spectroscopic parameters by repeating the same procedure for 49 stars with multiple spectra, resulting in $\left< \Delta \Teff/\Teff \right> = 0.024$, $\left< \Delta \logg \right> = 0.152$\,dex, $\left< \Delta \Mwd \right> = 0.086\,\Msun$ and $\left< \Delta \logg \right> = 0.2$\,dex. \citet{genest-beaulieu19-2} then concluded that both techniques yield the \Teff\ with similar accuracy, but stated that the photometric method is better suited for white dwarf mass determinations. 

The last effort to assess the systematic effects in the characterisation of DB atmospheres was carried out by \citet{cukanovaite21}, who presented a thorough study on the input microphysics, such as van der Waals line broadening or non-ideal effects, and convection models used in the computation of synthetic spectra. They demonstrated the need for 3D spectroscopic corrections\footnote{The simplistic ML theory employed in the treatment of convective energy transport was related to the DA high-\logg\ problem \citep{tremblay13b}. This issue was overcome by the use of 3D radiation-hydrodynamical models, which treat convection using first principles
and do not depend on any free parameters as the ML approximation.} by using the cross-matched DB and DBA sample of  \citeauthor{genest-beaulieu19-2} with the \textit{Gaia} DR2 white dwarf catalogue \citep{gentile19}, removing all spectra with $\mathrm{SNR} < 20$, which resulted in 126 DB and 402 DBA white dwarfs. In particular, they presented significant corrections for the spectroscopically derived \logg\ in the \Teff\ range where the high-\logg\ problem is found (DBs with $\Teff \leq 15\,000$\,K). Although these corrections represent a starting point towards solving the issues with the synthetic DB models due to their superior input physics, they have not yet accounted for the dramatic differences in the photospheric parameters of DBs derived from photometry and spectroscopy \citep[see e.g. figs. 9, 10, 14 and 15 in][]{cukanovaite21}.

\begin{table*}
\caption{White dwarf sample, including the WD\,J names from \citet{gentile19}, the short names used in this paper, the \textit{Gaia} $G$ magnitude, the distance $D$ of the source \citep[derived as $D$~(pc)~$=1000/\varpi$, being $\varpi$ the parallax in mas;][]{gaia-edr3}, the spectral classification of \citet{gentile15} (in italics) and the updated one based on our X-shooter spectra, the log of the X-Shooter spectroscopy and the signal-to-noise ratio of the UVB and VIS X-shooter, BOSS and SDSS spectra (the last four columns). }
\setlength{\tabcolsep}{1.2ex}
\label{tab:obslog}
\centering
\begin{tabular}{lccccccccccc}
\hline
Star & Short name  & \textit{Gaia G} & $D$ & \multicolumn{2}{c}{Spectral} & \multicolumn{4}{c}{X-shooter observations}  & \multicolumn{2}{c}{SDSS} \\
     &  &  & (pc) & \multicolumn{2}{c}{classification} & Date & Exposure time (s) & UVB & VIS & BOSS & SDSS\\ 
\hline
WD\,J003003.23$+$152629.34  & 0030$+$1526 & 17.6 & $175 \pm 3$  & \textit{DABZ} & DBAZ & 2018-07-11 & 2x(1250/1220/1300) & 54.9 & 40.0 & - & 29.1 \\
WD\,J025934.98$-$072134.29	& 0259$-$0721 & 18.2 & $222 \pm 7$  & \textit{DBZ} & DBAZ &  2018-01-12 & 4x(1221/1255/1298) & 48.0 & 40.9 & - & 19.5 \\
WD\,J082708.67$+$173120.52	& 0827$+$1731 & 17.8 & $127 \pm 2$  & \textit{DAZ} & DABZ &  2018-01-12 & 4x(1221/1255/1298) & 47.9 & 48.4 & 38.4 & 22.8 \\
WD\,J085934.18$+$112309.46	& 0859$+$1123 & 19.1 & $340 \pm 28$ & \textit{DABZ} & DBAZ & 2018-01-10 & 5x(1221/1255/1298) & 45.2 & 30.3 & 20.1 & - \\
WD\,J093031.00$+$061852.93	& 0930$+$0618 & 17.9 & $227 \pm 7$  & \textit{DABZ} & DBAZ & 2018-01-12 & 4x(1221/1255/1298) & 36.6 & 30.8 & - & 36.0 \\
WD\,J094431.28$-$003933.75	& 0944$-$0039 & 17.8 & $160 \pm 3$  & \textit{DBZ} & DBAZ &  2018-01-11 & 4x(1221/1255/1298) & 54.5 & 49.7 & 44.0 & 26.1 \\
WD\,J095854.96$+$055021.50  & 0958$+$0550 & 17.8 & $182 \pm 6$  & \textit{DBZ} & DBAZ &  2018-01-12 & 4x(1221/1255/1298)  & 48.4 & 44.7 & 27.0 & -\\
WD\,J101347.13$+$025913.82  & 1013$+$0259 & 18.2 & $202 \pm 9$  & \textit{DABZ} & DABZ & 2018-01-10 & 4x(1221/1255/1298)  & 48.3 & 42.6 & 25.2 & 27.1\\
WD\,J110957.82$+$131828.07	& 1109$+$1318 & 18.7 & $298 \pm 20$ & \textit{DABZ} & DBAZ & 2018-01-11 & 4x(1221/1255/1298) & 37.0 & 27.8 & 20.2 & 13.7\\
WD\,J135933.24$-$021715.16	& 1359$-$0217 & 17.8 & $217 \pm 6$  & \textit{DABZ} & DBAZ & 2018-07-12 & 2x(1250/1220/1300) & 41.3 & 31.5 & 43.1 & 24.5\\
WD\,J151642.97$-$004042.50	& 1516$-$0040 & 17.3 & $143 \pm 2$  & \textit{DABZ} & DBAZ & 2018-07-10 & 4x(1200/1200/1200) & 60.0 & 60.8 & 43.3 & -\\
WD\,J162703.34$+$172327.59	& 1627$+$1723 & 18.6 & $278 \pm 13$ & \textit{DBZ} & DBAZ &  2018-07-12 & 4x(1450/1420/1450) & 33.0 & 16.3 & 28.5 & 12.9\\
WD\,J232404.70$-$001813.01	& 2324$-$0018 & 18.9 & $329 \pm 33$ & \textit{DABZ} & DABZ & 2018-07-10 & 5x(1250/1220/1300) & 45.5 & 36.0 & 22.9 & -\\\hline
\end{tabular}
\end{table*}

\begin{figure*}
    \centering
    \includegraphics[width=0.95\textwidth]{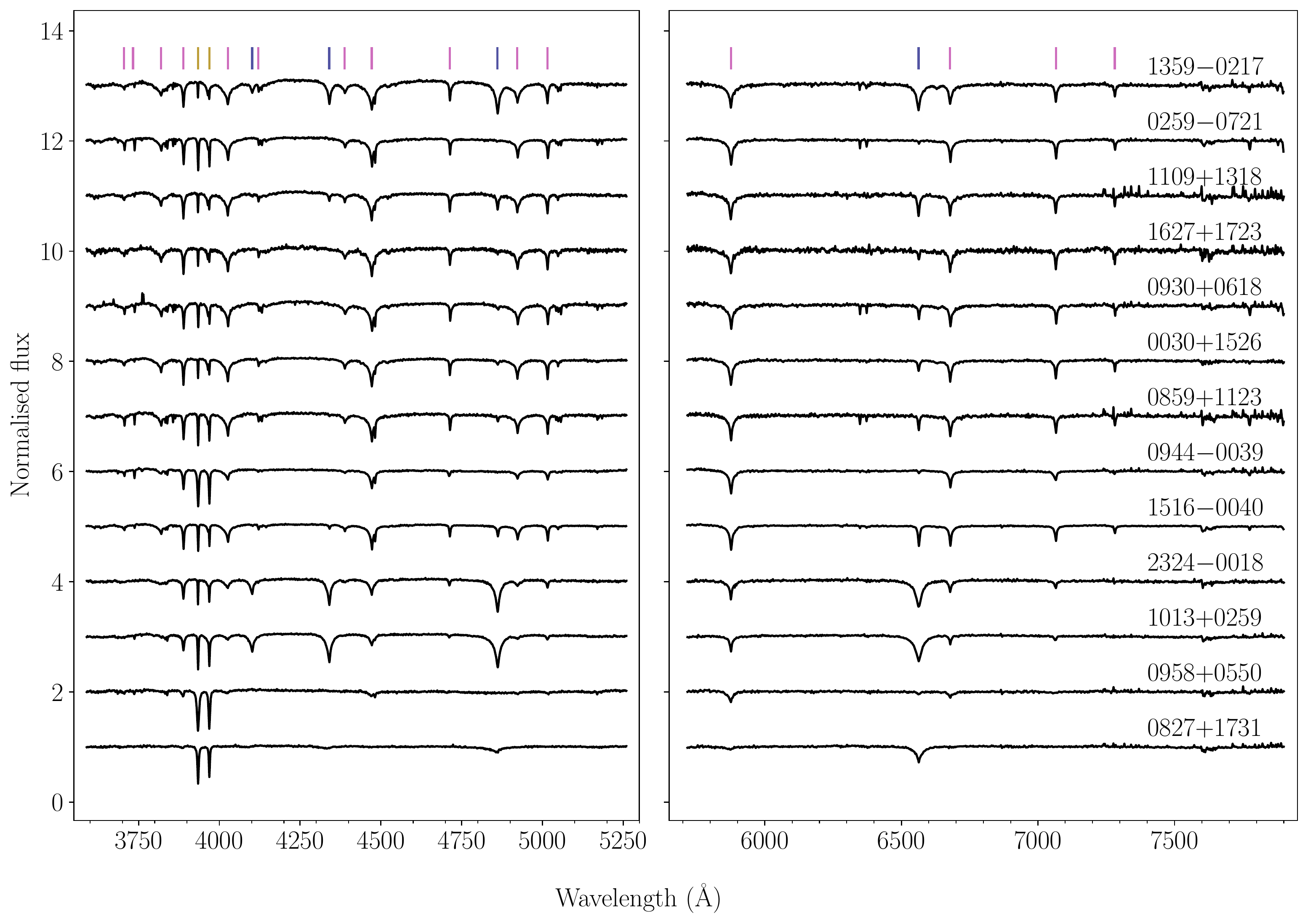}
    \caption{Normalised X-shooter spectra of the 13 metal-polluted white dwarfs. Hydrogen, helium and \Ion{Ca}{ii} H and K absorption lines are marked with pink, blue and yellow vertical
lines, respectively. The effective temperature increases from bottom to top. The spectra are offset vertically for display purposes.}
    \label{fig:XS_spectra}
\end{figure*}

\newcommand{\mc}[1]{\multicolumn{2}{c}{#1}}

\begin{table*}
\label{tab:phot_data}
\caption{Photometry of the 13 white dwarfs. We list the point spread function (PSF) SDSS magnitudes \citep{SDSSpass}, the mean PSF Pan--STARRS1 magnitudes along with their standard deviations \citep[PS1;][]{PanSTARRS2} and the broad-band photometry of \textit{Gaia} eDR3 \citep{gaia-edr3}.}
\begin{tabular}{lccccccr}
\hline
Star  &  $u$ & $g$ & $r$ & $i$ & $z$ &     & SDSS \\
      &      & $g$ & $r$ & $i$ & $z$ & $y$ & PS1 \\
            & \mc{$G_\mathrm{BP}$} 
            & \mc{$G$} 
            & \mc{$G_\mathrm{RP}$} & \textit{Gaia} \\
\hline
0030+1526 & $17.317 \pm 0.016$ & $17.431 \pm 0.022$ & $17.742 \pm 0.014$ & $17.952 \pm 0.017$ & $18.241\pm0.025$ \\
&& $17.481\pm0.017$ & $17.746\pm0.017$ & $17.981\pm0.014$ & $18.193\pm0.027$ & $18.317\pm0.047$ \\
& \mc{$17.529\pm0.006$} & \mc{$17.5752\pm0.0029$} & \mc{$17.6731\pm0.0143$} \\
\hline
0259$-$0721 & $18.031\pm0.018$ & $18.062\pm0.014$ & $18.326\pm0.015$ & $18.552\pm0.018$ & $18.823\pm0.054$ \\
& & $18.093\pm0.022$ & $18.328\pm0.019$ & $18.565\pm0.048$ & $18.784\pm0.041$ & $18.921\pm0.070$\\
& \mc{$18.1484\pm0.0139$} & \mc{$18.1763\pm0.0035$} & \mc{$18.2509\pm0.0491$}\\
\hline
0827+1731 & $17.848\pm0.019$ & $17.800\pm0.018$ & $17.964\pm0.015$ & $18.143\pm0.016$ & $18.324\pm0.028$\\
& & $17.820\pm0.020$ & $17.959\pm0.023$ & $18.153\pm0.022$ & $18.337\pm0.072$ & $18.438\pm0.054$\\
& \mc{$17.8475\pm0.0102$} & \mc{$17.8405\pm0.0030$} & \mc{$17.8321\pm0.0159$}\\
\hline
0859+1123 & $18.878\pm0.042$ & $18.979\pm0.017$ & $19.213\pm0.020$ & $19.555\pm0.036$ & $19.775\pm0.073$\\
&  & $18.994\pm0.033$ & $19.255\pm0.066$ & $19.523\pm0.047$  & $19.722\pm0.088$ & $19.790\pm0.226$\\
& \mc{$19.0889\pm0.0224$} & \mc{$19.0886\pm0.0035$} & \mc{$19.1602\pm0.0460$}\\
\hline
0930+0618& $17.775\pm 0.017$ & $17.838\pm 0.019$ & $18.135\pm 0.016$ & $18.380\pm 0.022$ & $18.765\pm 0.041$\\
& & $17.910\pm 0.019$ & $18.181\pm 0.018$ & $18.414\pm 0.034$ & $18.658\pm 0.041$ & $18.800\pm 0.085$\\
& \mc{$18.0020\pm0.0030$} & \mc{$17.9364\pm0.0115$} & \mc{$18.1420\pm0.0201$}\\
\hline
0944$-$0039 & $17.717\pm0.014$ & $17.749\pm0.015$ & $17.973\pm0.019$ & $18.187\pm0.019$ & $18.407\pm0.028$\\
& & $17.783 \pm0.034$ & $18.005 \pm0.024$ & $18.212 \pm0.045$ & $18.424 \pm0.029$ & $18.551 \pm0.123$\\
& \mc{$17.8396\pm0.0097$} & \mc{$17.8452\pm0.0029$} & \mc{$17.9183\pm0.0183$}\\
\hline
0958+0550 & $18.293\pm0.022$ & $18.215\pm0.015$ & $18.385\pm0.018$ & $18.524\pm0.021$ & $18.763\pm0.033$\\
& & $18.222 \pm0.025$ & $18.391 \pm0.022$ & $18.549 \pm0.034$ & $18.743 \pm0.032$ & $18.851 \pm0.143$\\
& \mc{$18.2631\pm0.0033$} & \mc{$18.2750\pm0.0281$} & \mc{$18.2012\pm0.0172$}\\
\hline
1013+0259 & $18.064\pm0.022$ & $18.146\pm0.018$ & $18.353\pm0.020$ & $18.546\pm0.018$ & $18.748\pm0.043$\\
& & $18.144\pm0.011$ & $18.361\pm0.020$ & $18.560\pm0.030$ & $18.773\pm0.041$ & $18.892\pm0.101$\\
& \mc{$18.1782\pm0.0157$} & \mc{$18.2165\pm0.0034$} & \mc{$18.1847\pm0.0468$}\\
\hline
1109+1318 & $18.493\pm0.022$ & $18.622\pm0.026$ & $18.902\pm0.021$ & $19.145\pm0.032$ & $19.357\pm0.059$\\
& & $18.625\pm0.017$ & $18.909\pm0.034$ & $19.148 \pm0.064$ & $19.388 \pm0.049$ & $19.490 \pm0.137$\\
& \mc{$18.7296\pm0.0037$} & \mc{$18.7042\pm0.0341$} & \mc{$18.9108\pm0.0624$}\\
\hline
1359$-$0217 & $17.664\pm0.019$ & $17.724\pm0.022$ & $17.993\pm0.014$ & $18.234\pm0.019$ & $18.481\pm0.036$\\
& & $17.758 \pm0.019$ & $18.007 \pm0.017$ & $18.238 \pm0.017$ & $18.464 \pm0.018$ & $18.601 \pm0.099$\\
& \mc{$17.8120\pm0.0146$} & \mc{$17.8457\pm0.0031$} & \mc{$18.0034\pm0.0257$}\\
\hline
1516$-$0040 & $17.152\pm0.015$ & $17.209\pm0.016$ & $17.454\pm0.014$ & $17.636\pm0.013$ & $17.899\pm0.023$\\
& & $17.242\pm0.019$  & $17.454\pm0.016$  & $17.658\pm0.022$  & $17.849\pm0.032$  & $18.001\pm0.056$\\
& \mc{$17.2784\pm0.0106$} & \mc{$17.3047\pm0.0031$} & \mc{$17.3011\pm0.0208$}\\
\hline
1627+1723 & $18.455\pm0.021$ & $18.468\pm0.017$ & $18.780\pm0.015$ & $19.027\pm0.018$ & $19.253\pm0.049$\\
& & $18.531\pm0.028$ & $18.784\pm0.043$ & $19.042\pm0.051$ & $19.260\pm0.075$ & $19.358\pm0.134$\\
& \mc{$18.5881\pm0.0169$} & \mc{$18.6155\pm0.0032$} & \mc{$18.7338\pm0.0256$}\\
\hline
2324$-$0018 & $18.808\pm0.019$ & $18.842\pm0.020$ & $19.017\pm0.019$ & $19.229\pm0.021$ & $19.387\pm0.050$\\
& & $18.857 \pm 0.028$  & $19.057 \pm 0.030$  & $19.246 \pm 0.058$  & $19.488 \pm 0.042$  & $19.476 \pm 0.134$\\
& \mc{$18.9313\pm0.0222$} & \mc{$18.9126\pm0.0038$} & \mc{$18.9019\pm0.0451$}\\
\hline
\end{tabular}
\label{tab:phot_points_alternative}
\end{table*}

\section{The white dwarf sample}
\label{sec:wd_13sample}

\citet{gentile15} presented the spectral classification of 8701 white dwarfs brighter than $g=19$ with at least one SDSS DR10 spectrum. We visually inspected all the spectra flagged by \citeauthor{gentile15} as metal-contaminated  and selected 13 stars that (1) had moderately strong \Ion{Ca}{ii} H and K absorption lines, and (2) were either confirmed, via the detection of helium absorption lines, or suspected helium-atmosphere white dwarfs (because of shallow and asymmetric Balmer line profiles). The selected white dwarfs are presented in Table\,\ref{tab:obslog}.


Additionally, we obtained X-shooter spectra for each target and collected the available SDSS and Pan-STARRS1 (PS1) photometry, and \textit{Gaia} eDR3 astrometry plus photometry for all of them (Fig.\,\ref{fig:XS_spectra} and Table\,\ref{tab:phot_data}).

\subsection{SDSS spectroscopy}

As mentioned above, our target selection is based on SDSS DR10. However, SDSS sometimes reobserves the same object, so we inspected the DR16 database \citep{ahumada20} and retrieved all available spectra of our 13 targets. Several white dwarfs were observed with both the original SDSS spectrograph (3800$-$9200\,\AA\ wavelength range and $R\simeq1850-2200$ spectral resolution) and the BOSS spectrograph (3600$-$10\,400\,\AA, $R\simeq1560-2650$; \citealt{smee13}; see Table\,\ref{tab:obslog}).

\begin{figure*}
    \centering
    \includegraphics[width=0.8\textwidth]{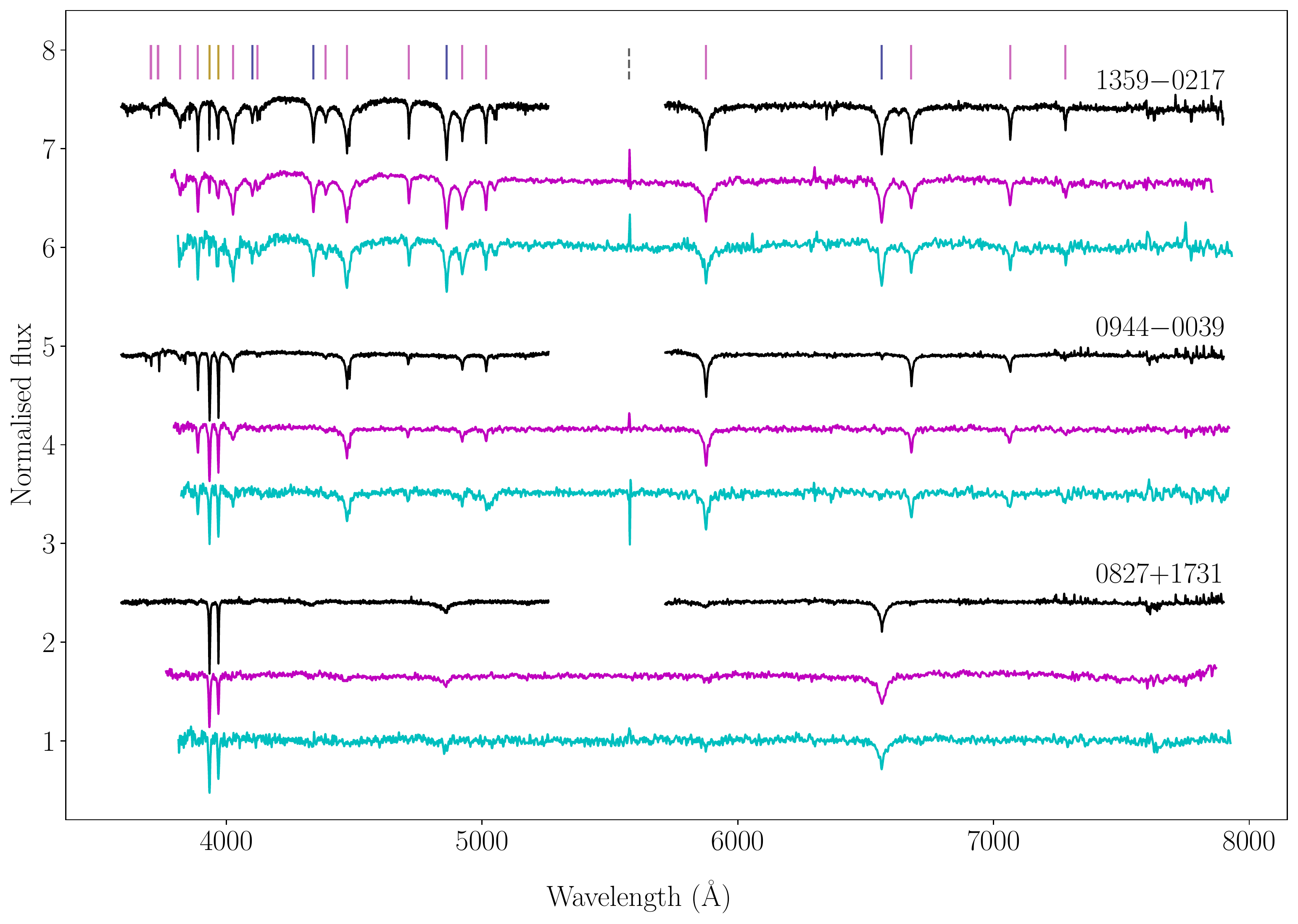}
    \caption{Comparison between the UVB+VIS X-shooter (spectral resolution $R=5400$, $8900$; black), BOSS ($R\simeq1850-2200$; magenta) and SDSS ($R\simeq1850-2200$; cyan) spectra of three white dwarfs in our sample. Hydrogen, helium and \Ion{Ca}{ii} H and K absorption lines are marked with blue, pink and yellow vertical lines, respectively. The effective temperature increases from bottom to top. The spectra are offset vertically for display purposes. We note that the spikes in the BOSS and SDSS spectra (marked with a dashed vertical grey line) are artefacts derived from the data calibrations.}
    \label{fig:SDSS_XS_spectra}
\end{figure*}

\subsection{VLT/X-shooter spectroscopy}

We obtained intermediate resolution spectroscopy of the 13 white dwarfs using the X-shooter spectrograph \citep{vernetetal11-1} mounted on the UT2 Kueyen telescope of the 8.2-m Very Large Telescope at Cerro Paranal, Chile, in January and July 2018 (ESO programmes 0100.C$-$0500 and 0101.C$-$0646). X-shooter is a three arm \'echelle spectrograph that simultaneously covers the ultraviolet-blue (UVB, $3000-5600$\,\AA), visible (VIS, $5500-10\,200$\,\AA) and near-infrared (NIR, $10\,200-24\,800$\,\AA) wavelength ranges. We used slit widths of 1.0 (UVB), 0.9 (VIS) and 0.9\,arcsec (NIR) to achieve spectral resolutions  $R=5400$, 8900 and 5600, respectively. However, the NIR spectra were of insufficient SNR for a quantitative analysis and were discarded. Depending on the target brightness and the observing conditions, we obtained between two and six exposures per star. Details on the observations are given in Table\,\ref{tab:obslog}, and a comparison between the X-shooter and SDSS/BOSS spectra for three white dwarfs of our sample is shown in Fig.~\ref{fig:SDSS_XS_spectra}.

We reduced the data within the ESO \textsc{Reflex} environment \citep{freudling13}. In brief, we removed the bias level and dark current, flat-fielded the images, identified and traced the \'echelle orders and established a dispersion solution. Then, we corrected for the instrument response  and  atmospheric  extinction using observations of a spectrophotometric standard star observed with the same instrumental setup, merged the individual orders and applied a barycentric velocity correction to the wavelength scale. Telluric absorptions were corrected for using \textsc{molecfit} \citep{kausch15,smette15}. Finally, we computed the UVB and VIS averages from the individual spectra of each white dwarf using the inverse of their variance as weights.

The X-shooter spectra of the 13 white dwarfs (Fig.\,\ref{fig:XS_spectra}) display at least the \Ion{Ca}{ii} H and K lines, H$\alpha$, and different helium absorption lines. Particular cases are 0827+1731, where the low $\Teff \approx 10500$\,K of the white dwarf only allows a really shallow helium line (\hel{i}{5876}) to be identified in addition to H$\alpha$ and H$\beta$ and a few shallow \Ion{Ti}{ii} absorption lines (in the $3300-3400$\,\AA\ range), and $0958+0550$, whose spectra display He and shallow metallic lines of Mg, Ca, Ti, Cr, Mn or Fe, but only a hint of H$\alpha$ due to the small hydrogen abundance.

\section{Methodology}
\label{sec:method_DBAZ13}

In order to explore the underlying systematic uncertainties in the determination of the atmospheric parameters of helium-dominated white dwarfs with traces of hydrogen and metals, we tested the spectroscopic and photometric techniques using the different data sets available for each star and synthetic spectra computed for several chemical compositions.

The spectroscopic analyses were performed using at least two different spectra per star: SDSS/BOSS and X-shooter (a few targets have both SDSS and BOSS spectra, in which case we also tested the level of agreement between those two data sets). For the photometric approach we used three catalogues: SDSS, PS1 and \textit{Gaia} eDR3.

For both techniques we used model spectra with three different chemical compositions: (1)~pure He, (2)~He with variable H contents and (3)~He with variable H and Z contents. We first employed (1)~pure He atmosphere models, and hence the spectroscopic method only considered helium absorption lines. This approach was historically applied for white dwarfs for which only a limited amount of spectroscopic information is available, e.g. H$\alpha$ is not covered at all or at poor SNR. We then fitted the spectroscopic data with (2)~mixed H/He atmosphere models (He+H henceforth) that were hydrogen-blanketed, now including \htohe\ as the third free parameter after \Teff\ and \logg, and also using the Balmer lines present in the observed spectra. Notice that we fix the \htohe\ at the spectroscopic value to perform these photometric fits. The final approach was performed with (3)~mixed H/He + metals atmosphere models (hydrogen- and metal-blanketed). These synthetic grids, He+H+Z henceforth, which are computed individually for each white dwarf (see Fig.\,\ref{fig:flow_chart_metals}), take into account the relative abundances of the metals estimated from the X-shooter spectra\footnote{Reliable metal abundances cannot be constrained from the SDSS/BOSS spectra due to their low SNR and resolution, which is insufficient to properly sample the narrow metallic lines. These have an average equivalent width of about 0.6 \AA, significantly smaller than the $\simeq 4$-\AA\ resolution of the BOSS/SDSS spectra.}. As in the case of the He+H analysis, the spectroscopic technique was performed first, in order to estimate the chemical composition [$\htohe + \ztohe$] of each star, which is then fixed in the photometric fits.


\subsection{Model atmospheres and fitting procedure}

We used the latest version of the \cite{koester10-1} code to generate all the synthetic model spectra. The substantial convection zones of helium-dominated white dwarfs were accounted for using a 1D ML prescription. In particular, we adopted the ML2 parametrisation and fixed the convective efficiency, $\alpha$. A more realistic line fitting would need 3D spectral synthesis, with a range of $\alpha$ values that describe the different spectral lines of the white dwarf \citep{cukanovaiteetal19-1}. These 3D models are still too computationally expensive and, for the scope of this paper, we are using 1D models and have fixed the convective efficiency at $\alpha=1.25$, which is the canonical and most extensively used value in the characterisation of DB white dwarfs \citep{bergeronetal11-1}.

Our pure He and He+H grids spanned $\Teff =5\,000-20\,000$\,K in steps of $250$\,K and  $\log g=7.0-9.5~ \rm dex$ in steps of 0.25\,dex. For the He+H grid we explored the \htohe\ range from $-7.0$ to $-3.0~ \rm dex $ in steps of 0.25\,dex. Notice that these two grids were computed with no metals, thus neglecting any metal line blanketing. 

\begin{figure}
\centering
\includegraphics[width=0.3\textwidth]{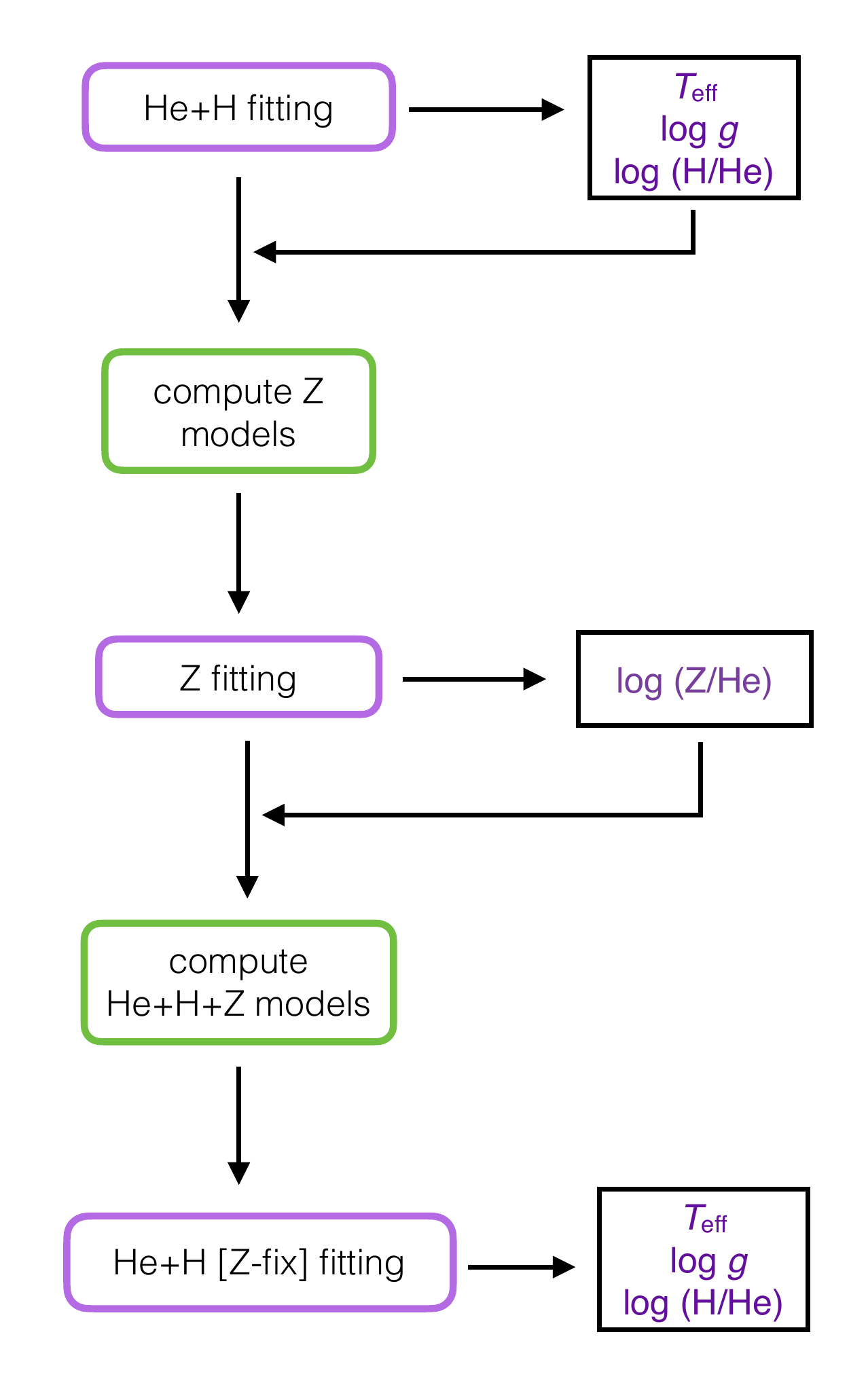}
\caption{Flow chart of the procedure used to add metals to the synthetic spectra of He+H white dwarfs.}
\label{fig:flow_chart_metals}
\end{figure}

The He+H+Z grids are computed in various steps (see the flowchart in Fig.\,\ref{fig:flow_chart_metals}). First, we performed an iterative analysis starting with a photometric fit to determine $\Teff_{\rm{phot}}$\ and $\logg_{\rm{phot}}$, with $\htohe$ fixed at $-5.0$\,dex. Then, a spectroscopic fit is performed with \logg\ fixed at $\logg_{\rm{phot}}$, which yields $\Teff_{\rm{spec}}$\ and \htohe. This \htohe\ is then used in the photometric fit and the procedure is iterated until convergence is achieved. As a result, we obtain the $\Teff_{\rm{phot}}$, $\logg_{\rm{phot}}$\footnote{We chose $\Teff_{\rm{phot}}$ because it is not affected by the dubious implementation of the resonance and van der Waals broadening in the computation of the synthetic models, and $\logg_{\rm{phot}}$ because it is well constrained by a reliable parallax estimate.} and \htohe, which we fix to compute 1D grids for \textit{each} metal identified in the X-shooter spectra of each star. The only parameter that varies throughout these 1D grids is \ztohe, and the synthetic models are centred at the Solar values and sampled in steps of $0.2$\,dex. Then, the normalised absorption lines of each metal are fitted individually to obtain the \ztohe\ relative abundances. These are then included in the computation of the He+H+Z model grid for each star. The \Teff, \logg\ and \htohe\ steps of the He+H+Z model grids are the same as used for the He+H grid, but probe a smaller parameter space centred on the He+H best-fit values obtained.

We fit the synthetic model spectra to the different data subsets using the Markov Chain Monte Carlo (MCMC) {\sc emcee} package within {\sc python} \citep{foreman13}. The parameter space was explored and the logarithmic function maximised using 16 different seeds and $10\,000$ steps per seed. We employed flat priors for all the parameters within the grid boundaries provided above, except for the \textit{Gaia} parallax $\varpi$, for which we used Gaussian priors (with a Gaussian width set to the published parallax uncertainty). 

\subsection{Spectroscopic fits}
\label{subsec:spec_fits}


We first degraded the synthetic spectra to the resolution of the observed ones (see Section~\ref{sec:wd_13sample} for details). Then, we continuum-normalised each of the relevant absorption lines in both the observed and synthetic spectra (helium, Balmer or metal lines, as appropriate) by fitting low-order polynomial functions to the surrounding continuum. Metal lines that are superimposed on helium or Balmer lines were masked out in the pure He and He+H fits. For the fits obtained with the He+H+Z models, we did not mask the narrow metal lines contained in the much broader helium or Balmer lines. However, the metal abundances were fixed at the values obtained by the 1D metal fits (see Fig.\,\ref{fig:flow_chart_metals} and Table~\ref{tab:Metal_abs_stars}).  

For all the spectroscopic fits we used the neutral helium lines $\lambda$3820, $\lambda$3889, $\lambda$4026, $\lambda$4120, $\lambda$4388, $\lambda$4471, $\lambda$4713, $\lambda$4922, $\lambda$5876, $\lambda$6678 and $\lambda$7066 (except for 0827$+$1731, see Appendix A:~\ref{sec:0827+1731} for further details). For the He+H and He+H+Z spectroscopic fits, we modelled H$\alpha$ for all the stars, and H$\beta$, H$\gamma$ and H$\delta$ when present. To obtain the estimates of the metal abundances we considered the absorption lines listed in Table~\ref{tab:metals_lines_DBAZ} that were present in the individual X-shooter spectra of each star. 

For the three chemical composition grids, \Teff\ and \logg\ were treated as free parameters, with the addition of \htohe\ when using the He+H and He+H+Z grids, exploring the parameter space with flat priors in all the cases.

\begin{table}
\small
\caption{Spectral lines used in the determination of the metal chemical abundances.}
\vspace{0.2cm}
\begin{tabularx}{0.48\textwidth}{ll}
\hline\noalign{\smallskip}
Ion & Air wavelength (\AA)\\
\noalign{\smallskip}\hline
\noalign{\smallskip}
O\,{\sc i} & 7771.94, 7774.17, 7775.39\\
Na\,{\sc i} & 5889.95 , 5895.92\\
Mg\,{\sc i} & 3829.36, 3832.30, 5167.32, 5172.68, 5183.60\\
Mg\,{\sc ii} & 3838.29, 4384.64, 4390.56, 4481.33\\
Al\,{\sc i} & 3944.01\\
Al\,{\sc ii} & 3586.56, 3587.07, 3587.45, 4663.06\\
Si\,{\sc ii} & 3853.66, 3856.02, 3862.60, 4128.07, 4130.89, 5055.98\\
Ca\,{\sc ii} & 3179.33, 3181.28, 3736.90, 3933.66, 3968.47 \\
Ti\,{\sc ii} & 3321.70, 3322.94, 3372.79, 3380.28, 3383.76, 3387.83, 3394.57 \\
Cr\,{\sc ii} & 3216.55, 3402.40, 3403.32, 3408.77, 3421.21, 3422.74, 3585.29, \\
&  3585.50, 3677.68, 3677.84\\
Mn\,{\sc ii} & 3441.98, 3460.31, 3474.04, 3474.13, 3487.90, 3495.83, 3496.81,\\
& 3497.53\\
Fe\,{\sc i} & 3190.82, 3249.50\\  
Fe\,{\sc ii} & 3192.07, 3192.91, 3193.80, 3210.45, 3213.31, 3247.18, 3247.39, \\
& 3255.87, 3258.77, 3259.05, 4233.16, 4351.76, 4583.83 \\
Ni\,{\sc i} & 3465.6, 3471.3, 3524.54 \\
\noalign{\smallskip}\hline
\end{tabularx}
\label{tab:metals_lines_DBAZ}
\end{table}

\subsection{Photometric fits}
\label{sec:phot_DBAZ13}

As a first step of the photometric fitting technique, the synthetic spectra were scaled by the solid angle subtended by the star, $\pi(\Rwd/D)^{2}$, where $D$ was derived from the \textit{Gaia} eDR3 parallax $\varpi$ \citep[in mas,][]{gaia-edr3} as $D=1000/\varpi~\rm (pc)$. We account for the interstellar extinction by reddening the synthetic spectra with the $E(B-V)$ values determined from the 3D dust map produced by \textsc{stilism}\footnote{\url{https://stilism.obspm.fr/}} using the distances. The white dwarf radii were calculated using the mass-radius relation\footnote{\url{http://www.astro.umontreal.ca/~bergeron/CoolingModels}} derived with the last evolutionary models of \cite{bedard20}. This mass-radius relation is appropriate for helium-dominated white dwarfs with C/O cores and thin hydrogen layers ($\sim 10^{-10} M_\mathrm{H}/\Mwd$, with $M_\mathrm{H}$ the mass of the H layer).

The comparison of the actual photometric data with the computed brightness from the scaled and reddened model spectra in each photometric passband was carried out in flux space. Hence, we converted the observed magnitudes into fluxes using the corresponding zero points and computed the integrated synthetic fluxes in all the filters using their transmission curves. The zero points and passbands of the SDSS, PS1 and \textit{Gaia} were obtained from the Spanish Virtual Observatory (SVO) Filter Profile Service\footnote{\url{http://svo2.cab.inta-csic.es/theory/fps/}}.

In all the photometric fits we fixed the chemical composition of the grid, i.e. the \htohe\ for the He+H grid as well as the metal abundances for the He+H+Z grid, at the best-fit spectroscopic values, since photometry alone is hardly sensitive to these two parameters. Consequently, the photometric fits have \Teff, \logg\ and $\varpi$ as free parameters\footnote{The parallax was treated as a free parameter with boundaries extending to the uncertainties published in \textit{Gaia} eDR3.} and we explore the parameter space with flat priors for the former two and a Gaussian prior for the latter. Note that we tested by how much the reddening changed given the parallax and its uncertainty and, for our sample, the variation in $E(B-V)$ was negligible, which validates our fixed reddening approach.

\section{Results and discussion}

All the available photometric and spectroscopic data for the 13 white dwarfs in our sample were analysed following the methods outlined above. We used model spectra computed for three different atmospheric compositions: pure He, He with traces of H (He+H), and He with traces of H and metals (He+H+Z). This work resulted in a very large number of solutions for the atmospheric parameters, which we will discuss in the following. 

We begin by investigating the overall trends from different sets of observational data (Section\,\ref{sec:diff_datasets}), providing an assessment of the associated systematic uncertainties. As a second test, we inspect the effects of using synthetic model spectra with different chemical compositions (Section\,\ref{sec:err_diff_grids}). Then, we compare our spectroscopic and photometric solutions (Section~\ref{sec:spec_phot_comp}) and contrast them with previously published works (Section~\ref{sec:previous_results_DBAZ}).

The individual results of the spectroscopic and photometric fits for the 13 helium-dominated white dwarfs using the pure He, He+H and He+H+Z grids are presented in full detail in Appendix\,\ref{app:appen_indiv} (Tables~\ref{tab:0030+1526}--\ref{tab:2324-0018}), along with notes on individual stars.

\begin{figure*}
    \centering
    \includegraphics[width=0.8\textwidth]{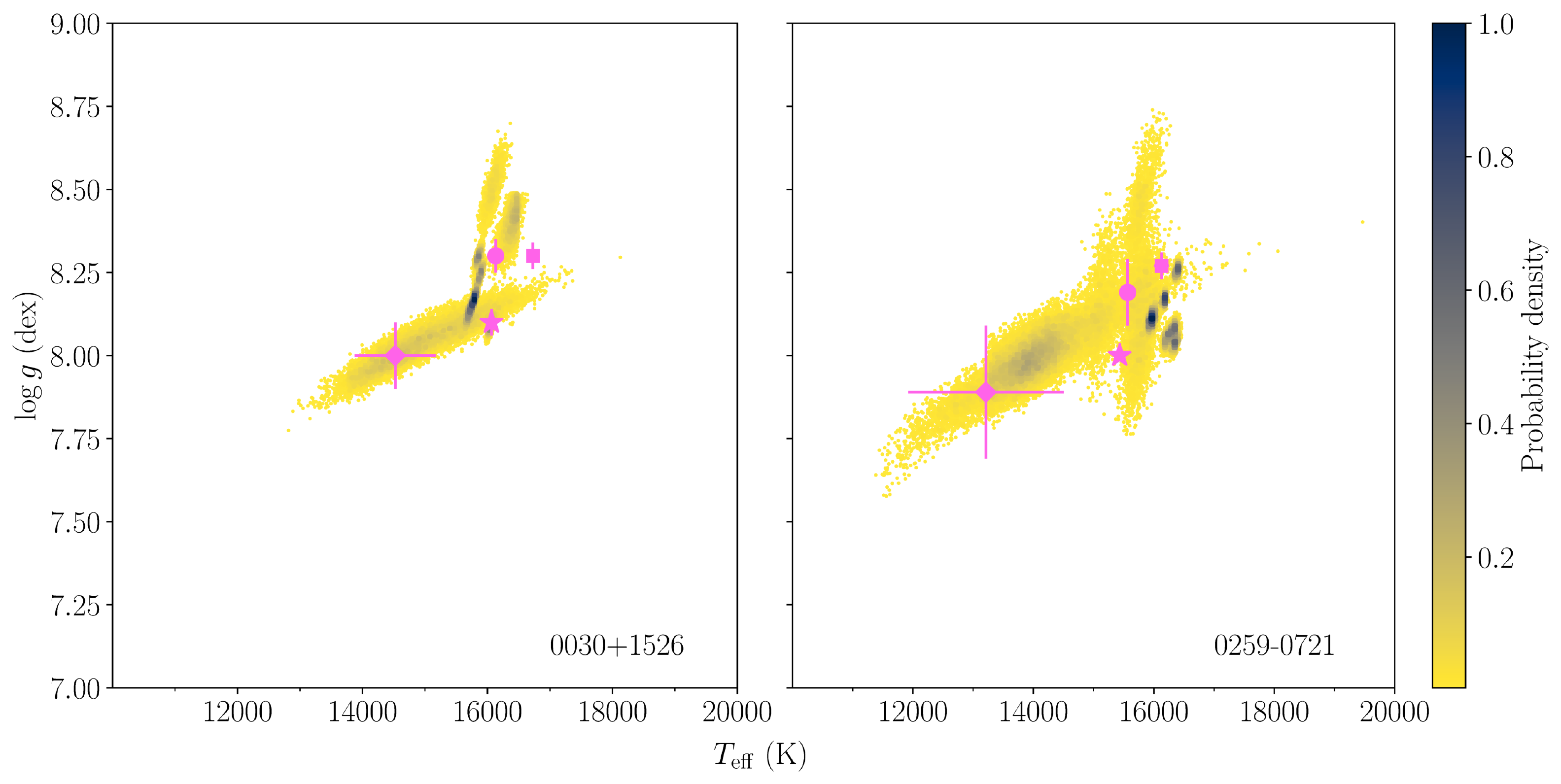}
    \includegraphics[width=0.8\textwidth]{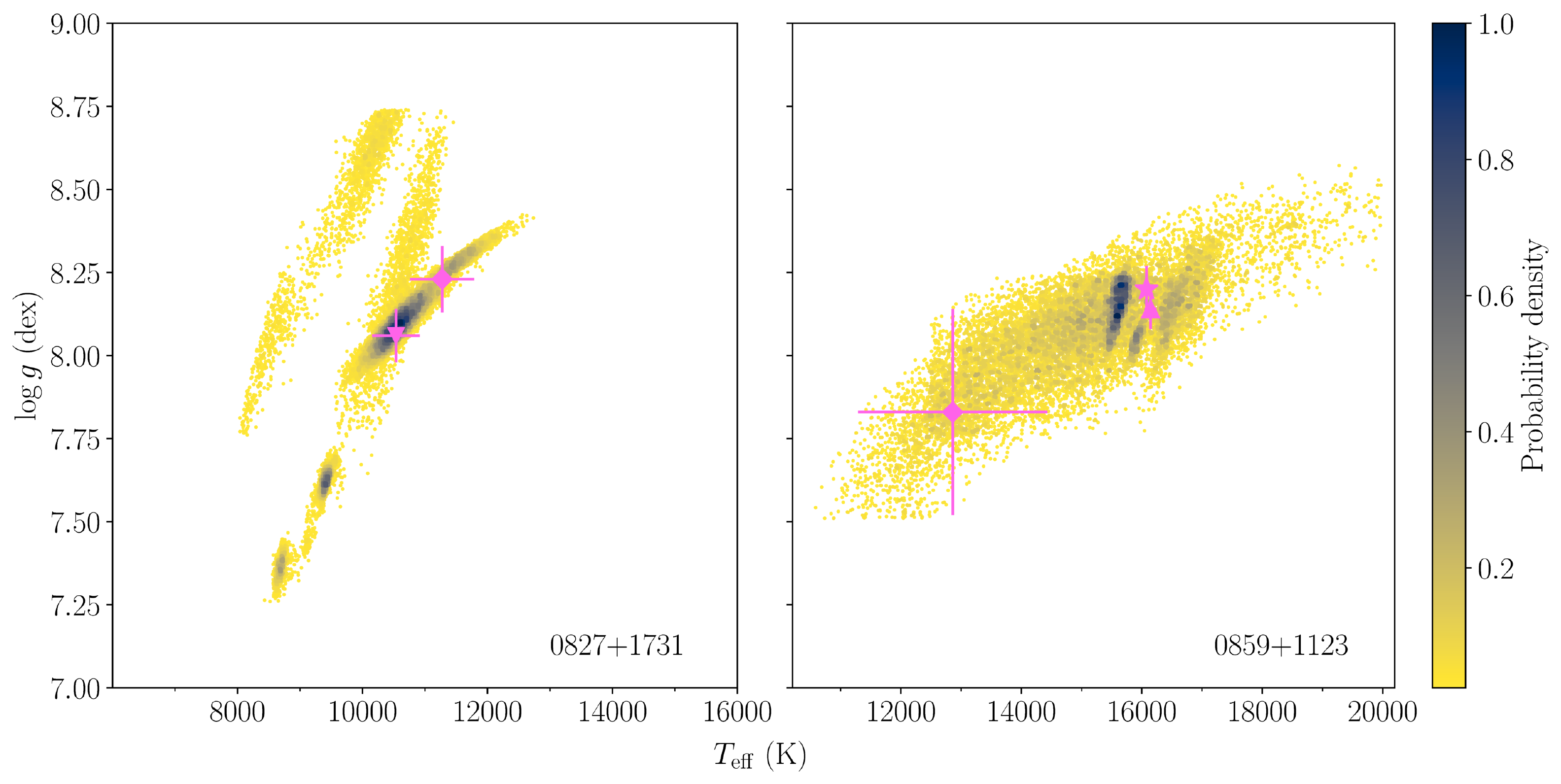}
    \includegraphics[width=0.8\textwidth]{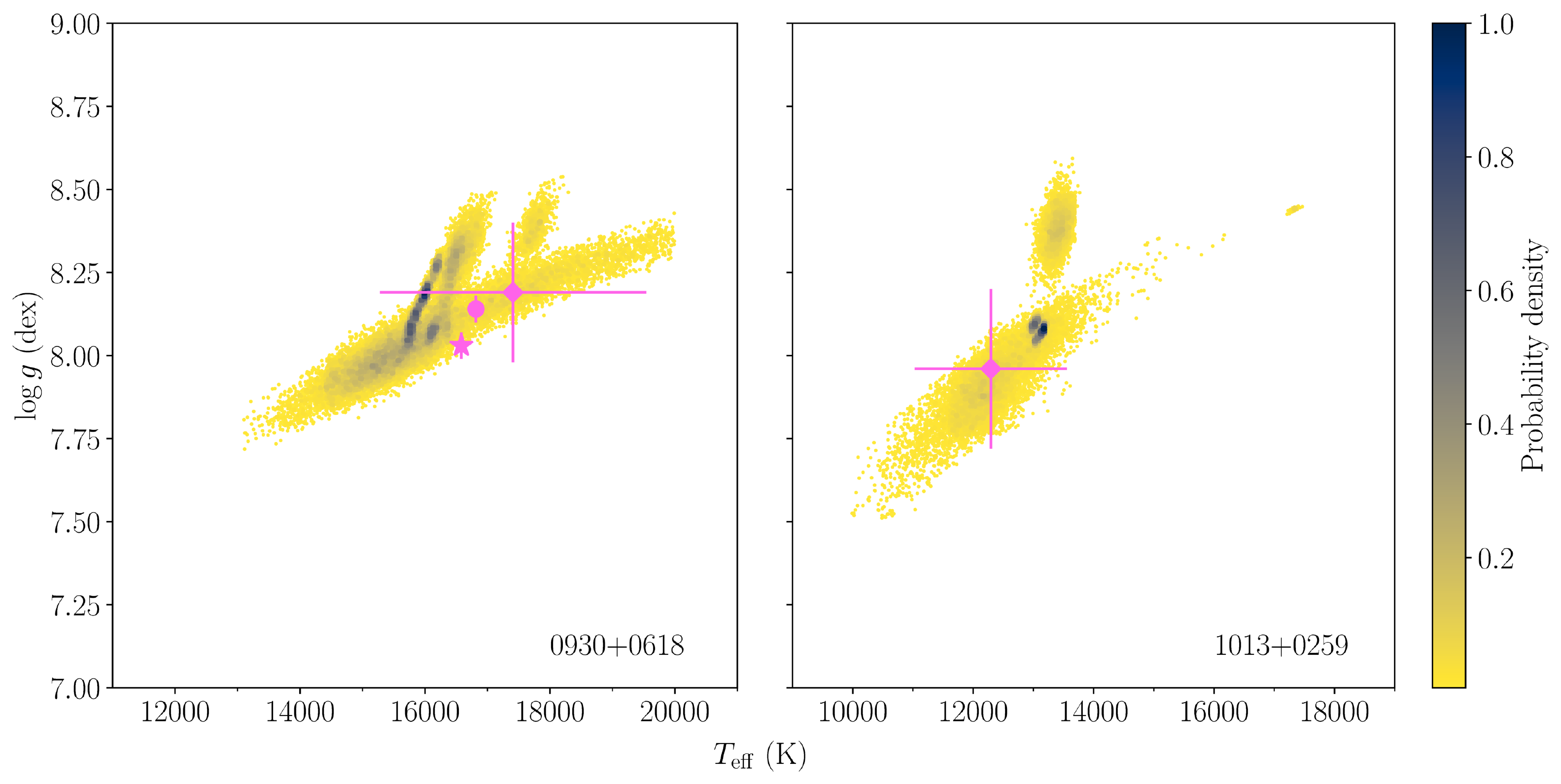}
    \caption{Probability distributions of the \logg\ as a function of the \Teff\ for the different spectroscopic and photometric fits. The distributions are normalised to the same number of samples. The previously published results (Tables~\ref{tab:prev_DBAZresults1} and~\ref{tab:prev_DBAZresults2}) are displayed in pink: \citet{eisenstein06a} as squares, \citet{kleinman13} as circles, \citet{koester15} as stars, \citet{kepler15} as triangles, \citet{coutu19} as inverted triangles and \citet{gentile21}  as diamonds. Note that only literature results within our plotting regions are shown.}
    \label{fig:Prob_distrib_DBAZ13_1}
\end{figure*}

\begin{figure*}
    \centering
    \includegraphics[width=0.8\textwidth]{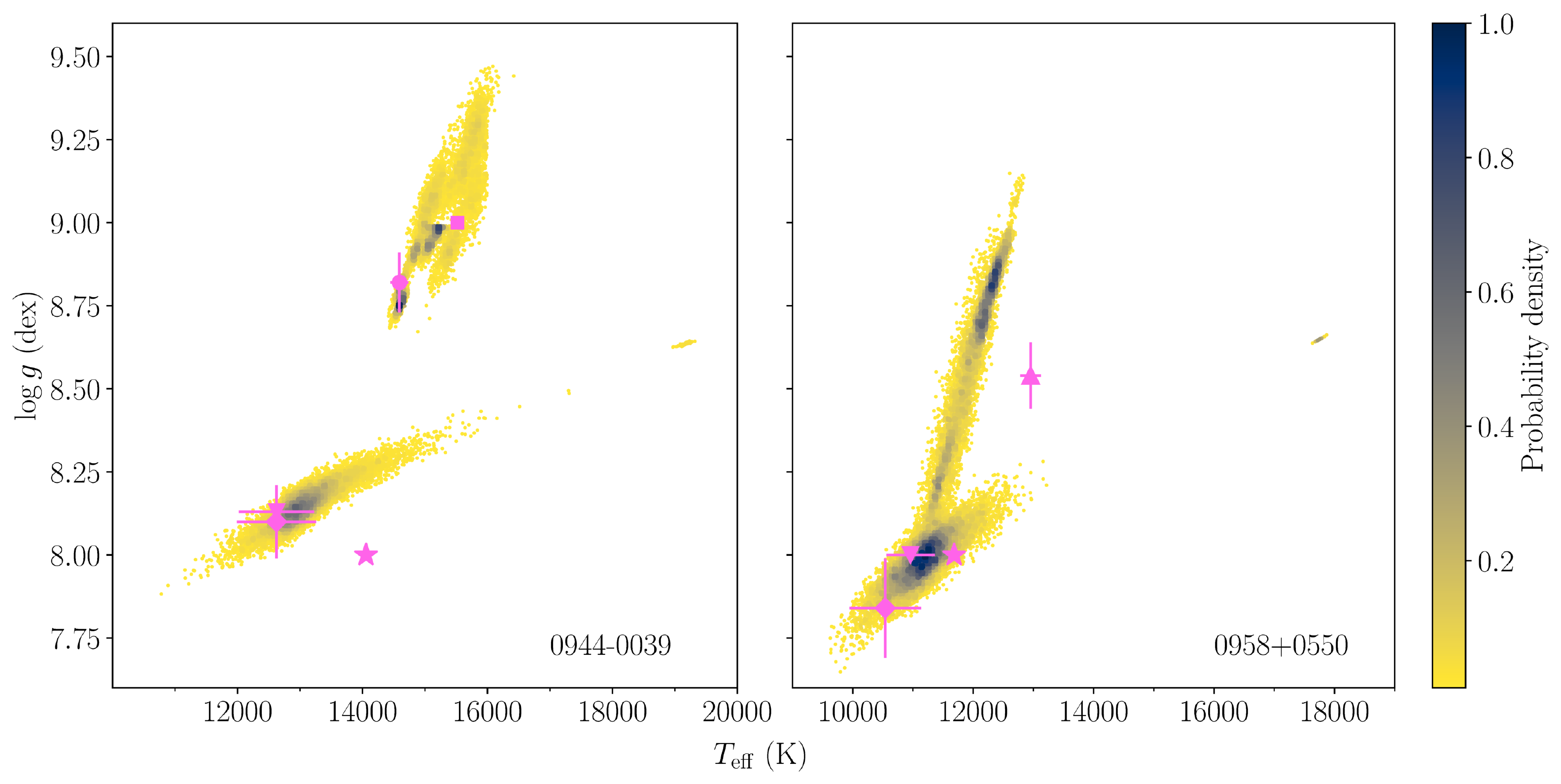}
    \includegraphics[width=0.8\textwidth]{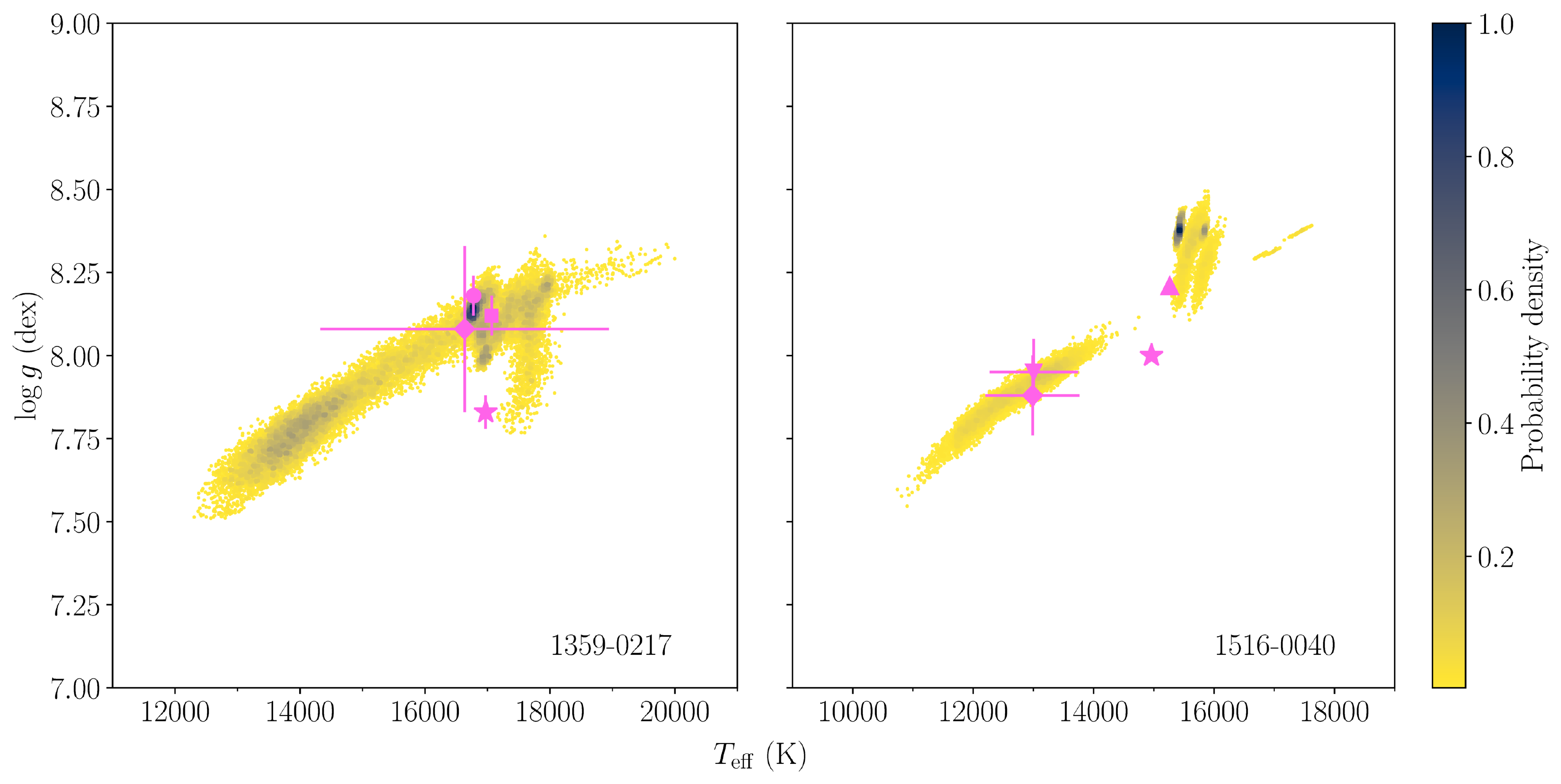}
    \includegraphics[width=0.8\textwidth]{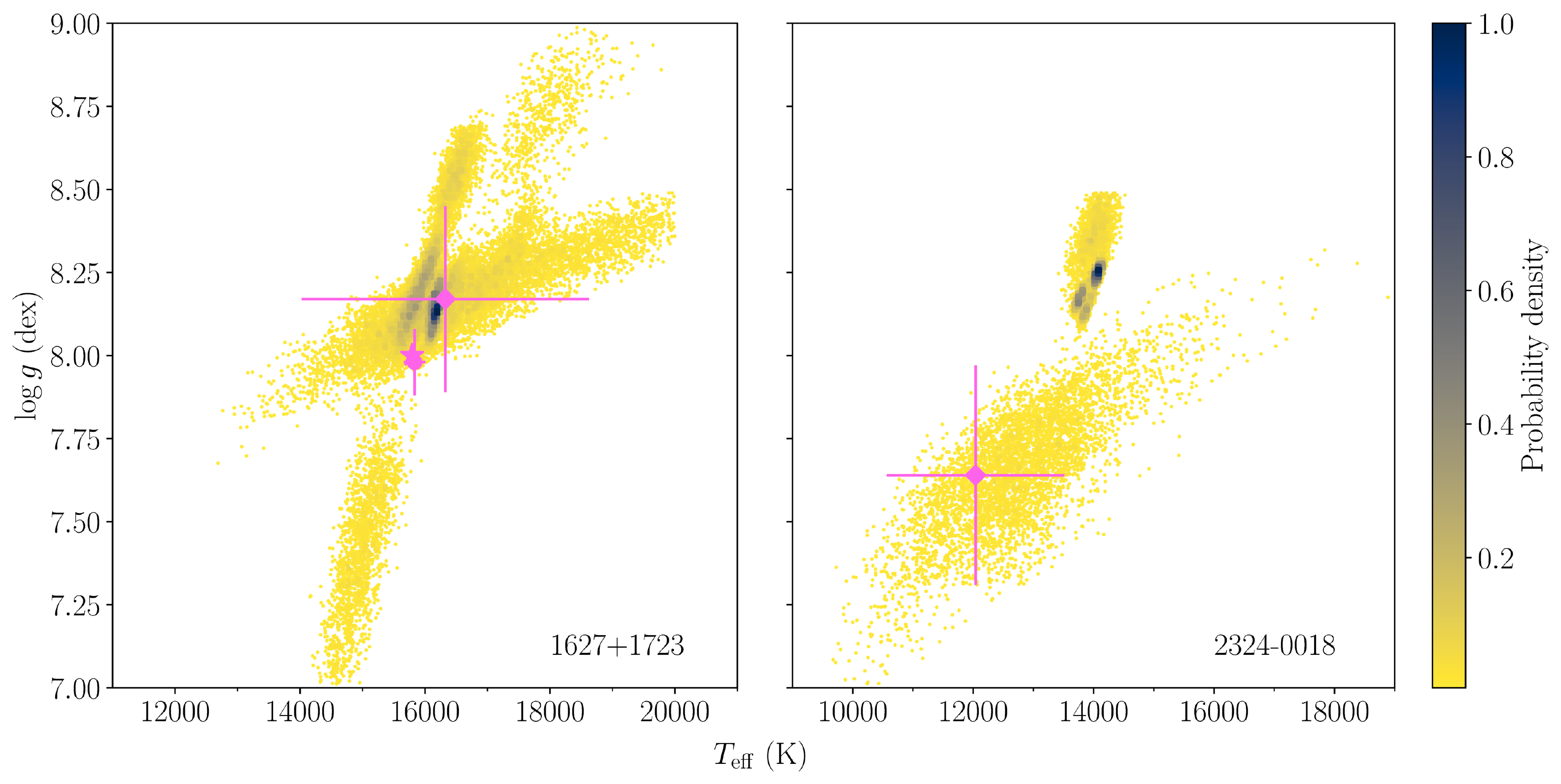}
    \caption{Same as Fig.~\ref{fig:Prob_distrib_DBAZ13_1}}
    \label{fig:Prob_distrib_DBAZ13_2}
\end{figure*}

\begin{figure*}
    \centering
    \includegraphics[width=0.5\textwidth]{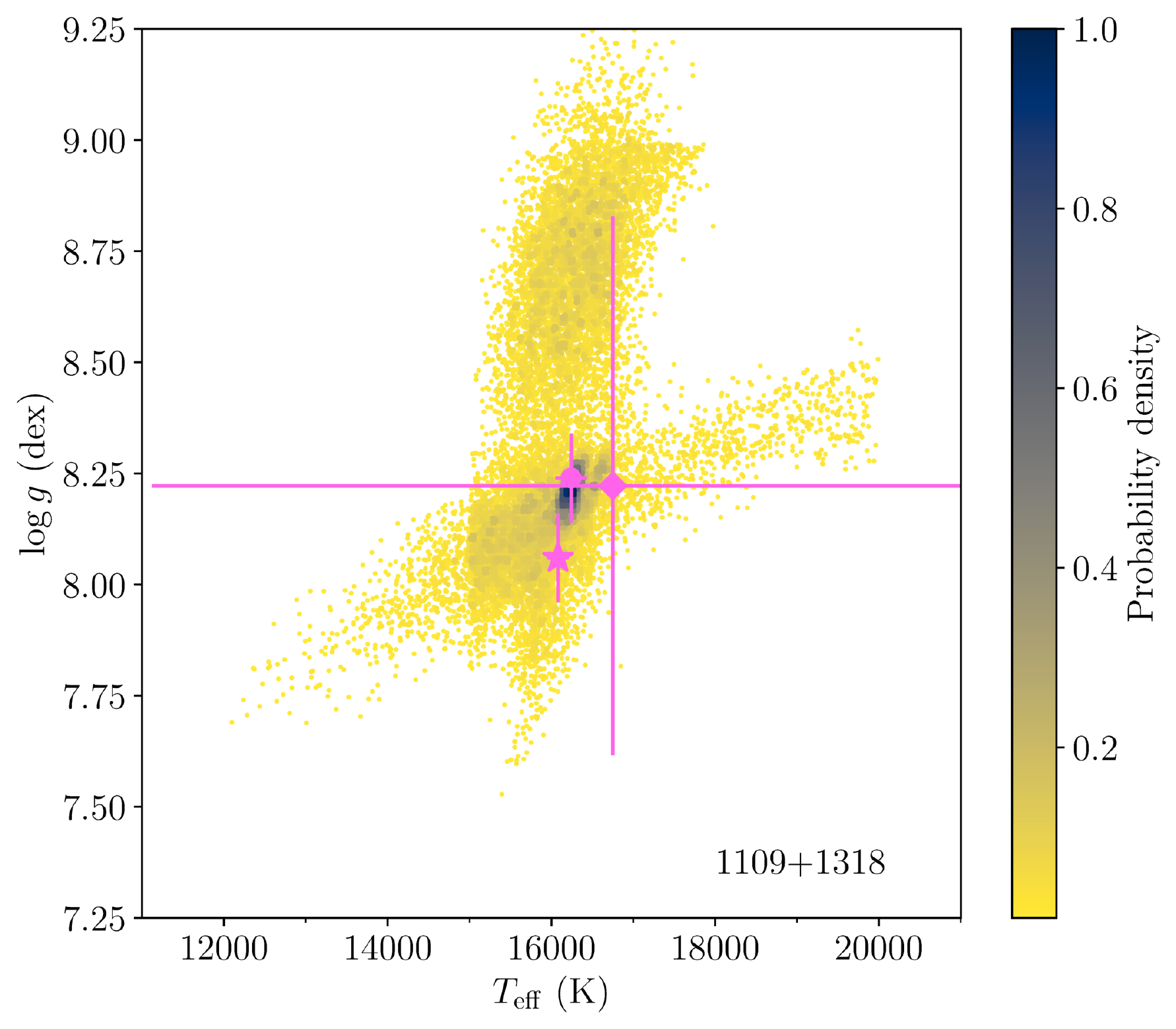}
    \caption{Same as Fig.~\ref{fig:Prob_distrib_DBAZ13_1}}
    \label{fig:Prob_distrib_DBAZ13_3}
\end{figure*}

The probability distributions in the $\Teff-\logg$ plane are shown for each star in Figs.~\ref{fig:Prob_distrib_DBAZ13_1}--\ref{fig:Prob_distrib_DBAZ13_3}, illustrating the results obtained with different data sets, chemical compositions and fitting techniques. The distributions are downsampled to match that with the minimum number of samples and then are normalised to the region with maximum probability. 

\subsection{Systematic uncertainties: different data sets}
\label{sec:diff_datasets}

\subsubsection{Spectroscopy}

\begin{figure}
    \centering
    \includegraphics[width=0.49\textwidth]{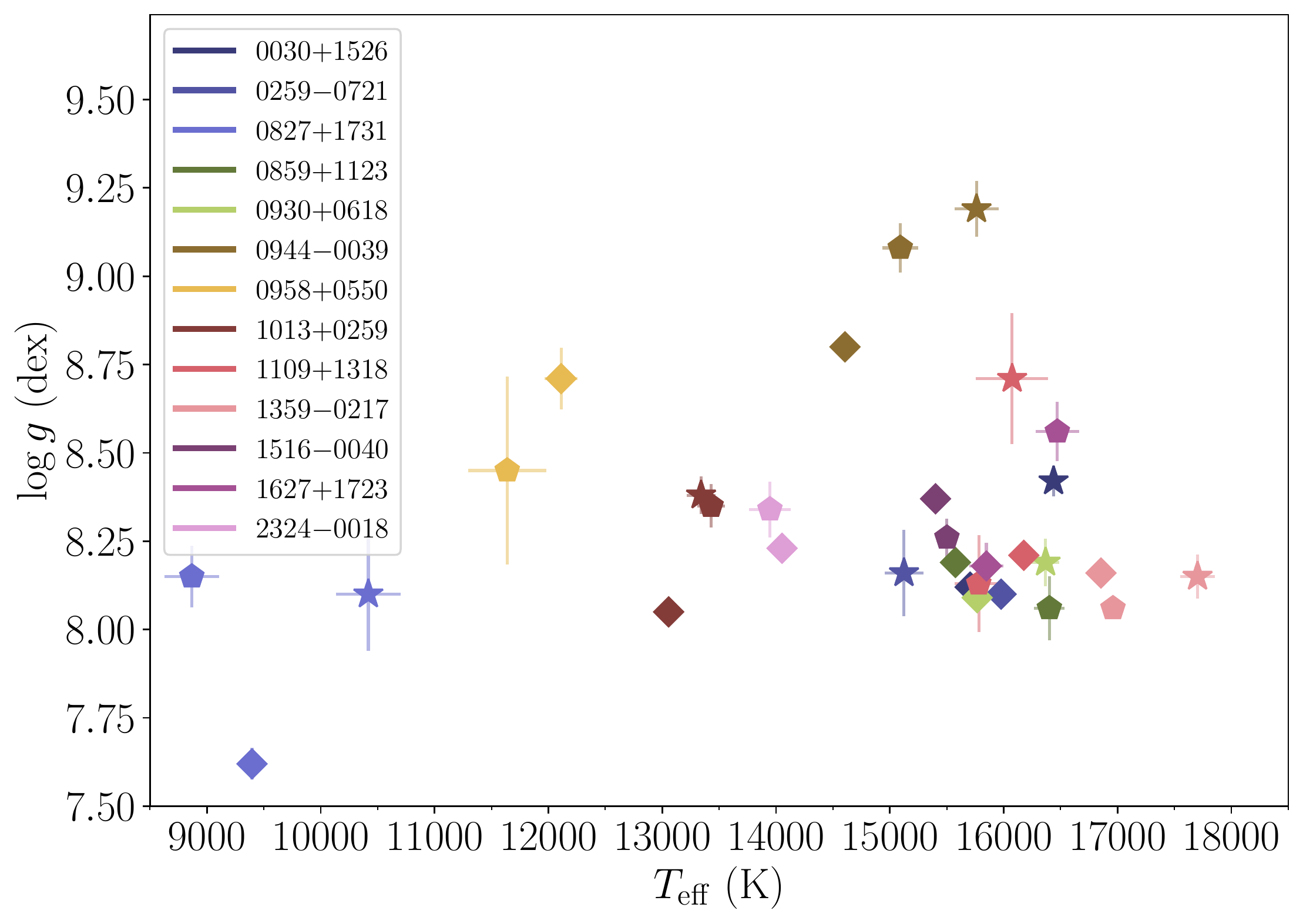}
    \caption{Atmospheric parameters of the 13 white dwarfs in our sample obtained by fitting the X-shooter (diamonds), BOSS (pentagons) and SDSS (stars) spectra with He+H+Z synthetic models (only six stars have three spectroscopic data sets; see Table~\ref{tab:obslog}). The metal abundances of the models were estimated from the metallic absorption lines identified in the X-shooter spectra. Note that the systematic differences between the parameters based on the individual spectra clearly exceed the statistical uncertainties (displayed as error bars in the figure).}
    \label{fig:3spec_DBAZ}
\end{figure}

\begin{figure}
    \centering
    \includegraphics[width=0.45\textwidth]{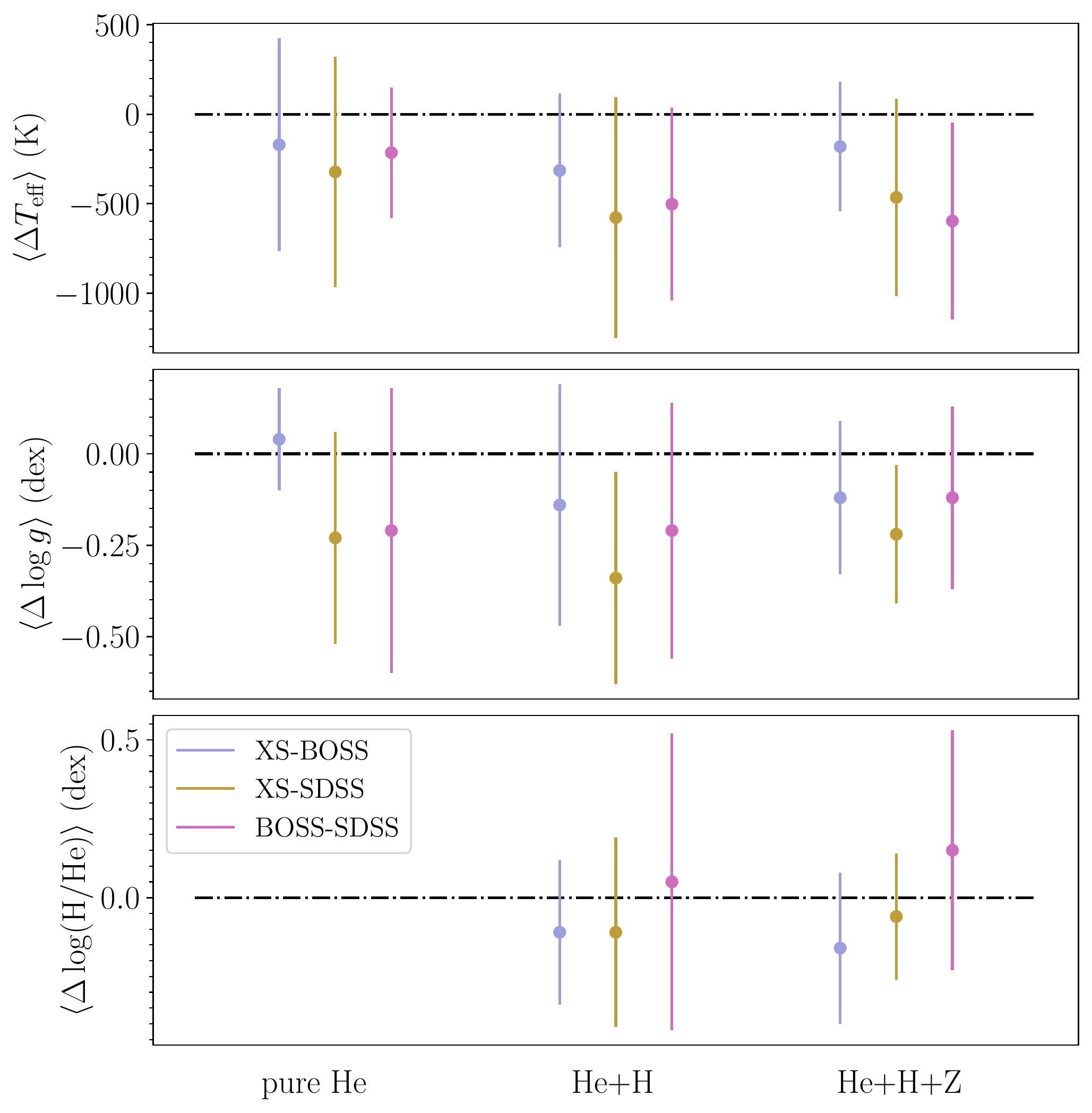}
    \caption{The average differences in \Teff\ (top), \logg\ (middle) and \htohe\ (bottom panel) between the X-shooter (XS), SDSS and BOSS spectroscopic fits for the pure He, He+H and He+H+Z synthetic grids (left to right) are used to check for general trends between the different data sets. There is no hydrogen in the pure He models, and thus no \htohe\ estimate (bottom panel). Note that the uncertainties are the standard deviations and hence show how dispersed are the data related to the mean value.}
    \label{fig:aver_differ_spec}
\end{figure}

We estimated the systematic uncertainties arising from the use of diverse spectroscopic data sets (X-shooter, BOSS and SDSS) by means of the differences in the best-fit \Teff, \logg\ and \htohe\ determined from the different observations. The spectroscopic results obtained from the He+H+Z fitting of the three data sets are shown in Fig.~\ref{fig:3spec_DBAZ} and the \Teff, \logg\ and \htohe\ average differences are computed to probe for systematic trends between the three data sets (see~Fig.\,\ref{fig:aver_differ_spec}). Note that the effect of using different chemical composition models is not discussed here, but will be presented in detail in Section~\ref{sec:err_diff_grids}.


On average, the X-shooter spectra provide smaller values of the atmospheric parameters than BOSS (X-shooter~--~BOSS) by $222$\,K, $0.07$\,dex and $0.14$\,dex for \Teff, \logg\ and \htohe, respectively. Even though multiple factors can play a role in these differences, the lower SNR of the BOSS spectra when compared to X-shooter ($\Delta \rm{SNR} \simeq 14$) may be decisive: the hydrogen lines, which are key in measuring the three atmospheric parameters, could be not fully resolved in the BOSS (and SDSS) spectra. One would expect the higher SNR and spectral resolution of X-shooter to provide more reliable \htohe\ estimates, translating in larger hydrogen abundances due to its ability to detect shallower lines. However, the BOSS \htohe\ values are on average larger than those measured in the X-shooter spectra with no clear explanation. 

Comparing the X-shooter to the SDSS parameters we obtain average differences (X-shooter~--~SDSS) of $ -455$\,K, $ -0.26$\,dex and $ 0.03$\,dex, which follow the same trend as X-shooter-BOSS, with the exception of \htohe. The SNR fraction between the SDSS and X-shooter UVB spectra ($\Delta$SNR$=23$), which contains most of the absorption lines are, could again lead to less reliable results.

On average, (BOSS~--~SDSS) yields a \Teff\ difference of $-438$\,K, $-0.18$\,dex for \logg, and a larger \htohe\ in the BOSS spectra by $+0.10$\,dex. The reasons behind the differences between these two data sets are unclear, although it should be noted that systematic parameter offsets between SDSS spectra and data from other instruments have already been found, and are attributed to the data reduction procedure. However, no exact cause could be determined \citep{tremblayetal11-2}. 

Whereas the average of the parameter differences reflect systematic offsets between the results from different data sets, the standard deviation provides an estimation of the amount of variation of those values and hence represents the typical magnitude of the true systematic  uncertainties in the analysis.

We find X-shooter~--~BOSS mean standard deviations of $\left< \sigma \Teff \right> = 462$ K, $\left< \sigma \logg \right> = 0.23$ and $\left< \sigma \htohe \right> = 0.24$ dex. These differences are larger for X-shooter~--~SDSS and are very likely related to the bigger SNR disparity between the two data sets: $\left< \sigma \Teff \right> = 623$ K, $\left< \sigma \logg \right> = 0.26$ and $\left< \sigma \htohe \right> = 0.25$ dex. Finally, the BOSS~--~SDSS mean standard deviations are: $\left< \sigma \Teff \right> = 485$ K, $\left< \sigma \logg \right> = 0.33$ and $\left< \sigma \htohe \right> = 0.43$ dex. In the last case, the statistics are obtained with just five objects (we are not taking into account $1627+1723$ since the SNR of the SDSS spectra is below 13 and gives untrustworthy results; see Table\,\ref{tab:1627+1723} for more details), but still these numbers are dominated by the results obtained for $1109+1318$, with a SDSS spectra SNR of 14.


We conclude that the analysis of separate spectroscopic data sets, in particular if obtained with different instrumental setups can result in differences in the resulting atmospheric parameters that are significantly larger than the statistical uncertainties of the fits to the individual spectra.

We suggest these results to be taken into account to assess the actual uncertainties inherent to spectroscopic analyses for cool helium-dominated white dwarfs, in particular when employing spectra with similar SNR and resolution. From our analysis, we derive systematic uncertainties of the spectroscopic \Teff, \logg\ and \htohe\ of $524$\,K, $0.27$\,dex and $0.31$\,dex, respectively (the average of the X-shooter~--~BOSS, X-shooter~--~SDSS and BOSS~--~SDSS mean standard deviations).

\subsubsection{Photometry}

Here, we explore and compare the systematic differences in \Teff\ and \logg\ obtained from the photometric fits using the magnitudes of three independent catalogues: SDSS, PS1 and \textit{Gaia}, adopting different chemical compositions (we refer to Section\,\ref{sec:err_diff_grids} for the discussion on the use of different chemical composition models). 

\begin{figure}
    \centering
    \includegraphics[width=0.48\textwidth]{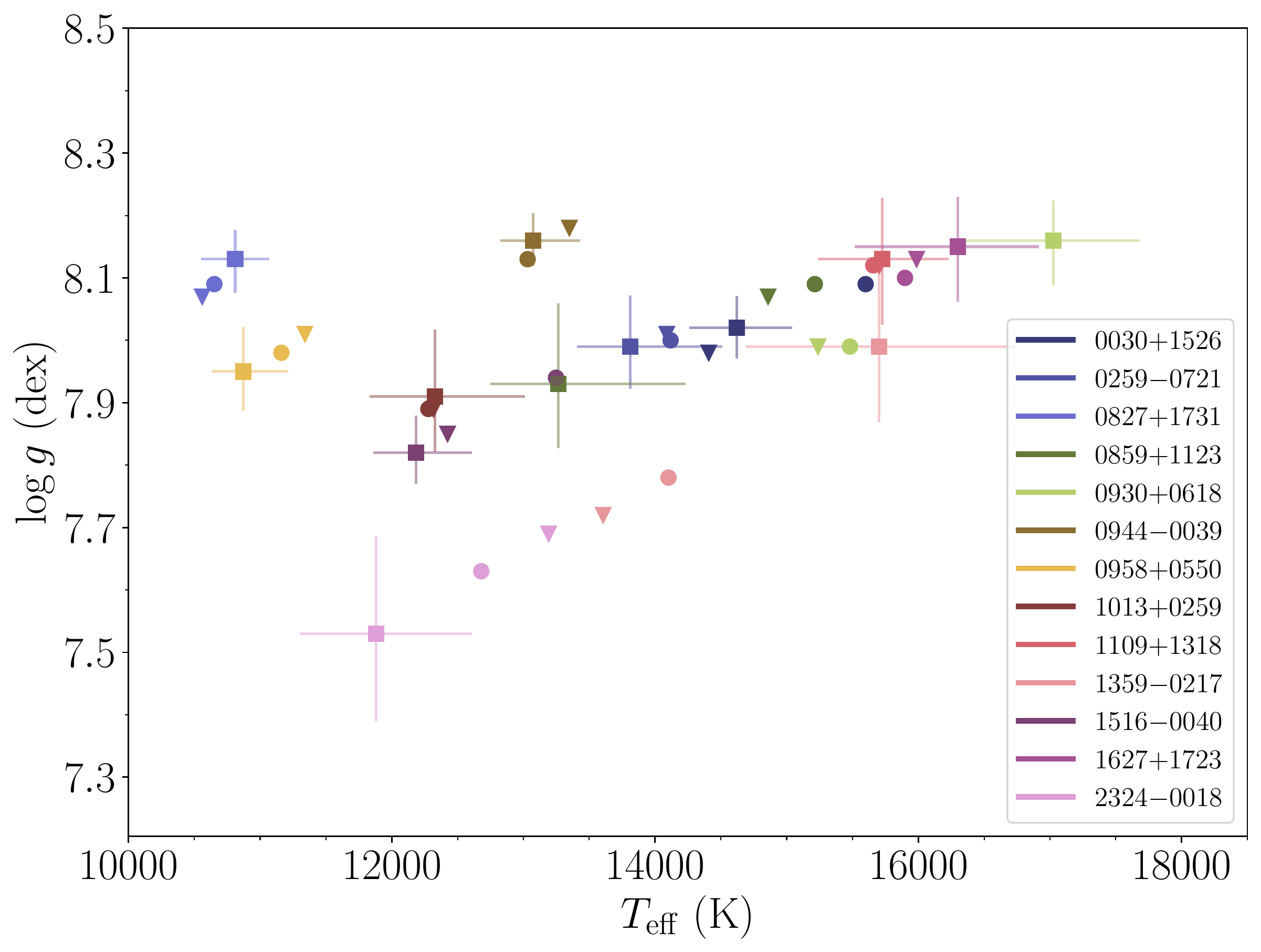}
    \caption{Atmospheric parameters obtained by fitting the SDSS (circles), PS1 (triangles) and \textit{Gaia} DR3 (squares) photometry with He+H+Z synthetic models (the $\htohe$ are fixed at the X-shooter spectroscopic values). Just the \textit{Gaia} uncertainties (the largest in all the cases) are displayed. The best-fit solutions for each target stray along a diagonal in $\Teff-\log g$, illustrating the correlation between these two parameters.}
    \label{fig:DBAZ_3cats_XScomp}
\end{figure}

\begin{figure}
    \centering
    \includegraphics[width=0.48\textwidth]{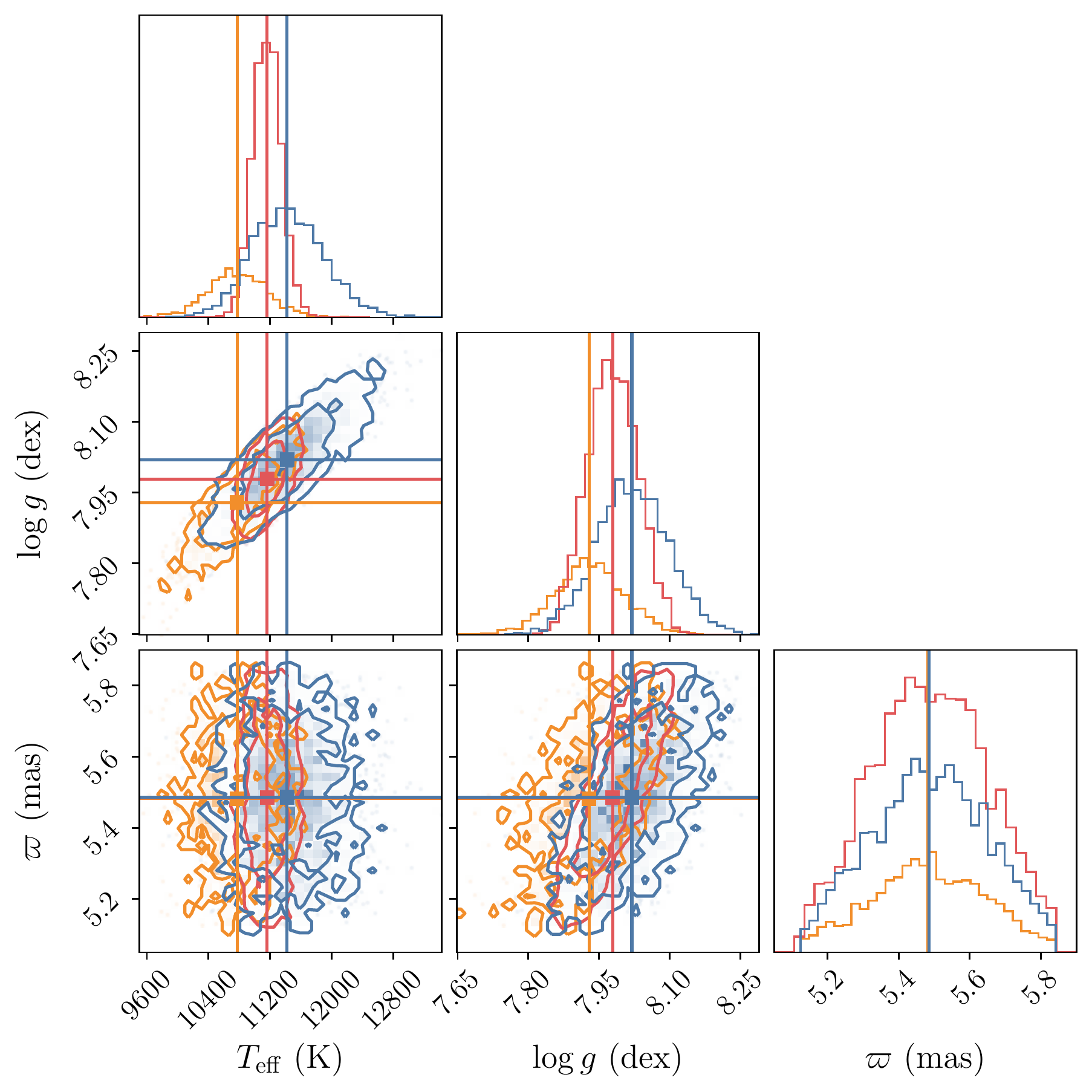}
    \caption{Corner plot for the white dwarf 0958+0550 using He+H models with fixed $\htohe = -5.7$\,dex, showing the probability distribution of the parameters obtained by fitting the SDSS (red), PS1 (blue) and \textit{Gaia} eDR3 photometry (orange). It illustrates the compatible values between the three catalogues and the correlation between \Teff\ and \logg: the published fluxes of the three catalogues are similar and scaled by the same distance (provided by the \textit{Gaia} eDR3 parallax) and hence, even a small change in \Teff\ produces a readjustment of the radius (and thus the \logg) to conserve the flux.}
    \label{fig:j0958_corner_plot}
\end{figure}

In Fig.\,\ref{fig:DBAZ_3cats_XScomp} we show the parameter differences for the He+H+Z model spectra, with \htohe\ fixed to the X-shooter best-fit spectroscopic value\footnote{This is just a choice to illustrate the general trend. The He+H+Z synthetic grids assess the full chemical composition of each photosphere and the X-shooter spectra have the highest spectral resolution, wavelength coverage and SNR.}. There is a steep correlation between \Teff\ and \logg: the published fluxes of the three catalogues are really similar for each star (e.g. an average $0.14$~per~cent difference in the SDSS-$g$ and PS1-$g$ bands) and scaled by the same distance (provided by the \textit{Gaia} eDR3 parallax) and hence, even a slight increase in \Teff\ translates to a smaller radius to conserve the flux, which ultimately leads to larger \logg\ (see Fig.\,\ref{fig:j0958_corner_plot}).


\begin{figure}
    \centering
    \includegraphics[width=0.48\textwidth]{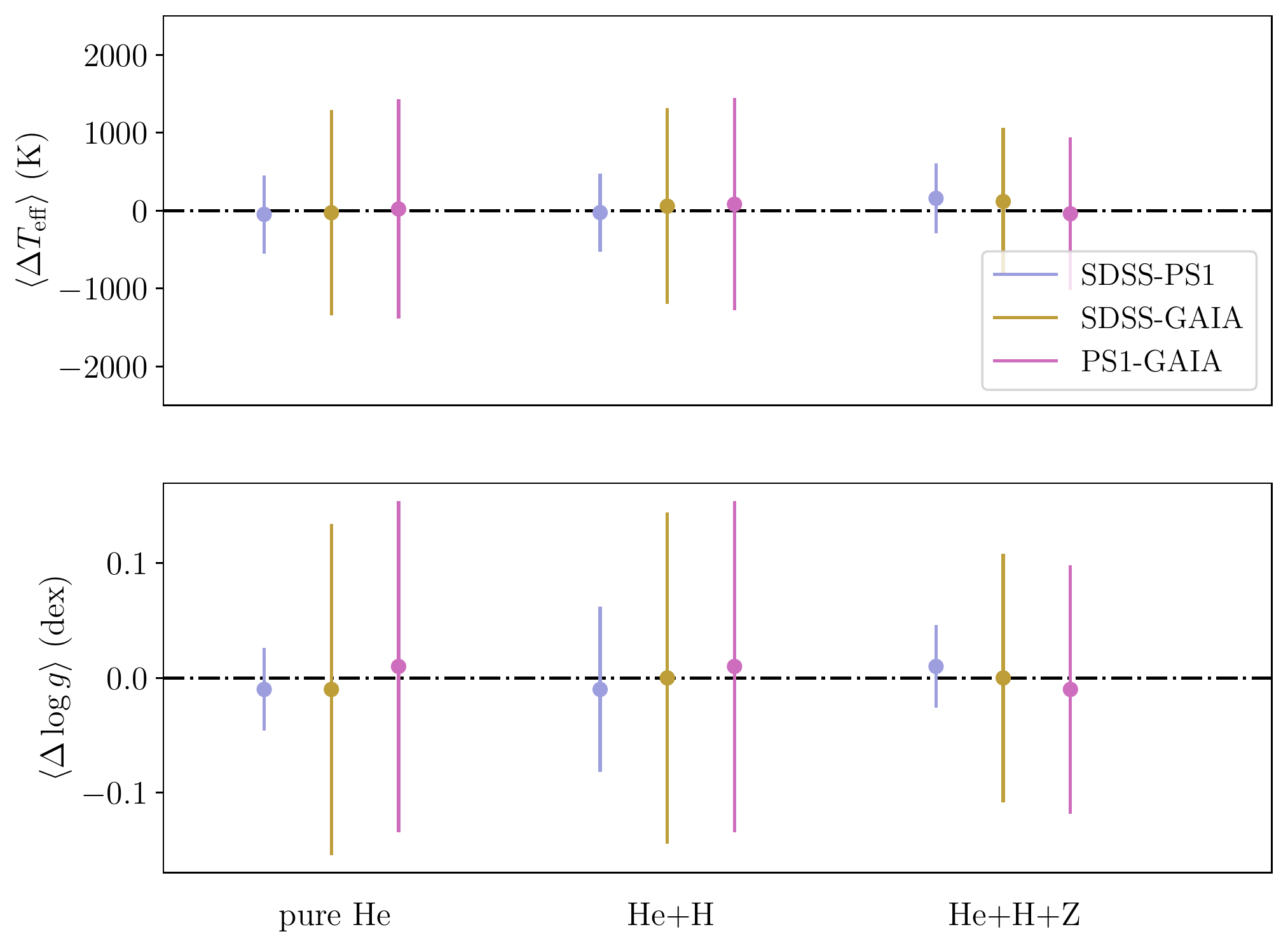}
    \caption{The average differences in \Teff\ and \logg\ between the SDSS, Pan-STARRS1 (PS1) and \textit{Gaia} eDR3 photometric results for the pure He, He+H and He+H+Z synthetic grids (left to right). No overall trend between the three catalogues is observed. Note that the uncertainties are the standard deviations, i.e. how dispersed is the data related to the mean value.}
    \label{fig:aver_differ_phot}
\end{figure}

The average photometric differences in \Teff\ and \logg\ are displayed in Fig.\,\ref{fig:aver_differ_phot}, displaying no systematic trends between the three data sets. 

The \Teff\ and \logg\ derived from all the SDSS and PS1 photometric fits are consistent with each other for the 13 white dwarfs except for 0030+1526 (see Appendix~\ref{app:appen_indiv} for comments on individual stars). However, we find mean standard deviations between the results derived from these two surveys of $\left < \sigma \Teff \right > = 485$\,K and $\left < \sigma \logg \right > = 0.05$\,dex, which could be related to the SDSS $u$-band, with no analogous in the PS1 survey and a measure that adds important constraints to the SED. Since no systematic offset between these two catalogues has been reported they should lead to the same set of parameters and thus we suggest these differences to be taken into account when quoting uncertainties derived from each of this data sets, being considerably larger than those usually published in the literature. 

The \textit{Gaia} atmospheric parameters are, in general, inconsistent with the SDSS and PS1 sets of solutions, leading to average standard deviations of $\left < \sigma \Teff \right > = 1210$\,K and $\left < \sigma \logg \right > =  0.13$\,dex. This might be related to the extremely broad \textit{Gaia} passbands, but the smaller number of filters cannot be discarded. We suggest these mean standard deviations to be the minimum uncertainty quoted when retrieving atmospheric parameters from \textit{Gaia} photometry for relatively cool helium-dominated white dwarfs.

We conclude that, as already found for the spectroscopic method, the analysis of different photometric data sets can result in atmospheric parameters that are discrepant by more than the statistical uncertainties. Underlying reasons include the use of different band-passes, and systematic uncertainties in the zero-points \citep[e.g.][]{PanSTARRS2}.


\subsection{Systematic uncertainties: atmospheric models with different chemical abundances}
\label{sec:err_diff_grids}

\subsubsection{Spectroscopy}

In this section we assess the systematic uncertainties in \Teff, \logg\ and \htohe\ when fitting spectroscopic data with atmospheric models of different chemical compositions. This situation may be encountered when having spectra with insufficient SNR to sample narrow or shallow lines or when having just a limited wavelength coverage, not including transitions of all relevant chemical elements. In those cases, we might fit the available observed spectra with synthetic models that do not take into account the complete chemical composition of the white dwarf.

\begin{figure}
    \centering
    \includegraphics[width=0.48\textwidth]{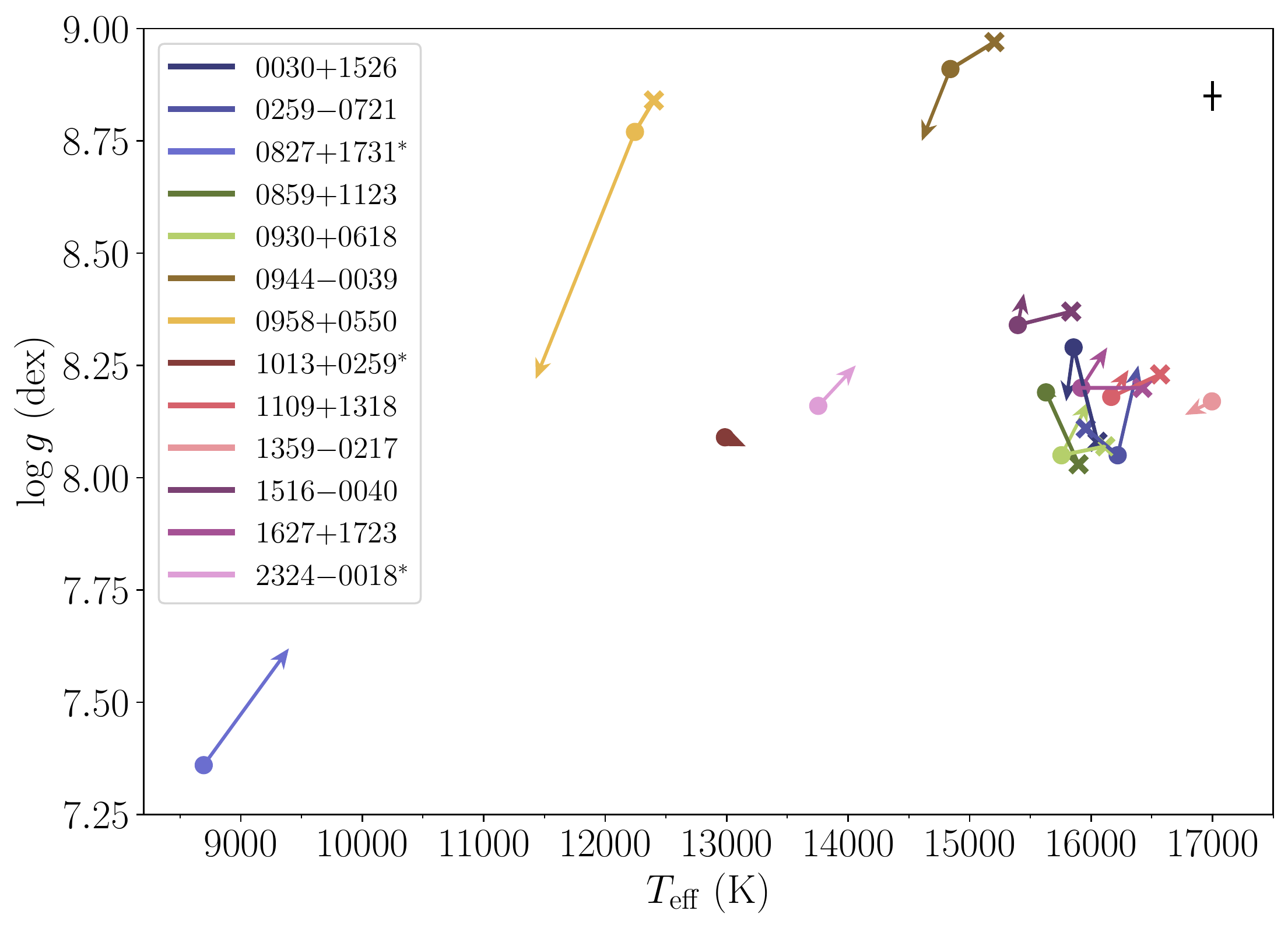}
    \caption{Spectroscopic X-shooter results using pure He (crosses), He+H (circles) and He+H+Z (arrow head) synthetic models. Metal absorption lines superimposed on the hydrogen and helium lines have been included in the He+H+Z fits (see Section\,\ref{sec:method_DBAZ13}). The stars identified with an asterisk lack a pure He analysis since their spectra are fully dominated by Balmer lines (see Fig.\,\ref{fig:XS_spectra} and Table\,\ref{tab:obslog}). The average error bars are displayed in the top right corner. Note that in some cases the pure He and He+H+Z results are not visible due to their similarity to the He+H values. The inclusion of hydrogen in the models (pure He $\rightarrow$ He+H) produces a drop in \Teff\ of $\simeq 300$\,K and a slight increase in \logg\ ($\simeq 0.02$\,dex). The addition of metals to the models (He+H $\rightarrow$ He+H+Z) suggests a small increase of $60$\,K in \Teff, while \logg\ remains, on average, equal.}
    \label{fig:pDB_DBA_DBAZ_spec_comparisons}
\end{figure}

The spectroscopic \logg\ as a function of \Teff\ obtained from the fits to the X-shooter spectra (the only set with spectra for all 13 white dwarfs) using pure He, He+H and He+H+Z synthetic models is displayed in Fig.\,\ref{fig:pDB_DBA_DBAZ_spec_comparisons}. The metallic lines blended with the helium and hydrogen lines were included in the He+H+Z fit since metals are implemented in those models, but the metal abundances were fixed to the values derived from the 1D metal fits (see Table~\ref{tab:Metal_abs_stars}).


We explored the likely errors introduced when fitting helium-dominated white dwarfs with traces of hydrogen and metals with pure He models. To do so, we determined the average $\Delta \Teff = \Teff^{\rm{He+H}} - \Teff^{\rm{pure He}}$ and $\Delta \logg = \logg^{\rm{He+H}} - \logg^{\rm{pure He}}$ differences for the X-shooter, SDSS and BOSS spectra for each star\footnote{Note that the differences between the parameters derived from the pure He and He+H analysis are greater the more hydrogen content is present in the photosphere.} to be $\left <\Delta \Teff \right> = -335 $\,K and $\left <\Delta \logg \right>= 0.01$\,dex for X-shooter, $\left <\Delta \Teff \right> = -251$\,K and $\left <\Delta \logg \right> = 0.02$\,dex for SDSS and $\left <\Delta \Teff \right> = -317$\,K and $\left <\Delta \logg \right> = 0.03$\,dex for BOSS. We see thus a generic trend when adding hydrogen: \Teff$^{\rm{He+H}} <$ \Teff$^{\rm{pure He}}$ and \logg$^{\rm{He+H}} >$ \logg$^{\rm{pure He}}$ ($\simeq -300$\,K, $\simeq +0.02$\,dex, respectively). This result is expected from the hydrogen-line blanketing: the addition of hydrogen increases the opacity (most noticeably in the UV) and thus produces a back-warming effect in the optical, which translates in an overall lower \Teff\ to match the \textit{unblanketed} model. However, we note this phenomenon has commonly been discussed for a fixed \logg, which is different from our analysis where \Teff\ and \logg\ are free parameters. Regarding the trend seen in \logg\ we highlight that, for the majority of cases, \logg\ decreases, and thus this average increase ($\simeq +0.02$\,dex) is dominated by the outliers.

\begin{figure*}
    \centering
    \includegraphics[width=\textwidth]{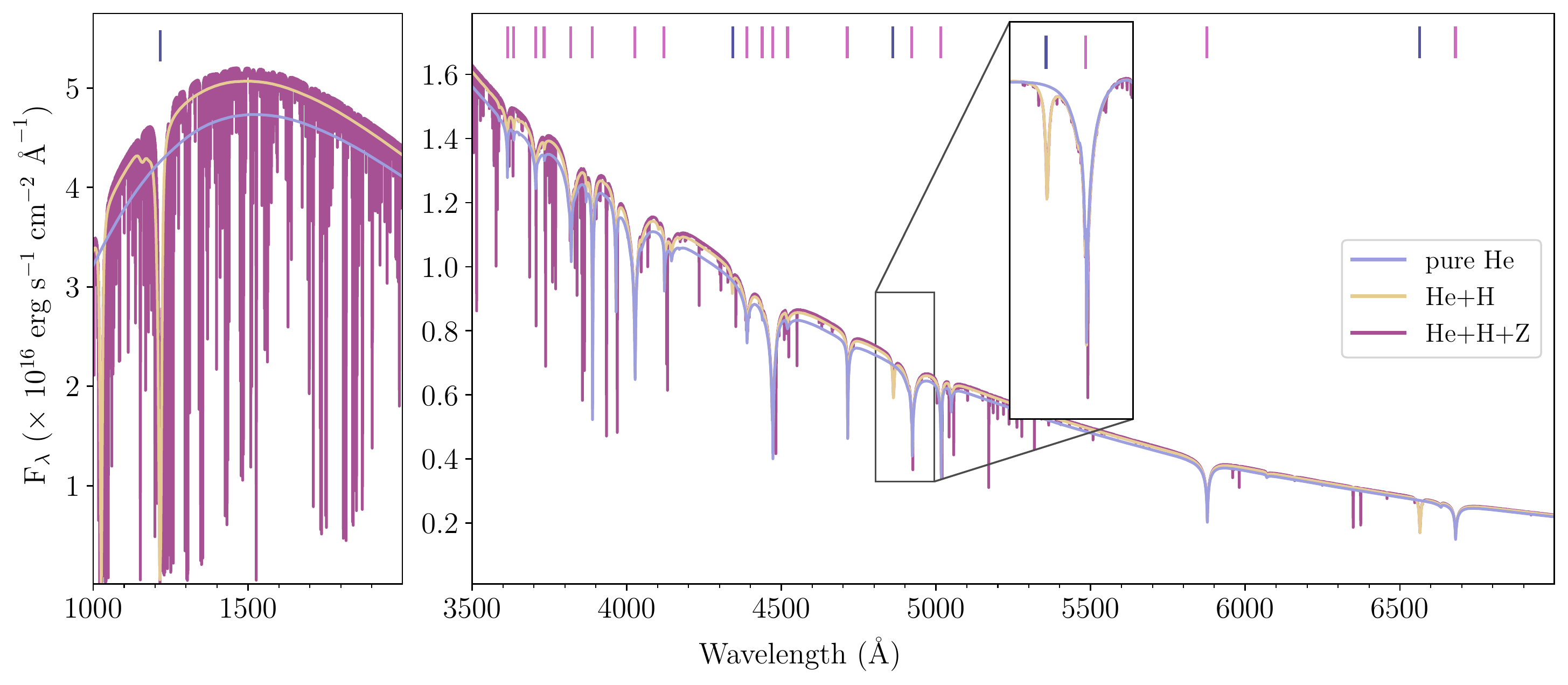}
    \caption{Synthetic spectra of a white dwarf with $\Teff = 16\,000$\,K and $\logg=8.0$\,dex. The \htohe\ is fixed to $-4.5$\,dex for the He+H and He+H+Z spectra and the relative metal abundances of the latter are fixed to those of 0930+0618 (see Table~\ref{tab:Metal_abs_stars}). The H$\beta$ and \hel{i}{4922} absorption lines have been zoomed-in and continuum-normalised to illustrate the slight increase in line width and depth as a result of the inclusion of hydrogen and metals. The hydrogen and helium lines are indicated by the blue and pink vertical lines, respectively.}
    \label{fig:HZ_plot_blank}
\end{figure*}

We carried out the same analysis to assess the systematic differences in \Teff, \logg\ and \htohe\ that may arise when fitting helium-dominated white dwarfs with traces of hydrogen and metals neglecting the presence of the latter in the photosphere. We found $\left<\Teff^{\rm{He+H+Z}}-\Teff^{\rm{He+H}} \right >=60$\,K, no \logg\ difference and $\left< \htohe^{\rm{He+H+Z}} - \htohe^{\rm{He+H}} \right > = -0.01$\,dex for X-shooter\footnote{The BOSS and SDSS data were also fitted with He+H+Z synthetic spectra but using the metal abundances estimated from the X-shooter spectra (see Section\,\ref{subsec:spec_fits} for more details and Tables~\ref{tab:0030+1526} to \ref{tab:2324-0018} for those fits).}. The inclusion of metals in the models produces an small overall increase in \Teff\ (i.e. metal-line blanketing) even though the change in the helium/hydrogen absorption lines is not noticeable (Fig.~\ref{fig:HZ_plot_blank}).

\subsubsection{Photometry}

Despite the rapid increase of spectroscopically characterised white dwarfs, the largest parameter analyses still rely on candidates retrieved from photometric surveys (e.g. \citealt{gentile21}). In these cases, but also for white dwarfs with poor SNR spectra, the chemical compositions might be unknown or unreliable, which might translate in inaccurate photospheric parameters. 

\begin{figure}
    \centering
    \includegraphics[width=0.48\textwidth]{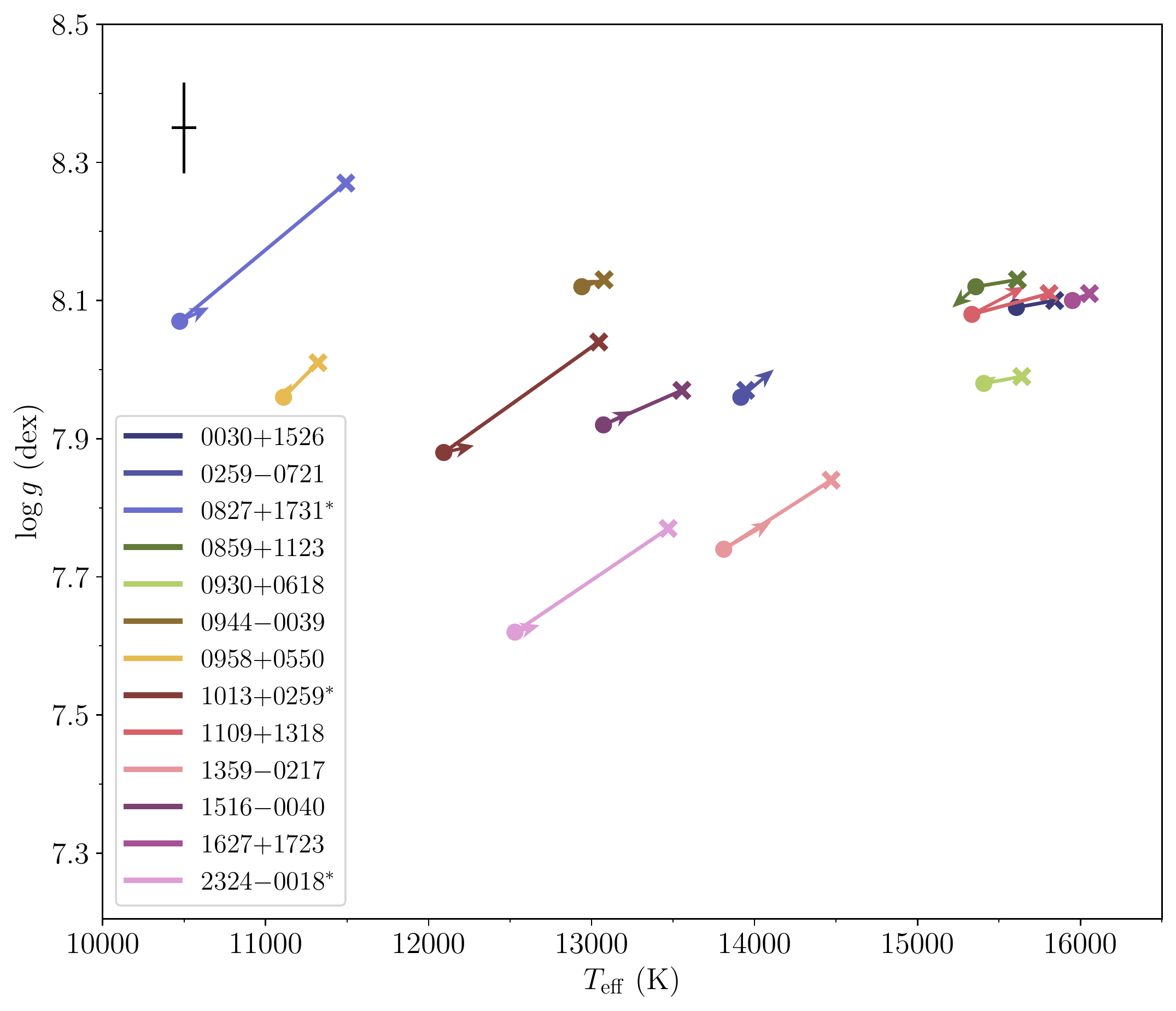}
    \caption{Photometric fits of the SDSS photometry data using pure He (crosses), He+H (circles) and He+H+Z (arrow head) synthetic models. For each star, the \htohe\ has been fixed to the X-shooter value for the He+H and He+H+Z spectroscopic fits. The average uncertainties are shown in the top left corner. The stars identified with an asterisk are clearly dominated by Balmer absorption lines and hence the difference between pure He and He+H results is larger (see Fig.~\ref{fig:XS_spectra} and Table~\ref{tab:obslog}).}
    \label{fig:phot_pDB_DBA_DBAZ}
\end{figure}

We explore this situation by investigating the differences in the best-fit photometric \Teff\ and \logg\ for different chemical compositions of the model spectra (pure He, He+H and He+H+Z), illustrating the miscalculations/uncertainties that arise from the use of incorrect chemical composition models. These differences are presented in Fig.~\ref{fig:phot_pDB_DBA_DBAZ} for the three grids best fits to the SDSS photometric data\footnote{Both the SDSS and PS1 photometry lead to consistent parameters and this is just a choice to show the general trend. All the individual results can be found in  Appendix~\ref{app:appen_indiv}.}. 

The addition of hydrogen to the model spectra (pure He $\rightarrow$ He+H) produces an overall drop in the best-fit \Teff\ and \logg\ (on average, $440$\,K and $0.06$\,dex, respectively and thus \Teff,\logg$^{\rm{He+H}} <$ \Teff,\logg$^{\rm{pure He}}$). The addition of hydrogen introduces line-blanketing from this species (mostly from Ly$\alpha$), which translates into a rise of the emitted flux in the optical range to compensate for the blocked flux in the UV. Considering that we only have optical data, this might explain the drop in \Teff\ and \logg\ (these are positively correlated). The stars with larger hydrogen abundances (0827+1731, 1013+0259, 2324-0018) clearly stand out with bigger deviations between the pure He and He+H results.

However, we see the opposite trend after adding metals (He+H $\rightarrow$ He+H+Z): both \Teff\ and \logg\ increase (on average, $117$\,K and $0.01$\,dex, respectively and thus \Teff, \logg$^{\rm{He+H+Z}} >$ \Teff, \logg$^{\rm{He+H}}$). This trend is at odds with the one obtained for the metal-polluted helium-dominated white dwarf GD\,424 \citep{izquierdo20}, where a He+H and He+H+Z analysis was performed and the results showed an increase of both \Teff\ and \logg\ when adding metals. For this sample, a further analysis focused on this matter will be needed to disentangle the behaviour of \Teff\ from that of \logg. The blanketing effect that the metals produce, which dominates in the UV where most metallic absorption lines reside, is expected to increase the emitted radiation towards redder wavelengths and hence rise the \Teff. However, in our analysis, there is an additional free parameter, \logg, which is strongly correlated to the \Teff.

We note that the differences obtained by comparing SDSS, PS1 and \textit{Gaia} eDR3 are significantly smaller with the addition of metals to the models, i.e. for the He+H+Z fits (see the standard deviations in Fig.~\ref{fig:aver_differ_phot}). This highlights the more reliable estimate of the white dwarf parameters when the chemical composition of the photosphere is fully characterised.

\subsection{Comparison between spectroscopic and photometric results}
\label{sec:spec_phot_comp}

\begin{figure}
    \centering
    \includegraphics[width=0.48\textwidth]{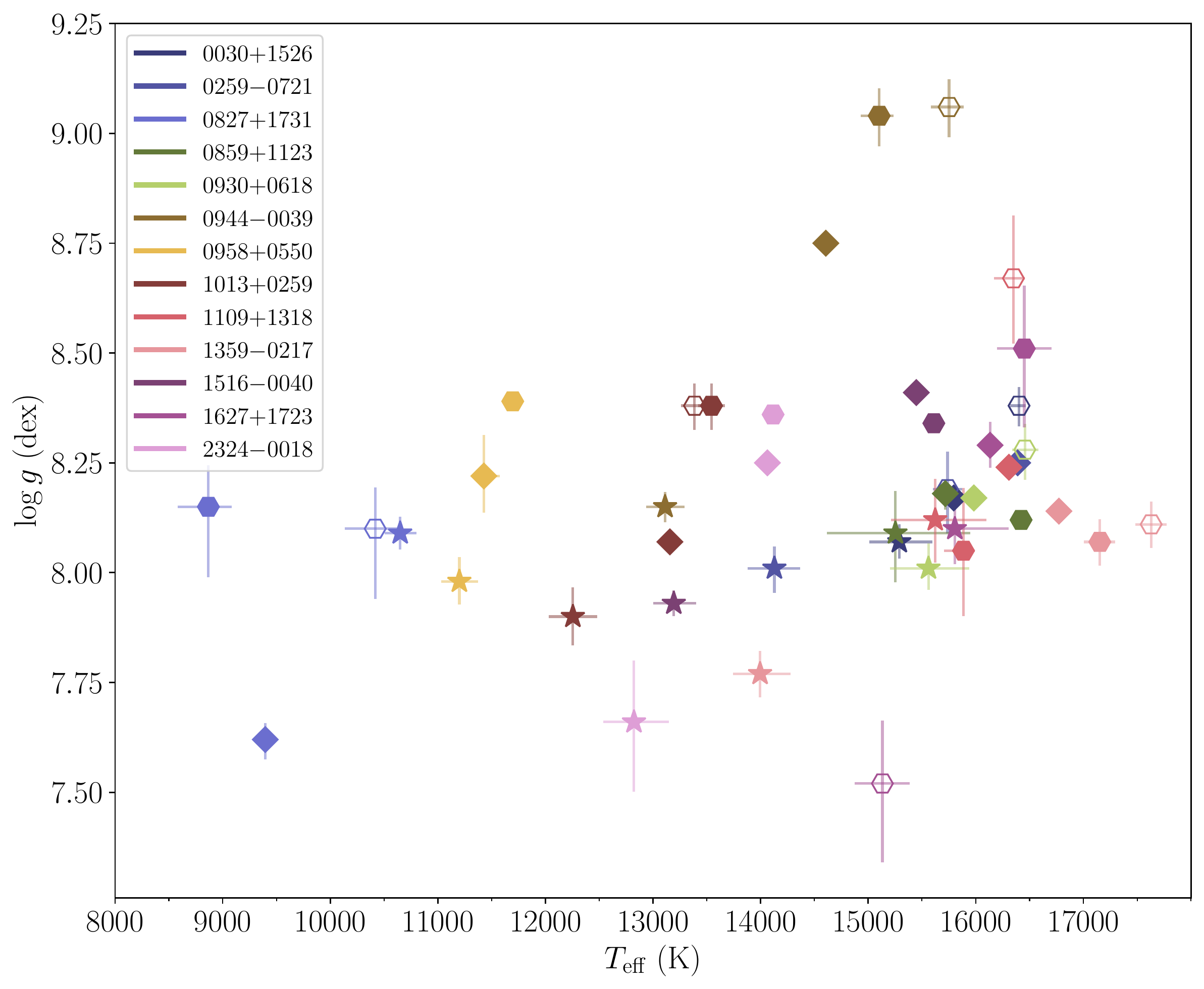}
    \caption{Atmospheric parameters obtained by fitting the SDSS+PS1 photometric data sets (stars), the X-shooter spectra (diamonds) and the BOSS and SDSS spectra (filled and open hexagons, respectively) with He+H+Z synthetic models.}
    \label{fig:spec_phot_DBA}
\end{figure}

In Fig.\,\ref{fig:spec_phot_DBA}, the \Teff\ and \logg\ obtained from the best fits to the X-shooter, BOSS and SDSS\footnote{The inclusion of BOSS and SDSS spectroscopic results in Fig.\,\ref{fig:spec_phot_DBA} highlights the important differences obtained between distinct methods and data sets, but note that only the numerical comparison between the X-shooter spectroscopic and SDSS+PS photometric parameters is calculated.} spectra are compared to the photometric results using the SDSS+PS1 fluxes and the He+H+Z synthetic models.

Comparing the X-Shooter spectroscopic results with those retrieved by fitting the SDSS+PS1 photometry shows that $\Teff_{\rm{spec}}$ is, on average, $950$\,K larger than $\Teff_{\rm{phot}}$. The same behaviour is obtained for the surface gravity, where $\logg_{\rm{spec}}$ is $0.22$\,dex larger than $\logg_{\rm{phot}}$. Despite the large overall differences between the parameters provided by the spectroscopic and photometric fits, we note an important decrease in these deviations for white dwarfs with $T_\mathrm{eff,phot} \geq 15\,000$\,K: $\left < \Teff_{\rm{spec}} - \Teff_{\rm{phot}} \right> = 480$\,K and $\left < \logg_{\rm{spec}} - \logg_{\rm{phot}} \right> = 0.13$\,dex. This fact reflects the yet unsolved issues with the broadening mechanisms of the neutral helium lines, which notably affects the spectroscopic method (the \Teff\ and \logg\ are measured from the width and depth of the absorption lines), but do not affect the photometric analysis. These significant differences between the spectroscopic and photometric results have been previously highlighted in the literature (Section~\ref{sec:past_studies}) and a forthcoming analysis, with a different sample that just contains objects above 15\,000\,K, is necessary to test the suitability of the spectroscopic, photometric and hybrid techniques to determine what is the most reliable method to characterise the population of helium-dominated white dwarfs with traces of hydrogen (and metals).

The goal of this paper was to assess the magnitude of systematic errors~--~which are often overlooked~--~that arise from the characterisation of white dwarfs with helium-dominated photospheres. Whereas we demonstrated the discrepancy in the atmospheric parameters derived from different photometric and spectroscopic data sets, there is currently no straight-forward answer to the question ``\textit{which are the most reliable parameters}''. Based on our experience, the photometric method based on SDSS and PS1 data, when using the appropriate models for the given atmospheric composition of a star, provides consistent results for \Teff\ and \logg. Turning to the analysis of different spectroscopic data sets, one would ideally obtain multiple observations of each star, in the hope that the differences in the resulting parameters average out.

Looking beyond the topic of systematic uncertainties, there are a range of studies of individual white dwarfs that require \Teff\ and \logg\ as a starting point for more detailed analyses, such as measuring the photospheric metal abundances. We will present such an analysis for the 13 stars used here in a forthcoming paper. Given the characteristics of this sample (helium-dominated white dwarfs with $\Teff 	\lesssim 15\,000$\,K) the photospheric parameters are derived by means of an iterative method \citep[similar to that employed in][]{izquierdo20}, where the \Teff\ and \logg\ are obtained from the photometric fit of SDSS+PS1 photometry and the \htohe\ from the X-shooter spectroscopy. Then, we fix those parameters to measure the photospheric metal abundances and translate them into parent body planetesimal composition. 

\subsection{Previously published results}
\label{sec:previous_results_DBAZ}

The 13 white dwarfs presented in this work have been previously characterised by \citet{eisenstein06a}, \citet{kleinman13}, \citet{koester15}, \citet{kepler15}, \citet{coutu19} and/or \citet{gentile21}\footnote{Each star has been examined by at least four of the cited studies.}. Their atmospheric parameters are listed in Tables~\ref{tab:prev_DBAZresults1} and \ref{tab:prev_DBAZresults2} along with the ones obtained in this analysis. We chose the X-shooter spectroscopic results since this is the only data set common to the 13 white dwarfs and it has the highest spectral resolution and wavelength coverage. The selection of the SDSS+PS1 photometric results was based on the consistency of the parameter values between the two catalogues, the lack of photometry issues reported in the literature and our previous experience with the white dwarf GD\,424 \citep{izquierdo20}. As described earlier, the He+H+Z synthetic models most realistically treat the complex chemical composition of the studied white dwarfs. In what follows, we compare our spectroscopic and photometric results with the atmospheric parameters given in the literature in terms of average differences.


%

\citet{eisenstein06a} performed spectroscopic and photometric fits to SDSS DR4 data with the latest version available at the time of publication of D.~Koester's DA and DB synthetic models (ML2/$\alpha=0.6$). They used {\sc autofit} \citep{kleinman04}, an automatic fitting technique based on $\chi^{2}$ minimisation, where the model spectra can be freely re-fluxed to incorporate flux calibration errors and unreliable or unknown reddening. To overcome the degeneracies produced by similar strengths and profiles of the absorption lines, they calculated the synthetic SDSS colours of the best-fit models yielded by the spectroscopic fits and compared them to the observed colours. They adopted the parameters that delivered the lowest $\chi^{2}$. We found average differences from our X-shooter spectroscopic parameters and theirs of $\left < \Delta \Teff/\Teff \right >_{\rm{spec}} = 0.03$ and $\left < \Delta \logg \right >_{\rm{spec}} = -0.21$\,dex, while the comparison of their parameters with our photometric SDSS+PS1 ones provide $\left < \Delta \Teff/\Teff \right >_{\rm{phot}} = -0.08$ and $\left < \Delta \logg \right >_{\rm{phot}}= -0.53$\,dex. The large differences found for the photometric fits are expected since \citeauthor{eisenstein06a}'s analysis relied mostly on the spectroscopic data, and our photometric fits largely benefit from knowledge of the distances (unknown at the time). Besides, these results are in agreement with our findings presented in Section~\ref{sec:spec_phot_comp}, where spectroscopy leads to much higher \Teff\ and \logg\ than those derived from photometric data.

\citet{kleinman13} carried out the same analysis as \citeauthor{eisenstein06a} but with SDSS DR7 spectroscopy and photometry data. \citeauthor{kleinman13} used improved model atmospheres \citep[we refer the reader to][for further details]{koester09-1,koester10-1} and $\alpha=1.25$. In this case, we find $\left < \Delta \Teff/\Teff \right >_{\rm{spec}} = 0.01$ and $\left < \Delta \logg \right >_{\rm{spec}} = -0.31$\,dex and $\left < \Delta \Teff/\Teff \right >_{\rm{phot}} = -0.04$ and $\left < \Delta \logg \right >_{\rm{phot}} = -0.48$\,dex. The increase in deviation in the spectroscopic \logg\ with respect to \citeauthor{eisenstein06a}'s sample is due to the new member additions, in particular 0827+1731, for which \citeauthor{kleinman13} obtained $\logg = 9.59 \pm 0.3$\,dex, very far from our $\logg = 7.62 \pm 0.04$\,dex.

We have 11 white dwarfs in common with \citet{koester15}'s sample, but they only estimated the \logg\ for five of them\footnote{We refer the reader to Section\,\ref{sec:past_studies} and \cite{koester15} for details on their model atmospheres and fitting techniques.}. The derived differences are $\left < \Delta \Teff/\Teff \right >_{\rm{spec}} = 0.04$ and $\left < \Delta \logg \right >_{\rm{spec}} = -0.08$\,dex, and $\left < \Delta \Teff/\Teff \right >_{\rm{phot}} = -0.03$ and $\left < \Delta \logg \right >_{\rm{phot}} = -0.32$\,dex. Although the synthetic spectra are similar (we used an updated, improved version of D.~Koester's models), our fitting techniques differ considerably as described in Sections~\ref{sec:past_studies} and \ref{sec:method_DBAZ13}, which may explain the deviations. The large discrepancy between \citeauthor{koester15}'s \logg\ and our photometric \logg\ is completely dominated by the object $2324-0018$, for which they derived $\logg=9.43$\,dex.

The third white dwarf catalogue based on SDSS DR10 spectra was published by \citet{kepler15}. They used {\sc autofit} to characterise three of the 13 white dwarfs of our sample. We find $\left < \Delta \Teff/\Teff \right >_{\rm{spec}} = -0.05$ and $\left < \Delta \logg \right >_{\rm{spec}} = -0.03$\,dex, and $\left < \Delta \Teff/\Teff \right >_{\rm{phot}} = -0.12$ and $\left < \Delta \logg \right >_{\rm{phot}} = -0.30$\,dex. As previously outlined, the smaller deviations between their results and our spectroscopic parameters compared to our photometric ones are the result of similar techniques.

\citet{coutu19} presented an iterative analysis of spectroscopic and photometric data of 1023 DBZ/DZ(A) white dwarfs, which contains four of the 13 white dwarfs in our sample. Briefly, their atmospheric parameter determination relied on  a first photometric fit to SDSS photometry, if available, and alternatively PS1 or \textit{Gaia} DR2 data, in that priority order, with \Teff\ and the solid angle as free parameters and fixed \logg, \htohe\ and $\log{\rm{(Ca/He})}$. From the best-fit solid angle value and the known $D$, they computed the \logg\ from interpolation of the evolutionary models by \cite{fontaineetal01-1} and performed the photometric fit with this new \logg\ fixed. This photometry fitting process is repeated until convergence is achieved. Then, they fit the available spectra (mainly retrieved from SDSS DR14, but also from \citealt{bergeron97,bergeron01}, \citealt{subasavage07}, \citealt{limoges13,limoges15} or archival data
obtained by the Montreal group) with the solid angle, \htohe\ and $\log{\rm{(Ca/He})}$ as free parameters and \Teff\ and \logg\ fixed to the best photometric fit values. The resulting \htohe, $\log{\rm{(Ca/He})}$ and spectroscopic \logg\ (as derived from the spectroscopic solid angle and $D$ by interpolation of evolutionary models) is then fixed in a subsequent photometric fit. This whole photometric-spectroscopic sequential process is repeated until \Teff, \logg, \htohe\ and $\log{\rm{(Ca/He})}$ arrived at steady solutions. 

The comparison of \citeauthor{coutu19}'s results with our best-fit parameters led to $\left < \Delta \Teff/\Teff \right >_{\rm{spec}} = 0.05$ and $\left < \Delta \logg \right >_{\rm{spec}} = 0.21$\,dex and $\left < \Delta \Teff/\Teff \right >_{\rm{phot}} = 0.02$ and $\left < \Delta \logg \right >_{\rm{phot}} = 0.01$\,dex. The large difference in the spectroscopic \logg\ is probably related to our spectroscopic method, since, as previously mentioned, this technique fails to deliver reliable \logg\ values for \Teff\ below $15\,000$\,K, which happens to be the case for the white dwarfs in common with \citet{coutu19}.

\citet{gentile21} compiled a catalogue of potential white dwarfs retrieved from \textit{Gaia} eDR3, which contains our 13 helium-dominated stars. Their white dwarf candidates were characterised by means of \textit{Gaia} eDR3 photometry in a similar way as described in Section~\ref{sec:phot_DBAZ13}: they computed the synthetic magnitudes using DA, DB and mixed hydrogen-helium models \citep{bergeronetal11-1,tremblayetal11-2,tremblay14,mccleery20} and the $G_{\rm{RP}}$, $G$ and $G_{\rm{BP}}$ passbands, scaling the model spectra to the solid angle of the star using the evolutionary models of \cite{bedard20} and comparing with the published dereddened \textit{Gaia} eDR3 magnitudes\footnote{\citet{gentile21} used an unpublished 3D map of Galactic interstellar dust to derive the extinction of the sources (Vergely et al., in preparation).}. A comparison of their photometric parameters with our spectroscopic ones leads to $\left < \Delta \Teff/\Teff \right > = 0.05$ and $\left < \Delta \logg \right > = 0.21$\,dex; and with our SDSS+PS1 photometric ones to $\left < \Delta \Teff/\Teff \right > = -0.002$ and $\left < \Delta \logg \right > = -0.01$\,dex. Since we have also performed photometric fits to the \textit{Gaia} eDR3 data, we can compare our results with theirs and find $\left < \Delta \Teff/\Teff \right > = -0.01$ and $\left < \Delta \logg \right > = -0.01$\,dex. The differences may arise from the use of different synthetic models with different chemical composition, but the use of distinct reddening values are also a possibility.

\begin{table*}
\caption{Literature results from: (1) \citet{eisenstein06a}, (2) \citet{kleinman13}, (3) \citet{koester15}, (4) \citet{kepler15}, (5) \citet{coutu19}, (6) \citet{gentile21}, (7) and (8) X-shooter spectroscopic and SDSS+PS1 photometric fits presented in this paper, respectively.  The sixth column states the synthetic spectra composition used in the fitting, where bracketed letters mark the estimation of those elements by independent fits (we refer to Section\,\ref{sec:past_studies} and the main text for further details).}
\begin{tabular}{c|cccccc}
\label{tab:prev_DBAZresults1}
Star & $\Teff$ (K) & $\logg$ (dex) & $\htohe$ (dex) & $\log{\rm{(Ca/He)}}$ (dex) & Synthetic spec & Refs.\\
\hline
          &                 &                  &     &     &    &     \\  
0030+1526 & $16728 \pm 72$  & $8.30 \pm 0.04$  & $-$ & $-$ & He & (1) \\  
          & $16133 \pm 77$  & $8.30 \pm 0.05$  & $-$ & $-$ & He & (2) \\  
          & $16065 \pm 47$  & $8.10 \pm 0.04$  & $-4.62 \pm 0.15$ & $-7.01 \pm 0.08$ & He(+H+Z) & (3) \\
          & $14621 \pm 664$ & $8.00	\pm 0.10$  & $-$ & $-$ & He & (6) \\  
          & $14524 \pm 649$ & $8.00	\pm 0.10$  & $-5.0$ & $-$ & He+H & (6) \\  
          & $15795 \pm 27$ & $8.18 \pm 0.02$ & $-5.01 \pm 0.02$ & $-7.60$& He+H+Z & (7)\\
          & $15285 \pm 300$ & $8.07 \pm 0.04$ & $-5.01$ &          $-7.60$ & He+H+Z & (8)\\[0.15cm]

0259--0721 & $16128 \pm 124$ & $8.27 \pm 0.08$  & $-$ & $-$ & He & (1) \\ 
          & $15565 \pm 139$ & $8.19 \pm 0.10$  & $-$ & $-$ & He & (2) \\ 
          & $15433 \pm 74$ & $8.0$  & $<-5.37$ & $-6.77 \pm 0.22$ & He(+H+Z) & (3) \\
          & $13298 \pm 1263$ & $7.89 \pm 0.19$ & $-$ & $-$ & He & (6) \\ 
          & $13211 \pm 1293$ & $7.89 \pm 0.20$ & $-5.0$ & $-$ & He+H & (6) \\
          & $16390 \pm 28$ & $8.26 \pm 0.02$ & $-6.04 \pm 0.08$ & $-6.24$ & He+H+Z & (7)\\
          & $14128 \pm 250$ & $8.01 \pm 0.06$ & $-6.14$ &         $-6.24$ & He+H+Z & (8)\\ [0.15cm] 

0827+1731 & $12003 \pm 329$ & $	9.59 \pm 0.3$ & $-$ & $-$ & He & (2)\\
          & $10537 \pm 382$ & $	8.06 \pm 0.08$ & $-4.27 \pm 0.07$ & $-$ & He+H+Z  & (5)\\
          & $11544 \pm 453$ & $8.27 \pm	0.08$ & $-$ & $-$ & He & (6) \\ 
          & $11276 \pm 513$ & $8.23 \pm	0.10$ & $-5.0$ & $-$ & He+H & (6) \\
          & $9397^{+96}_{76}$ & $7.62 \pm 0.04$ & $-4.17 \pm 0.03$ & $-9.93$ & He+H+Z & (7) \\
          & $10651 \pm 154$ & $8.09 \pm	0.04$ & $-4.17$ &            $-9.93$ & He+H+Z & (8) \\[0.15cm]

0859+1123 & $16078 \pm 93$ & $8.20 \pm 0.07$ & $-4.39 \pm 0.23$ & $-6.35 \pm 0.27$ & He(+H+Z) & (3) \\
          & $16145 \pm 99$ & $8.14 \pm 0.06$ & $-$ & $-$ & He & (4) \\
          & $12964 \pm 1505$ & $7.84 \pm 0.29$ & $-$ & $-$ & He & (6) \\ 
          & $12861 \pm 1573$ & $7.83 \pm 0.31$ & $-5.0$ & $-$ & He+H & (6) \\
          & $15717 \pm 63$ & $8.19 \pm 0.04$ & $-4.84 \pm 0.04$ & $-6.71$ & He+H+Z & (7) \\
          & $15253 \pm 698$ & $8.09 \pm 0.10$ & $-4.86$ &         $-6.71$ & He+H+Z & (8) \\[0.15cm]

0930+0618 & $16817 \pm 73$ & $8.14 \pm 0.04$ & $-$ & $-$ & He & (2) \\
          & $16583 \pm 56$ & $8.03 \pm 0.04$ & $-4.72 \pm 0.26$  & $-6.55 \pm 0.10$ & He(+H+Z) & (3) \\ 
          & $17474 \pm 2092$ & $8.18 \pm 0.21$ & $-$ & $-$ & He & (6) \\ 
          & $17409 \pm	2132$ & $8.19 \pm 0.21$ & $-5.0$ & $-$ & He+H & (6) \\
          &$15982 \pm 41$ & $8.18 \pm 0.02$ & $-4.87 \pm 0.04$ & $-7.11$ & He+H+Z & (7) \\
          & $15560 \pm 380$ & $8.01 \pm 0.06$ & $-4.9$ &           $-7.11$ & He+H+Z & (8) \\ [0.15cm]

0944--0039 & $15522 \pm 76$ & $9.00 \pm 0.01$ &  & & He & (1) \\
          & $14592 \pm 144$ & $8.82 \pm 0.09$ &  & & He & (2)\\ 
          & $14057 \pm 62$ & $8.00$ & $<-5.75$ & $-7.14 \pm 0.10$ & He(+H+Z) & (3)\\
          & $12625 \pm 604$ & $	8.13 \pm 0.07$ & $<-6.08$ & $-$ & He+H+Z & (5)\\
          & $12744 \pm 598$ & $8.11 \pm 0.10$ & $-$ & $-$ & He & (6) \\ 
          & $12623 \pm 634$ & $8.10 \pm 0.11$ & $-5.0$ & $-$ & He+H & (6) \\
          & $14607 \pm 45$ & $8.76 \pm 0.02$ & $-5.87 \pm 0.05$ & $-7.58$ & He+H+Z & (7) \\
          & $13113 \pm 180$ & $8.15 \pm 0.04$ & $-5.81$ &         $-7.58$ & He+H+Z & (8) \\[0.15cm]

0958+0550 & $11684 \pm 83$ & $8.0$ & $-5.62 \pm  0.40$ & $-8.75 \pm 0.11$ & He(+H+Z)  & (3) \\ 
          & $12955 \pm 171$ & $8.54 \pm 0.1$ & $-$ & $-$ & He & (4) \\ 
          & $10960 \pm 402$ & $8.0$ & $-5.84 \pm 0.25$ & $-8.66 \pm 0.09$ & He+H+Z  & (5) \\
          & $10861 \pm 558$ & $7.92 \pm 0.13$ & $-$ & $-$ & He & (6) \\ 
          & $10540 \pm 597$ & $7.84 \pm 0.15$ & $-5.0$ & $-$ & He+H & (6) \\
          & $11428 _{-110}^{+149}$ & $8.22 \pm 0.09$ & $-5.82 \pm 0.07$ & $-8.89$ & He+H+Z & (7) \\
          & $11201 \pm 176$ & $7.99 \pm 0.06$ & $-5.64$ &                $-8.89$ & He+H+Z & (8) \\[0.15cm]

1013+0259 & $8512 \pm 24$ & $9.00 \pm 0.01$ & $-$  & $-$ & He & (1)\\
          & $8351 \pm 42$ & $9.09 \pm 0.06$ & $-$ & $-$ & He & (2)\\ 
          & $12428 \pm 1154$ & $7.97 \pm 0.21$ & $-$ & $-$ & He & (6) \\ 
          & $12294 \pm 1263$ & $7.96 \pm 0.24$ & $-5.0$ & $-$ & He+H & (6) \\
          & $13158 \pm 27$ & $8.08 \pm 0.02$ & $-3.13 \pm 0.01$ & $-8.37$ & He+H+Z & (7) \\
          & $12255 \pm 225$ & $7.90 \pm 0.07$ & $-3.13$ &         $-8.37$ & He+H+Z & (8) \\[0.15cm]
          
\hline
\end{tabular}
\end{table*}

\begin{table*}
\caption{Literature results from: (1) \citet{eisenstein06a}, (2) \citet{kleinman13}, (3) \citet{koester15}, (4) \citet{kepler15}, (5) \citet{coutu19}, (6) \citet{gentile21}, (7) and (8) X-shooter spectroscopic and SDSS+PS1 photometric fits presented in this paper, respectively.  The sixth column states the synthetic spectra composition used in the fitting, where bracketed letters mark the estimation of those elements by independent fits (we refer to Section\,\ref{sec:past_studies} and the main text for further details).}
\begin{tabular}{c|cccccc}
\label{tab:prev_DBAZresults2}
Star & $\Teff$ (K) & $\logg$ (dex) & $\htohe$ (dex) & $\log{\rm{(Ca/He)}}$ (dex) & Synthetic spec & Refs.\\
\hline
          &                 &                  &     &     &    &     \\  
1109+1318 & $16242.0 \pm 194$ & $8.24 \pm 0.10$ & $-$ & $-$ & He & (2) \\ 
          & $16081 \pm 130$ & $8.06 \pm 0.10$ & $-3.85 \pm 0.33$ & $-6.46 \pm 0.50$ & He(+H+Z) & (3) \\
          & $16722 \pm 5342$ & $8.21 \pm 0.59$ & $-$ & $-$ & He & (6) \\ 
          & $16751 \pm 5632$ & $8.22 \pm 0.61$ & $-5.0$ & $-$ & He+H & (6) \\
          & $16308 \pm 62$  & $8.25 \pm 0.03$ & $-4.01 \pm 0.03$ & $-7.51$ & He+H+Z & (7) \\
          & $15623 \pm 480$ & $8.12 \pm 0.10$ & $-4.05$ &         $-7.51$ & He+H+Z & (8) \\[0.15cm]

1359--0217 & $17067 \pm 104$ & $8.12 \pm 0.06$ & $-$ & $-$ & He & (1) \\ 
          & $16778 \pm 123$ & $8.18 \pm 0.06$ & $-$ & $-$ & He & (2) \\ 
          & $16973 \pm 60$ & $ 7.83 \pm 0.05$ & $-3.33 \pm 0.11$ & $-6.49 \pm 0.30$ & He(+H+Z) & (3) \\
          & $16701 \pm 2238$ & $8.07 \pm 0.25$ & $-$ & $-$ & He & (6) \\ 
          & $16634 \pm 2309$ & $8.08 \pm 0.25$ & $-5.0$ & $-$ & He+H & (6) \\
          & $16773 \pm 55$ & $8.14 \pm 0.02$ & $-3.15 \pm 0.02$ & $-7.23$ & He+H+Z & (7) \\
          & $13995 \pm 285$ & $7.78 \pm 0.05$ & $-3.16$ &         $-7.23$ & He+H+Z & (8) \\[0.15cm]
          
1516--0040 & $14961 \pm 28$ & $8.0$ & $-4.47 \pm 0.10$ & $-7.38 \pm 0.20$ & He(+H+Z) & (3) \\
          & $15264 \pm 50$ & $8.21 \pm 0.01$ & $-$ & $-$ & He & (4) \\
          & $13006 \pm 735$ & $7.95 \pm 0.10$ & $-4.83 \pm 0.08$ & $-8.59 \pm 0.10$ & He+H+Z & (5) \\
          & $13081 \pm 751$ & $7.89 \pm 0.12$ & $-$ & $-$ & He & (6) \\ 
          & $12987 \pm 779$ & $7.88 \pm 0.12$ & $-5.0$ & $-$ & He+H & (6) \\
          & $15448 \pm 20$  & $8.42 \pm 0.01$ & $-4.50 \pm 0.01$ & $-7.59$ & He+H+Z & (7) \\
          & $13193 \pm 207$ & $7.94 \pm 0.03$ & $-5.0$ &          $-7.59$ & He+H+Z & (8) \\[0.15cm]

1627+1723 & $15834 \pm 174$ & $7.98 \pm 0.1$& $-$ & $-$ & He & (2)\\
          & $15795 \pm 112$ & $8.0$ & $<-5.02$ & $<-6.66$ & He(+H+Z) & (3)  \\
          & $16407 \pm 2233$ & $8.17 \pm 0.27$ & $-$ & $-$ & He & (6) \\ 
          & $16326 \pm2299$ & $8.17 \pm 0.28$ & $-5.0$ & $-$ & He+H & (6) \\
          & $16134 \pm 102$ & $8.29 \pm 0.05$ & $-5.05 \pm 0.07$ & $-7.73$ & He+H+Z & (7) \\
          & $15903 \pm 503$ & $8.11 \pm 0.09$ & $-5.13$ &          $-7.73$ & He+H+Z & (8) \\[0.15cm]

2324--0018 & $23431 \pm 697$ & $5.01 \pm 0.02$ & $-$ &  $-$ & $-$ & sdB (1) \\
          & $8231 \pm 39$ & $9.43 \pm 0.04$ & $-$ & $-$ & He & (3) \\
          & $12198 \pm 1303$ & $7.66 \pm 0.29$ & $-$ & $-$ & He & (6) \\ 
          & $12039 \pm 1473$ & $7.64 \pm 0.33$ & $-5.0$ & $-$ & He+H & (6) \\
          & $14063 \pm 53$ & $8.25 \pm 0.02$ & $-3.32 \pm 0.01$ & $-8.99$ & He+H+Z & (7) \\
          & $12823 \pm 325$ & $7.66 \pm 0.15$ & $-3.33$ &         $-8.99$ & He+H+Z & (8) \\
\hline
\end{tabular}
\end{table*}

\section{Conclusions}
\label{sec:conlus_DBAZ13}

In this paper we have determined the atmospheric parameters of 13 white dwarfs with helium-dominated photospheres, traces of hydrogen and metals from spectroscopy and photometry data and investigated the overall trends of the use of different data sets and chemical composition models. 

The use of different data sets leads to contrasting results both for spectroscopic and photometric data. The differences are in all the cases greater than the uncertainties published in individual studies. These discrepancies are most likely related to calibration issues, but differences in the spectral ranges and hence the use of different absorption lines, SNR or photometric filters cannot be ruled out. In particular:

\begin{itemize}
\item We find mean standard deviations of $524$\,K, $0.27$\,dex and $0.31$\,dex for \Teff, \logg\ and \htohe, respectively, when fitting model spectra to diverse spectroscopic data sets. These values are substantially larger than the purely statistical uncertainties usually reported in studies of helium-dominated white dwarfs (with or without traces of hyrogen/metals), and we consider them as a more realistic assessment of the overall uncertainties of the model atmosphere analysis of these stars. We suggest to quote them when spectroscopically characterising helium-dominated white dwarfs (with or without traces of hyrogen/metals), in particular, in the cool end ($\Teff \leq 15000$\,K) with just one spectroscopic data set. 

\item The photometric fits provide mean standard deviations between SDSS and PS1 data of $\left < \sigma \Teff \right > = 485$\,K and $\left < \sigma \logg \right > = 0.05$\,dex. We encourage these values to be adopted as the minimum uncertainties when publishing atmospheric parameters from SDSS or PS1 photometry for cool helium-dominated white dwarfs (with or without traces of hyrogen/metals). The mean standard deviations become larger when \textit{Gaia} eDR3 data are used: $\left < \sigma \Teff \right > = 1210$\,K and $\left < \sigma \logg \right > = 0.13$\,dex. This should be taken into account when quoting the uncertainties in the parameters derived from \textit{Gaia} eDR3 photometry data.
\end{itemize}


With the aim of investigating the effect of the assumed (often inaccurate) chemical composition on the best-fit atmospheric parameters, we carried out the data modelling using synthetic spectra of three different chemical compositions: (1) pure helium, (2) helium-dominated atmospheric models with traces of hydrogen (He+H) and (3) hydrogen plus metals in helium-dominated photospheres (He+H+Z). In general, pure helium model spectra result in larger \Teff\ than those derived from He+H, while the \logg\ differences are also notable but change from spectroscopic to photometric data. The addition of metals does also affect the best-fit parameters, but the change is less dramatic than in the previous case. In particular:

\begin{itemize}

\item The addition of hydrogen to the pure helium synthetic models (pure He $\rightarrow$ He+H) produces a drop in the derived spectroscopic \Teff\ of $300$\,K and a slight increase of $0.02$\,dex in the \logg, on average. Although the addition of metals does not translate into a significant absolute change in the average spectroscopic values ($\simeq 60$\,K, no change and $0.01$\,dex for \Teff, \logg\ and \htohe, respectively), we note it does affect the derived atmospheric parameters of each star and refer the reader to the individual results (Tables~\ref{tab:0030+1526}--\ref{tab:2324-0018}).

\item As for the photometric fits, the inclusion of hydrogen (pure He $\rightarrow$ He+H) produces a mean drop in the \Teff\ and \logg\ of $440$\,K and $0.06$\,dex, respectively, while the addition of metals (He+H $\rightarrow$ He+H+Z) results in an increase of $\simeq 120$\,K and $0.01$\,dex, on average.

\end{itemize}

The 13 white dwarfs in our sample have helium-dominated photospheres polluted with hydrogen and up to ten different metals (see Table~\ref{tab:Metal_abs_stars}). Therefore, a realistic characterisation must be based on model spectra that accurately reflect the actual chemical compositions. The above parameter differences illustrate the systematic uncertainties expected when the model grid chemical composition is not well suited for the actual data.

We also compared our spectroscopic and photometric results and find significant differences for those stars with $\Teff \leq 15\,000$\,K. This is a well-known issue due to the poor implementation of resonance and van der Waals theories for the helium atom (see Sections~\ref{sec:intro} and \ref{sec:past_studies} for more details), which affects the spectroscopic modelling but does not have an overall effect on the photometric fits, as the latter do not rely on the width and depth of the absorption lines. This can also be noticed in the literature of the white dwarfs in our sample. A future analysis, with a different sample that just contains white dwarfs above 15\,000~K, will be needed to test the suitability of the different techniques in order to find the best method to characterise helium-dominated white dwarfs (with or without hydrogen/metals).

Even though there is no straightforward recipe to obtain the most realistic parameters, based on our experience, the SDSS and PS1 photometry provide consistent results for \Teff\ and \logg\ when employing appropriate synthetic models. For the analysis of cool helium-dominated white dwarfs with spectroscopic data, we suggest to ideally obtain multiple observations to test for systematic uncertainties in the hope that such differences in the parameters average out.


\section*{Acknowledgements}
Based on observations collected at the European Southern Observatory under ESO programmes 0100.C-0500(A) and 0101.C-0646(A). PI was supported by a Leverhulme Trust Research Project Grant. PI and BTG were supported by grant ST/T000406/1 from the Science and Technology Facilities Council (STFC). This project has received funding from the European Research Council (ERC) under the European Union’s Horizon 2020 research and innovation programme (Grant agreement No. 101020057). OT was supported by a FONDECYT project 321038. This research was supported in part by the National Science Foundation under Grant No. PHY-1748958.

\section*{Data Availability Statement}

The data underlying this article will be shared on reasonable request to the corresponding author.

%
%
%
%
%
%
%
%
%
%



\bibliographystyle{mnras}
\bibliography{aa} 




\appendix

\section{Individual results}\label{app:appen_indiv}

\subsection{0030+1526}

The best-fit \Teff\ values found from the PS1 and \textit{Gaia} eDR3 photometry are consistent with each other, but differ by $\simeq -1200$ and $-1000$\,K, respectively, from the SDSS \Teff\ ($\Teff_{\rm{SDSS}}$ is larger), as derived from the He+H+Z fits. Despite the fact that \logg\ is usually consistent for the three data sets, we obtained larger SDSS values by $0.11$ and $0.08$\,dex, respectively. We have visually inspected the surrounding field of this star and did not found any contamination due to other targets nearby. We performed the SDSS photometric fits neglecting the SDSS $u$-band filter (the only one in the near-UV, and hence most affected by the hydrogen content due to the Balmer jump) and arrived at more consistent results, which points to this band being the source of the difference, favouring lower \Teff\ if we neglect it.

\subsection{0827+1731}
\label{sec:0827+1731}

The optical spectrum of this star is dominated by H$\alpha$ and H$\beta$, the strong and deep \Ion{Ca}{ii} H and K lines and a shallow \Line{He}{i}{5875} absorption line (bottom three spectra in Fig.~\ref{fig:SDSS_XS_spectra}). This is the result of its low \Teff\ ($\simeq 10500$\,K), which makes the presence of helium almost undetectable despite being the main constituent (see footnote 1). The small number of absorption lines available, the shallowness of the only helium absorption line and the low \Teff\ (note the large uncertainty of the line-broadening theory for neutral helium) yield unreliable results (Table~\ref{tab:0827+1731}). This is illustrated by the large average differences up to $\Delta \Teff \simeq 1900$\,K and $\Delta \logg \simeq 0.9$\,dex, $\Delta \htohe \simeq 0.9$\,dex between the atmospheric parameters derived from the X-shooter, SDSS and BOSS spectra for He+H and He+H+Z compositions. The photometric fits are unaffected by the dubious implementation of the helium lines broadening and show consistent results, also with those reported in the literature.



\subsection{0859+1123}

The fits to the X-shooter and BOSS spectra yield atmospheric parameters that differ from each other by $\Delta \Teff \simeq 1000$\,K, $\Delta \logg$ up to $0.07$\,dex and $\Delta \htohe$ up to $0.41$\,dex (see Table~\ref{tab:0859+1123}). This may be due to the SNR difference between the X-shooter ($\simeq 38$) and the BOSS spectra ($\simeq 20$).





\subsection{1109+1318}

Both our best-fit spectroscopic and photometric parameters are consistent with those previously reported in the literature, except for the ones inferred from the SDSS spectra. The low SNR of the spectra ($\simeq 14$) could be the source of these differences.




\subsection{1627+1723}


We find significant differences between the X-shooter and BOSS spectroscopic results, with the latter always delivering higher \Teff\ and \logg\ and smaller \htohe\ (up to $\Delta \Teff = 1020$\,K, $\Delta \logg = 0.32$\,dex and $\Delta \htohe = 0.1$\,dex, although these differences vary with the assumed chemical composition). Both the comparison between the X-shooter and SDSS spectroscopic results and  between BOSS and SDSS do not show a clear trend, with the parameter offsets considerably varying with the assumed chemical composition. All these differences are most likely originated from the lower SNR of the SDSS spectra ($\simeq 13$, while ${\rm SNR}\simeq 33$ and $29$ for X-shooter and BOSS, respectively). The spectroscopic values we obtained also differ considerably from the ones of \citet{kepler15} and \citet{koester15}, but these authors used different methodologies. 


The spectroscopic values we obtained also differ considerably from the ones of \citet{kepler15} and \citet{koester15}, but these authors used different methodologies. A further analysis of this white dwarf will be needed to explain these differences. 

\subsection{2324--0018}

Previous works on this star report inconsistent parameters, with \Teff\ ranging from $23431$ to $8231$\,K and \logg\ from $5.01$ to $9.43$\,dex \citep{eisenstein06a,koester15}, making any comparison with our parameters useless. We obtain compatible results from our two spectroscopy data sets (X-shooter and BOSS) and then among our photometric ones. On the other hand, the spectroscopic and photometric solutions display significant differences that we attribute to the reported issues with the cool models for helium-dominated white dwarfs (with or without hydrogen/metals). 

\begin{table*}
\caption{Relative metal abundances measured for the 13 white dwarfs from the analysis of the X-shooter spectra. These abundances are fixed to generate new metal-blanketed He+H+Z models.}
\vspace{0.1cm}
\footnotesize
\setlength{\tabcolsep}{1.0ex}
\begin{center}
\begin{tabular}{c|cccccccccc}
\hline
     &  &  &  &  &  &  &  & &  &  \\
Star & \multicolumn{10}{c}{$\log{\rm{(He/Z)}}$ (dex)} \\[0.15cm]
     & O & Mg & Al & Si & Ca & Ti & Cr & Mn & Fe & Ni \\[0.15cm]
\hline
     &  &  &  &  &  &  &  & &  &  \\
0030+1526 & $5.85 \pm 0.08$ & $6.99 \pm 0.04$ & $-$ & $7.03 \pm 0.10$ & $7.60 \pm 0.02$ & $-$ & $-$ & $-$ & $7.27 \pm 0.18$ & $-$ \\[0.2cm]  
0259$-$0721 & $4.87 \pm 0.05$ & $5.61 \pm 0.03$ & $6.88 \pm 0.18$ & $6.05 \pm 0.04$ & $6.24 \pm 0.02$ & $8.45 \pm 0.07$ & $8.15 \pm 0.09$ & $8.51 \pm 0.05$ & $6.38 \pm 0.13$ & $7.72 \pm 0.10$ \\[0.2cm] 
0827+1731 & $-$ & $-$ & $-$ & $-$ & $9.93 \pm 0.02$ & $10.95 \pm 0.30$ & $-$ & $-$ & $-$ & $-$ \\[0.2cm]
0859+1123 & $5.0 \pm 0.09$ & $5.92 \pm 0.04$ & $6.65 \pm 0.17$ & $6.02 \pm 0.04$ & $6.71 \pm 0.05$ & $9.05 \pm 0.11$ & $8.03 \pm 0.28$ & $8.85 \pm 0.19$ & $6.66 \pm 0.16$ & $-$ \\[0.2cm]       
0930+0618 & $4.72 \pm 0.05$ & $5.90 \pm 0.03$ & $6.98 \pm 0.21$ & $5.98 \pm 0.04$ & $7.11 \pm 0.03$ & $9.05 \pm 0.08$ & $8.26 \pm 0.16$ & $8.39 \pm 0.04$ & $6.29 \pm 0.09$ & $6.22 \pm 0.19$ \\[0.2cm]   
0944$-$0039 & $5.94 \pm 0.07$ & $6.96 \pm 0.03$ & $7.83 \pm 0.40$ & $7.18 \pm 0.13$ & $7.58 \pm 0.02$ & $9.58 \pm 0.05$ & $8.86 \pm 0.07$ & $9.42 \pm 0.29$ & $7.22 \pm 0.07$ & $-$\\[0.2cm] 
0958+0550 & $-$ &  $6.99 \pm 0.05$ & $-$ & $-$ & $8.89 \pm 0.02$ & $10.21 \pm 0.06$ & $9.09 \pm 0.19$ & $9.93 \pm 0.23$ & $7.70 \pm 0.30$ & $-$\\[0.2cm]                 
1013+0259 & $6.64 \pm 0.37$ & $7.54 \pm 0.05$ & $-$  & $-$ & $8.37 \pm 0.01$ & $10.17 \pm 0.08$ & $9.35 \pm 0.09$ & $-$ & $8.09 \pm 0.17$ & $-$ \\[0.2cm] 
1109+1318 & $5.54 \pm 0.14$ & $6.73 \pm 0.09$ & $-$ & $6.77 \pm 0.17$ & $7.51 \pm 0.03$ & $9.28 \pm 0.14$ & $8.51 \pm 0.19$ & $-$ & $6.77 \pm 0.13$ & $-$ \\[0.2cm] 
1359$-$0217 & $5.20 \pm 0.12$ & $6.32 \pm 0.08$ & $6.99 \pm 0.25$ & $6.30 \pm 0.05$ & $7.23 \pm 0.04$  & $-$ & $8.11 \pm 0.25$  & $-$ & $6.86 \pm 0.14$ & $-$ \\[0.2cm] 
1516$-$0040 & $5.89 \pm 0.04$ & $6.82 \pm 0.03$ & $7.50 \pm 0.31$ & $7.04 \pm 0.06$ & $7.59 \pm 0.02$ & $9.86 \pm 0.11$ & $9.03 \pm 0.21$ & $9.63 \pm 0.14$ & $7.00 \pm 0.08$ & $-$ \\[0.2cm] 
1627+1723 & $5.96 \pm 0.29$ & $6.85 \pm 0.15$ & $7.18 \pm 0.41$ & $7.07 \pm 0.35$ & $7.73 \pm 0.06$ & $9.2 \pm 0.31$ & $-$ & $-$ & $6.78 \pm 0.19$ & $-$\\[0.2cm] 
2324$-$0018 & $-$ & $8.09 \pm 0.25$ & $-$ & $-$ & $8.99 \pm 0.02$ & $10.79 \pm 0.15$ & $-$ & $-$ & $-$ & $-$ \\[0.2cm] 
\hline
\end{tabular}
\end{center}
\label{tab:Metal_abs_stars}
\end{table*}

\begin{table*}
\caption{Spectroscopic (Spec) and photometric (Phot) fit results for 0030+1526. Parameters without uncertainties have been fixed to the given value.}
\vspace{0.1cm}
\footnotesize
\setlength{\tabcolsep}{0.9ex}
\begin{center}
\begin{tabular}{l|cccccccc}
\hline
 & & & & & & & & \\[0.05cm]
0030+1526 & \multicolumn{2}{c}{He} & \multicolumn{3}{c}{He+H} & \multicolumn{3}{c}{He+H+Z}\\[0.12cm]
 & $\Teff$ (K) & $\logg$ (dex) & $\Teff$ (K) & $\logg$ (dex) & $\htohe$ (dex) & $\Teff$ (K) & $\logg$ (dex) & $\htohe$ (dex) \\[0.1cm]
 \hline
 & & & & & & & & \\
Spec XS & $16054 \pm 32$ & $8.08 \pm 0.02$ & $15857 \pm 32$ & $8.30 \pm 0.02$ & $-4.96 \pm 0.02$ & $15795 \pm 27$ & $8.18 \pm 0.02$ & $-5.01 \pm 0.02$\\[0.2cm]
Spec SDSS & $16355 \pm 97$ & $8.39 \pm 0.05$ & $16088 \pm 96$ & $8.50 \pm 0.06$ & $-4.91 \pm 0.05$ & $16402_{-85}^{+61}$ & $8.38 \pm 0.05$ & $-4.81 \pm 0.07$\\[0.2cm]
Phot PS1 & $14564^{+515}_{-472} $ & $7.99 \pm 0.05$ & $14329_{-499}^{+533}$ & $7.98 \pm 0.05$ & $-4.94$ & $14409_{-404}^{+520}$ & $7.99 \pm 0.05$ & $-5.01$ \\[0.2cm]
Phot Gaia & $14856^{+431}_{-396} $ & $8.03 \pm 0.05$ & $14623_{-407}^{+439}$ & $8.02 \pm 0.05$ & $-4.94$ & $14622_{-362}^{+422}$ & $8.02 \pm 0.05$ & $-5.01$ \\[0.2cm]
Phot SDSS & $15843^{+430}_{-393} $ & $8.11 \pm 0.04$ & $15608_{-407}^{+474}$ & $8.10 \pm 0.05$ & $-4.94$ & $15601 \pm 390$ & $8.10 \pm 0.05$ & $-5.01$ \\[0.2cm]
\hline
\end{tabular}
\end{center}
\label{tab:0030+1526}
\end{table*}

\begin{table*}
\caption{Same as Table~\ref{tab:0030+1526} but for 0259--0721.}
\vspace{0.1cm}
\footnotesize
\setlength{\tabcolsep}{0.9ex}
\begin{center}
\begin{tabular}{l|cccccccc}
\hline
 & & & & & & & & \\[0.05cm]
0259$-$0721 & \multicolumn{2}{c}{He} & \multicolumn{3}{c}{He+H} & \multicolumn{3}{c}{He+H+Z}\\[0.12cm]
& $\Teff$ (K) & $\logg$ (dex) & $\Teff$ (K) & $\logg$ (dex) & $\htohe$ (dex) & $\Teff$ (K) & $\logg$ (dex) & $\htohe$ (dex) \\[0.1cm]
 \hline
 & & & & & & & & \\
Spec XS & $15956 \pm 38$ & $8.11 \pm 0.02$ & $16220 \pm 36$ & $8.05 \pm 0.02$ & $-6.8 \pm 0.5$ & $16390 \pm 28$ & $8.26 \pm 0.02$ & $-6.04 \pm 0.08$\\[0.2cm]
Spec SDSS & $15769 \pm 150$ & $8.44 \pm 0.05$ & $15810 ^{+162}_{-152}$ & $8.06 \pm 0.11$ & $-6.7 \pm 0.8$ & $15738 \pm 138$ & $8.19 \pm 0.10$ & $-6.3 \pm 0.6$\\[0.2cm]
Phot PS1 & $14302^{+815}_{-700} $ & $8.02 \pm 0.09$ & $14297_{-682}^{+800}$ & $8.02 \pm 0.05$ & $-6.75$ & $14090_{-544}^{+780}$ & $8.02 \pm 0.09$ & $-6.04$ \\[0.2cm]
Phot Gaia & $13435^{+853}_{-820} $ & $7.93 \pm 0.10$ & $13273_{-750}^{+907}$ & $7.92 \pm 0.1$ & $-6.75$ & $13812_{-407}^{+700}$ & $8.00 \pm 0.08$ & $-6.04$ \\[0.2cm]
Phot SDSS & $13947^{+430}_{-393} $ & $7.97 \pm 0.04$ & $13916_{-272}^{+314}$ & $7.97 \pm 0.05$ & $-6.75$ & $14119 \pm 269$ & $8.01 \pm 0.06$ & $-6.04$ \\[0.2cm]
\hline
\end{tabular}
\end{center}
\label{tab:0259-0721}
\end{table*}

\begin{table*}
\caption{Same as Table~\ref{tab:0030+1526} but for 0827+1731.}
\vspace{0.1cm}
\footnotesize
\setlength{\tabcolsep}{0.9ex}
\begin{center}
\begin{tabular}{l|cccccccc}
\hline
 & & & & & & & & \\[0.05cm]
0827+1731 & \multicolumn{2}{c}{He} & \multicolumn{3}{c}{He+H} & \multicolumn{3}{c}{He+H+Z}\\[0.12cm]
& $\Teff$ (K) & $\logg$ (dex) & $\Teff$ (K) & $\logg$ (dex) & $\htohe$ (dex) & $\Teff$ (K) & $\logg$ (dex) & $\htohe$ (dex) \\[0.1cm]
 \hline
 & & & & & & & & \\
Spec XS & $-$ & $-$ & $8696^{+68}_{-57}$ & $7.37 \pm 0.04$ & $-3.93 \pm 0.03$ & $9397^{+96}_{-76}$ & $7.62 \pm 0.04$ & $-4.17 \pm 0.03$\\[0.2cm]
Spec BOSS & $-$ & $-$ & $9173 ^{+1077}_{-732}$ & $8.3 \pm 0.3$ & $-3.4 \pm 0.3$ & $8867 ^{+241}_{-217}$ & $8.15 \pm 0.09$ & $-3.38 \pm 0.08$\\[0.2cm]
Spec SDSS & $-$ & $-$ & $10591 ^{+334}_{-345}$ & $8.2 \pm 0.2$ & $-4.27 \pm 0.07$ & $10418^{+218}_{-283}$ & $8.10 \pm 0.16$ & $-4.29 \pm 0.08$\\[0.2cm]
Phot PS1 & $11334^{+388}_{-368} $ & $8.25 \pm 0.09$ & $10468_{-317}^{+350}$ & $8.07 \pm 0.07$ & $-3.93$ & $10561_{-346}^{+338}$ & $8.08 \pm 0.07$ & $-4.17$ \\[0.2cm]
           & $-$ & $-$                                & $10475_{-316}^{+340}$ & $8.07 \pm 0.07$ & $-3.4$ & $-$ & $-$ & $-$ \\[0.2cm]
           & $-$ & $-$                                & $10538_{-323}^{+370}$ & $8.10 \pm 0.07$ & $-4.27$ & $-$ & $-$ & $-$ \\[0.2cm]
Phot Gaia & $11700 \pm 298$ & $8.31 \pm 0.04$ & $10691_{-254}^{+272}$ & $8.12 \pm 0.05$ & $-3.93$ & $10811 \pm 260 $ & $8.13 \pm 0.05$ & $-4.17$ \\[0.2cm]
            & $-$ & $-$                         & $10628_{-257}^{+283}$ & $8.11 \pm 0.05$ & $-3.4$  & $-$ & $-$ & $-$ \\[0.2cm]
            & $-$ & $-$                         & $10794_{-263}^{+304}$ & $8.15 \pm 0.05$ & $-4.27$ & $-$ & $-$ & $-$ \\[0.2cm]
Phot SDSS & $11493^{+156}_{-165} $ & $8.27 \pm 0.03$ & $10474_{-139}^{+154}$ & $8.07 \pm 0.04$ & $-3.93$ & $10653 _{-163}^{+180}$ & $8.09 \pm 0.04$ & $-4.17$ \\[0.2cm]
            & $-$ & $-$                                & $10400_{-132}^{+142}$ & $8.05 \pm 0.04$ & $-3.4$  & $-$ & $-$ & $-$ \\[0.2cm]
            & $-$ & $-$                                & $10617_{-157}^{+168}$ & $8.11 \pm 0.04$ & $-4.27$ & $-$ & $-$ & $-$ \\[0.2cm]
\hline
\end{tabular}
\end{center}
\label{tab:0827+1731}
\end{table*}

\begin{table*}
\caption{Same as Table~\ref{tab:0030+1526} but for 0859+1123.}
\vspace{0.1cm}
\footnotesize
\setlength{\tabcolsep}{0.9ex}
\begin{center}
\begin{tabular}{l|cccccccc}
\hline
 & & & & & & & & \\[0.05cm]
0859+1123 & \multicolumn{2}{c}{He} & \multicolumn{3}{c}{He+H} & \multicolumn{3}{c}{He+H+Z}\\[0.12cm]
& $\Teff$ (K) & $\logg$ (dex) & $\Teff$ (K) & $\logg$ (dex) & $\htohe$ (dex) & $\Teff$ (K) & $\logg$ (dex) & $\htohe$ (dex) \\[0.1cm]
 \hline
 & & & & & & & & \\
Spec XS & $15898 \pm 77$ & $8.03 \pm 0.04$ & $15629 \pm 85$ & $8.20 \pm 0.05$ & $-4.84 \pm 0.04$ & $15717 \pm 63$ & $8.19 \pm 0.04$ & $-4.86 \pm 0.04$\\[0.2cm]
Spec BOSS & $16948 ^{+193}_{-152}$ & $8.19 \pm 0.11$ & $16709 ^{+140}_{-122}$ & $8.21 \pm 0.08$ & $-4.43 \pm 0.08$ & $16422 ^{+114}_{-118}$ & $8.13 \pm 0.06$ & $-4.50 \pm 0.07$\\[0.2cm]
Phot PS1 & $16051^{+1932}_{-1460} $ & $8.18 \pm 0.17$ & $15600_{-1381}^{+1646}$ & $8.14 \pm 0.17$ & $-4.84$ & $14860_{-1078}^{+1127}$ & $8.07 \pm 0.15$ & $-4.86$ \\[0.2cm]
           & $-$ & $-$                                  & $15662_{-1550}^{+1750}$ & $8.15 \pm 0.19$ & $-4.43$ & $-$ & $-$ & $-$ \\[0.2cm]
Phot Gaia & $13129 ^{+1131}_{-843}$ & $7.88 \pm 0.14$ & $13027_{-700}^{+1023}$ & $7.88 \pm 0.16$ & $-4.84$ & $13265_{-517}^{+967} $ & $7.94 \pm 0.12$ & $-4.86$ \\[0.2cm]
            & $-$ & $-$                                 & $13103_{-750}^{+1427}$ & $7.9 \pm 0.2$ & $-4.43$ & $-$ & $-$ & $-$ \\[0.2cm]
Phot SDSS & $15613^{+866}_{-722} $ & $8.13 \pm 0.10$  & $15359_{-745}^{+975}$ & $8.12 \pm 0.14$ & $-4.84$ & $15215 _{-645}^{+736}$ & $8.10 \pm 0.12$ & $-4.86$ \\[0.2cm]
            & $-$ & $-$                                 & $15377_{-876}^{+988}$ & $8.12 \pm 0.14$ & $-4.43$ & $-$ & $-$ & $-$ \\[0.2cm]
\hline
\end{tabular}
\end{center}
\label{tab:0859+1123}
\end{table*}

\begin{table*}
\caption{Same as Table~\ref{tab:0030+1526} but for 0930+0618.}
\vspace{0.1cm}
\footnotesize
\setlength{\tabcolsep}{0.9ex}
\begin{center}
\begin{tabular}{l|cccccccc}
\hline
 & & & & & & & & \\[0.05cm]
0930+0618 & \multicolumn{2}{c}{He} & \multicolumn{3}{c}{He+H} & \multicolumn{3}{c}{He+H+Z}\\[0.12cm]
& $\Teff$ (K) & $\logg$ (dex) & $\Teff$ (K) & $\logg$ (dex) & $\htohe$ (dex) & $\Teff$ (K) & $\logg$ (dex) & $\htohe$ (dex) \\[0.1cm]
 \hline
 & & & & & & & & \\
Spec (XS)   & $16117 \pm 53$ & $8.07 \pm 0.02$ & $15757 \pm 54$   & $8.05 \pm 0.02$ & $-4.86 \pm 0.03$ &$15982 \pm 41$ & $8.18 \pm 0.02$ & $-4.86 \pm 0.04$\\[0.2cm]
Spec (SDSS) & $17739 \pm 152$ & $8.38 \pm 0.05$ & $16753 \pm 117$ & $8.35 \pm 0.06$ & $-4.66 \pm 0.06$ & $16456 \pm 125 $ & $8.29 \pm 0.07$ & $-4.75 \pm 0.06$\\[0.2cm]
Phot (PS1) & $15407^{+1009}_{-825} $ & $7.99 \pm 0.10$  & $15219_{-873}^{+1032}$ & $7.99 \pm 0.10$ & $-4.86$   & $15238_{-534}^{+665}$ & $7.99 \pm 0.07$ & $-4.86$ \\[0.2cm]
           & $-$ & $-$                                  & $15136_{-820}^{+1024}$ & $7.98 \pm 0.10$ & $-4.66$   & $-$ & $-$ & $-$ \\[0.2cm]
Phot (Gaia) &$17685 ^{+1290}_{-1079}$ & $8.20 \pm 0.09$ & $17600_{-1040}^{+1265}$ & $8.20 \pm 0.09$ & $-4.86$  & $17025_{-739}^{+658} $ & $8.16 \pm 0.07$ & $-4.86$ \\[0.2cm]
            & $-$ & $-$                                 & $17535_{-1144}^{+1243}$ & $8.20 \pm 0.09$ & $-4.66$  & $-$ & $-$ & $-$ \\[0.2cm]
Phot (SDSS) & $15640^{+472}_{-445} $ & $7.99 \pm 0.05$  & $15407_{-447}^{+548}$ & $7.98 \pm 0.06$ & $-4.86$    & $15481 _{-396}^{+425}$ & $7.99 \pm 0.05$ & $-4.86$ \\[0.2cm]
            & $-$ & $-$                                 & $15358_{-458}^{+533}$ & $7.98 \pm 0.06$ & $-4.66$    & $-$ & $-$ & $-$ \\[0.2cm]
\hline
\end{tabular}
\end{center}
\label{tab:0930+0618}
\end{table*}

\begin{table*}
\caption{Same as Table~\ref{tab:0030+1526} but for 0944--0039.}
\vspace{0.1cm}
\footnotesize
\setlength{\tabcolsep}{0.9ex}
\begin{center}
\begin{tabular}{l|cccccccc}
\hline
 & & & & & & & & \\[0.05cm]
0944$-$0039 & \multicolumn{2}{c}{He} & \multicolumn{3}{c}{He+H} & \multicolumn{3}{c}{He+H+Z}\\[0.12cm]
& $\Teff$ (K) & $\logg$ (dex) & $\Teff$ (K) & $\logg$ (dex) & $\htohe$ (dex) & $\Teff$ (K) & $\logg$ (dex) & $\htohe$ (dex) \\[0.1cm]
 \hline
 & & & & & & & & \\
Spec (XS)   & $15204 \pm 44$ & $8.97 \pm 0.02$        & $14842^{+54}_{-63}$   & $8.91 \pm 0.03$ & $-5.78 \pm 0.07$ & $14607 \pm 45$ & $8.76 \pm 0.02$ & $-5.81 \pm 0.05$\\[0.2cm]
Spec (BOSS) & $15387 _{-111}^{+95}$ & $8.91 \pm 0.05$ & $15587 \pm 187$ & $9.17 \pm 0.96$ & $-5.49 \pm 0.14$        & $15102_{-154}^{+138} $ & $9.04 \pm 0.07$ & $-5.58 \pm 0.10$\\[0.2cm]
Spec (SDSS) & $15118 \pm 117$ & $8.97 \pm 0.02$       & $15810 ^{+125}_{-136}$ & $9.29 \pm 0.06$ & $-4.96 \pm 0.09$ & $15753^{+135}_{-172}$ & $9.06 \pm 0.07$ & $-5.43 \pm 0.04$\\[0.2cm]

Phot (PS1) & $13249_{-706}^{+840}$ & $8.17 \pm 0.08$    & $13176_{-688}^{+778}$ & $8.17 \pm 0.08$ & $-5.78$  & $13350_{-435}^{+679}$ & $8.19 \pm 0.07$ & $-5.81$ \\[0.2cm]
           & $-$ & $-$                                  & $13140_{-672}^{+820}$ & $8.16 \pm 0.08$ & $-5.49$  & $-$ & $-$ & $-$ \\[0.2cm]
           & $-$ & $-$                                  & $13021_{-734}^{+8.02}$ & $8.15 \pm 0.09$ & $-4.96$ & $-$ & $-$ & $-$ \\[0.2cm]
Phot (Gaia) &  $13137 ^{+399}_{-390}$ & $8.16 \pm 0.05$ & $12968_{-356}^{+395}$ & $8.15 \pm 0.05$ & $-5.78$  & $13074_{-248}^{+355} $ & $8.17 \pm 0.05$ & $-5.81$ \\[0.2cm]
            & $-$ & $-$                                 & $12934_{-360}^{+412}$ & $8.15 \pm 0.05$ & $-5.49$  & $-$ & $-$ & $-$ \\[0.2cm]
            & $-$ & $-$                                 & $12873_{-433}^{+532}$ & $8.14 \pm 0.07$ & $-4.96$  & $-$ & $-$ & $-$ \\[0.2cm]
Phot (SDSS) & $13077 \pm 197 $ & $8.13 \pm 0.03$        & $12940_{-180}^{+195}$ & $8.12 \pm 0.04$ & $-5.78$  & $13032_{-168}^{+180}$ & $8.14 \pm 0.04$ & $-5.81$ \\[0.2cm]
            & $-$ & $-$                                 & $12910_{-184}^{+197}$ & $8.12 \pm 0.04$ & $-5.49$  & $-$ & $-$ & $-$ \\[0.2cm]
            & $-$ & $-$                                 & $12784_{-194}^{+211}$ & $8.11 \pm 0.04$ & $-4.96$  & $-$ & $-$ & $-$ \\[0.2cm]
\hline
\end{tabular}
\end{center}
\label{tab:0944-0039}
\end{table*}

\begin{table*}
\caption{Same as Table~\ref{tab:0030+1526} but for 0958+0550.}
\vspace{0.1cm}
\footnotesize
\setlength{\tabcolsep}{0.9ex}
\begin{center}
\begin{tabular}{l|cccccccc}
\hline
 & & & & & & & & \\[0.05cm]
0958+0550 & \multicolumn{2}{c}{He} & \multicolumn{3}{c}{He+H} & \multicolumn{3}{c}{He+H+Z}\\[0.12cm]
& $\Teff$ (K) & $\logg$ (dex) & $\Teff$ (K) & $\logg$ (dex) & $\htohe$ (dex) & $\Teff$ (K) & $\logg$ (dex) & $\htohe$ (dex) \\[0.1cm]
 \hline
 & & & & & & & & \\
Spec (XS)   & $12401_{-125}^{+136}$ & $8.85 \pm 0.08$ & $12245 \pm 120 $  & $8.77 \pm 0.10$ & $-5.60 \pm 0.08$ & $11428 _{-110}^{+149}$ & $8.22 \pm 0.09$ & $-5.82 \pm 0.07$\\[0.2cm]
Spec (BOSS) & $11950_{-313}^{+296}$ & $8.55 \pm 0.24$ & $11897^{+279}_{-266}$  & $8.49 \pm 0.22$ & $-5.8 \pm 0.3$   & $11696 \pm 320 $ & $8.40 \pm 0.24$ & $-5.79 \pm 0.14$\\[0.2cm]
Phot (PS1) & $11571_{-434}^{+457}$ & $8.04 \pm 0.07$  & $11362_{-442}^{+488}$ & $8.01 \pm 0.08$ & $-5.7$           & $11340_{-455}^{+484}$ & $8.01 \pm 0.08$ & $-5.82$ \\[0.2cm]
Phot (Gaia) & $10969^{+378}_{-343}$ & $7.96 \pm 0.06$ & $10731 \pm 410$ & $7.92 \pm 0.08$       & $-5.7$           & $10872_{-243}^{+341} $ & $7.95 \pm 0.07$ & $-5.82$ \\[0.2cm]
Phot (SDSS) & $11321_{-168}^{+187}$ & $8.01 \pm 0.04$ & $11111 \pm 188$ & $7.97 \pm 0.06$       & $-5.7$           & $11161\pm 190$ & $7.98 \pm 0.06$ & $-5.82$ \\[0.2cm]
\hline
\end{tabular}
\end{center}
\label{tab:0958+0550}
\end{table*}

\begin{table*}
\caption{Same as Table~\ref{tab:0030+1526} but for 1013+0259.}
\vspace{0.1cm}
\footnotesize
\setlength{\tabcolsep}{0.9ex}
\begin{center}
\begin{tabular}{l|cccccccc}
\hline
 & & & & & & & & \\[0.05cm]
1013+0259 & \multicolumn{2}{c}{He} & \multicolumn{3}{c}{He+H} & \multicolumn{3}{c}{He+H+Z}\\[0.12cm]
& $\Teff$ (K) & $\logg$ (dex) & $\Teff$ (K) & $\logg$ (dex) & $\htohe$ (dex) & $\Teff$ (K) & $\logg$ (dex) & $\htohe$ (dex) \\[0.1cm]
 \hline
 & & & & & & & & \\
Spec (XS)   & $-$ & $-$                              & $12986\pm 39$   & $8.09 \pm 0.02$ & $-3.10 \pm 0.02$ & $13158 \pm 27$ & $8.08 \pm 0.02$ & $-3.13 \pm 0.01$\\[0.2cm]
Spec (BOSS) & $-$ & $-$                              & $13420 ^{+105}_{-115}$ & $8.37 \pm 0.06$ & $-3.02 \pm 0.02$ & $13544_{-107}^{+94}$ & $8.38 \pm 0.05$ & $-2.99 \pm 0.03$\\[0.2cm]
Spec (SDSS) & $-$ & $-$                              & $13300 ^{+124}_{-140}$ & $8.42 \pm 0.06$ & $-3.04 \pm 0.03$ & $13386 \pm 125$ & $8.38 \pm 0.05$ & $-3.06 \pm 0.02$\\[0.2cm]
Phot (PS1) & $13097^{+399}_{-390}$ & $8.06 \pm 0.07$ & $12147_{-418}^{+478}$ & $7.89 \pm 0.09$ & $-3.10$  & $12317_{-380}^{+414}$ & $7.90 \pm 0.09$ & $-3.13$ \\[0.2cm]
           & $-$ & $-$                               & $12162_{-400}^{+460}$ & $7.90 \pm 0.09$ & $-3.03$  & $-$ & $-$ & $-$ \\[0.2cm]
Phot (Gaia) & $12834^{+785}_{-740}$ & $8.04 \pm 0.1$ & $11847_{-758}^{+1014}$ & $7.86 \pm 0.14$ & $-3.10$ & $12328_{-497}^{+685} $ & $7.91 \pm 0.10$ & $-3.13$ \\[0.2cm]
            & $-$ & $-$                     & $11865_{-737}^{+1317}$ & $7.87 ^{+0.20}_{-0.13}$ & $-3.03$  & $-$ & $-$ & $-$ \\[0.2cm]
Phot (SDSS) & $13045 \pm 280 $ & $8.05 \pm 0.05$     & $12093 \pm 294$       & $7.89 \pm 0.07$ & $-3.10$  & $12279_{-255}^{+278}$ & $7.90 \pm 0.07$ & $-3.13$ \\[0.2cm]
            & $-$ & $-$                              & $12117_{-277}^{+304}$ & $7.89 \pm 0.07$ & $-3.03$  & $-$ & $-$ & $-$ \\[0.2cm]
\hline
\end{tabular}
\end{center}
\label{tab:1013+0259}
\end{table*}

\begin{table*}
\caption{Same as Table~\ref{tab:0030+1526} but for 1109+1318.}
\vspace{0.1cm}
\footnotesize
\setlength{\tabcolsep}{0.9ex}
\begin{center}
\begin{tabular}{l|cccccccc}
\hline
 & & & & & & & & \\[0.05cm]
1109+1318 & \multicolumn{2}{c}{He} & \multicolumn{3}{c}{He+H} & \multicolumn{3}{c}{He+H+Z}\\[0.12cm]
& $\Teff$ (K) & $\logg$ (dex) & $\Teff$ (K) & $\logg$ (dex) & $\htohe$ (dex) & $\Teff$ (K) & $\logg$ (dex) & $\htohe$ (dex) \\[0.1cm]
 \hline
 & & & & & & & & \\
Spec (XS)   & $16568 _{-95}^{+87}$  & $8.23 \pm 0.04$ & $16167 \pm 69$ & $8.19 \pm 0.03$ & $-4.05 \pm 0.03$ & $16308 \pm 62$  & $8.25 \pm 0.03$ & $-4.01 \pm 0.03$\\[0.2cm]
Spec (BOSS) & $16498_{-293}^{+315}$ & $8.16 \pm 0.20$ & $15890_{-181}^{+210}$ & $7.96 \pm 0.15$ & $-4.07 \pm 0.07$          & $15887 \pm 192$ & $8.06 \pm 0.13$ & $-4.02 \pm 0.05$\\[0.2cm]
Spec (SDSS) & $17113_{-392}^{+355}$ & $8.91_{-0.14}^{+0.06}$ & $16115 _{-310}^{+343}$ & $8.86 _{-0.14}^{+0.09}$ & $-4.06 \pm 0.10$ & $16350_{-182}^{+106}$ & $8.68 \pm 0.15$ & $-4.05 \pm 0.07$\\[0.2cm]
Phot (PS1) & $16220^{+1560}_{-1163} $ & $8.16 \pm 0.13$         & $15690_{-1190}^{+1450}$ & $8.12 \pm 0.15$ & $-4.05$       & $15705_{-455}^{+523}$ & $8.12 \pm 0.10$ & $-4.01$ \\[0.2cm]
Phot (Gaia) & $17013^{+1910}_{-1836} $ & $8.23_{-0.18}^{+0.14}$ & $16178_{-2275}^{2038}$ & $8.18_{-0.23}^{+0.19}$ & $-4.05$ & $15728\pm 502$ & $8.13 \pm 0.10$ & $-4.01$ \\[0.2cm]
Phot (SDSS) & $15807_{-590}^{+701}$ & $8.12 \pm 0.08$           & $15335_{-665}^{+803}$ & $8.09 \pm 0.11$ & $-4.05$         & $15659_{-409}^{+465}$ & $8.12 \pm 0.10$ & $-4.01$ \\[0.2cm]
\hline
\end{tabular}
\end{center}
\label{tab:1109+1318}
\end{table*}

\begin{table*}
\caption{Same as Table~\ref{tab:0030+1526} but for 1359--0217.}
\vspace{0.1cm}
\footnotesize
\setlength{\tabcolsep}{0.9ex}
\begin{center}
\begin{tabular}{l|cccccccc}
\hline
 & & & & & & & & \\[0.05cm]
1359$-$0217 & \multicolumn{2}{c}{He} & \multicolumn{3}{c}{He+H} & \multicolumn{3}{c}{He+H+Z}\\[0.12cm]
& $\Teff$ (K) & $\logg$ (dex) & $\Teff$ (K) & $\logg$ (dex) & $\htohe$ (dex) & $\Teff$ (K) & $\logg$ (dex) & $\htohe$ (dex) \\[0.1cm]
 \hline
 & & & & & & & & \\
Spec (XS)   & $17920^{+74}_{-164}$ & $8.20 \pm 0.03$ & $16995\pm 91$   & $8.18 \pm 0.02$ & $-3.11 \pm 0.03$ & $16773 \pm 55$ & $8.14 \pm 0.02$ & $-3.16 \pm 0.02$\\[0.2cm]
Spec (BOSS) & $17369^{+107}_{-89}$ & $8.14 \pm 0.04$ & $16912^{+71}_{-65}$ & $8.08 \pm 0.03$ & $-3.19 \pm 0.03$   & $17153 \pm 72$ & $8.07 \pm 0.03$ & $-3.13 \pm 0.03$\\[0.2cm]
Spec (SDSS) & $17671 \pm 170$ & $7.95 \pm 0.09$      & $17681^{+100}_{-122}$ & $8.10 \pm 0.05$ & $-3.04 \pm 0.03$ & $17630 \pm 147$ & $8.12 \pm 0.06$ & $-3.05 \pm 0.05$\\[0.2cm]

Phot (PS1) & $14456^{+526}_{-480}$ & $7.85 \pm 0.05$    & $13584_{-497}^{+602}$ & $7.72 \pm 0.09$ & $-3.11$   & $13607_{-455}^{+565}$ & $7.73 \pm 0.08$ & $-3.16$ \\[0.2cm]
           & $-$ & $-$                                  & $13557_{-487}^{+591}$ & $7.72 \pm 0.09$ & $-3.04$   & $-$ & $-$ & $-$ \\[0.2cm]
Phot (Gaia) & $16890^{+1447}_{-1083}$ & $8.10 \pm 0.12$ & $15943_{-1150}^{+1081}$ & $8.02 \pm 0.13$ & $-3.11$ & $15701_{-1011}^{+1040} $ & $7.99 \pm 0.12$ & $-3.16$ \\[0.2cm]
            & $-$ & $-$                                 & $15834_{-1101}^{+1160}$ & $8.00 \pm 0.13$ & $-3.04$ & $-$ & $-$ & $-$ \\[0.2cm]
Phot (SDSS) & $14471^{+355}_{-325}$ & $7.85 \pm 0.04$   & $13812_{-375}^{+436}$ & $7.75\pm 0.07$ & $-3.11$    & $14103_{-322}^{+376}$ & $7.79 \pm 0.06$ & $-3.16$ \\[0.2cm]
            & $-$ & $-$                                 & $13915_{-377}^{+396}$ & $7.76 \pm 0.06$ & $-3.04$   & $-$ & $-$ & $-$ \\[0.2cm]
\hline
\end{tabular}
\end{center}
\label{tab:1359-0217}
\end{table*}

\begin{table*}
\caption{Same as Table~\ref{tab:0030+1526} but for 1516--0040.}
\vspace{0.1cm}
\footnotesize
\setlength{\tabcolsep}{0.9ex}
\begin{center}
\begin{tabular}{l|cccccccc}
\hline
 & & & & & & & & \\[0.05cm]
1516$-$0040 & \multicolumn{2}{c}{He} & \multicolumn{3}{c}{He+H} & \multicolumn{3}{c}{He+H+Z}\\[0.12cm]
& $\Teff$ (K) & $\logg$ (dex) & $\Teff$ (K) & $\logg$ (dex) & $\htohe$ (dex) & $\Teff$ (K) & $\logg$ (dex) & $\htohe$ (dex) \\[0.1cm]
 \hline
 & & & & & & & & \\
Spec (XS)   & $15838 \pm 28$  & $8.37 \pm 0.02$ & $15397 \pm 22$ & $8.35 \pm 0.02$ & $-4.49 \pm 0.02$ & $15448 \pm 20$  & $8.42 \pm 0.01$ & $-4.50 \pm 0.01$\\[0.2cm]
Spec (BOSS) & $15854 \pm 97$ & $8.28 \pm 0.06$  & $15717_{-67}^{+85}$ & $8.37 \pm 0.04$ & $-4.44 \pm 0.03$  & $15611_{-72}^{+88}$ & $8.35 \pm 0.04$ & $-4.45 \pm 0.03$\\[0.2cm]
Phot (PS1) & $13006^{+487}_{-456} $ & $7.92 \pm 0.05$ & $12425_{-477}^{+548}$ & $7.85 \pm 0.07$ & $-4.46$ & $12424_{-437}^{+532}$ & $7.85 \pm 0.07$ & $-4.50$ \\[0.2cm]
Phot (Gaia) & $12668^{+574}_{-442}$ & $7.89 \pm 0.06$ & $11995_{-447}^{573}$ & $7.79\pm 0.08$ & $-4.46$   & $12184_{-325}^{+427}$ & $7.83 \pm 0.06$ & $-4.50$ \\[0.2cm]
Phot (SDSS) & $13554_{-202}^{+215}$ & $7.97 \pm 0.04$ & $13073_{-224}^{+229}$ & $7.92 \pm 0.03$ & $-4.46$ & $13248_{-206}^{+228}$ & $7.95 \pm 0.03$ & $-4.50$ \\[0.2cm]
\hline
\end{tabular}
\end{center}
\label{tab:1516-0040}
\end{table*}

\begin{table*}
\caption{Same as Table~\ref{tab:0030+1526} but for 1627+1723.}
\vspace{0.1cm}
\footnotesize
\setlength{\tabcolsep}{0.7ex}
\begin{center}
\begin{tabular}{l|cccccccc}
\hline
 & & & & & & & & \\[0.05cm]
1627+1723 & \multicolumn{2}{c}{He} & \multicolumn{3}{c}{He+H} & \multicolumn{3}{c}{He+H+Z}\\[0.12cm]
& $\Teff$ (K) & $\logg$ (dex) & $\Teff$ (K) & $\logg$ (dex) & $\htohe$ (dex) & $\Teff$ (K) & $\logg$ (dex) & $\htohe$ (dex) \\[0.1cm]
 \hline
 & & & & & & & & \\
Spec (XS)   & $16422_{-157}^{+141}$  & $8.21 \pm 0.07$    & $15920 \pm 115$  & $8.20 \pm 0.06$ & $-5.05 \pm 0.10$ & $16134 \pm 102$ & $8.29 \pm 0.05$ & $-5.05 \pm 0.07$\\[0.2cm]
Spec (BOSS) & $17442 _{-227}^{+210}$ & $8.33\pm 0.08$     & $16520^{+200}_{-177}$ & $8.52 \pm 0.09$ & $-5.15 \pm 0.18$ & $16451 \pm 193 $ & $8.51 \pm 0.09$ & $-5.15 \pm 0.17$\\[0.2cm]
Spec (SDSS) & $17865_{-454}^{+504}$  & $8.71 \pm 0.15$ & $14757^{+222}_{-193}$ & $7.26 \pm 0.18$ & $-5.70 \pm 0.26$ & $15132 \pm 255$ & $7.52 \pm 0.18$ & $-4.85 \pm 0.31$\\[0.2cm]

Phot (PS1) & $16537_{-1465}^{+1855}$ & $8.17 \pm 0.15$   & $16376_{-1460}^{+1790}$ & $8.16 \pm 0.15$ & $-5.10$ & $15987_{-773}^{+848}$ & $8.13 \pm 0.10$ & $-5.05$ \\[0.2cm]
           & $-$ & $-$                                   & $16516_{-1520}^{+1875}$ & $8.16 \pm 0.15$ & $-5.70$ & $-$ & $-$ & $-$ \\[0.2cm]
Phot (Gaia) & $17137 ^{+1385}_{-1277}$ & $8.22 \pm 0.12$ & $16760_{-1240}^{+1707}$ & $8.20 \pm 0.13$ & $-5.10$ & $16300_{-782}^{+618} $ & $8.15 \pm 0.09$ & $-5.05$ \\[0.2cm]
            & $-$ & $-$                                  & $16896_{-1210}^{+1540}$ & $8.20 \pm 0.12$ & $-5.70$ & $-$ & $-$ & $-$ \\[0.2cm]
Phot (SDSS) & $16056_{-468}^{+574}$ & $8.12 \pm 0.06$    & $15951_{-560}^{+670}$ & $8.11 \pm 0.08$   & $-5.10$ & $15890_{-496}^{+547}$ & $8.11 \pm 0.08$ & $-5.0$ \\[0.2cm]
            & $-$ & $-$                                  & $15978_{-535}^{+593}$ & $8.10 \pm 0.08$   & $-5.70$ & $-$ & $-$ & $-$ \\[0.2cm]
\hline
\end{tabular}
\end{center}
\label{tab:1627+1723}
\end{table*}

\begin{table*}
\caption{Same as Table~\ref{tab:0030+1526} but for 2324--0018.}
\vspace{0.1cm}
\footnotesize
\setlength{\tabcolsep}{0.7ex}
\begin{center}
\begin{tabular}{l|cccccccc}
\hline
 & & & & & & & & \\[0.05cm]
2324$-$0018 & \multicolumn{2}{c}{He} & \multicolumn{3}{c}{He+H} & \multicolumn{3}{c}{He+H+Z}\\[0.12cm]
& $\Teff$ (K) & $\logg$ (dex) & $\Teff$ (K) & $\logg$ (dex) & $\htohe$ (dex) & $\Teff$ (K) & $\logg$ (dex) & $\htohe$ (dex) \\[0.1cm]
 \hline
 & & & & & & & & \\
Spec (XS)   & $-$ & $-$  & $13754 \pm 45$  & $8.17\pm 0.03$ & $-3.32 \pm 0.02$  & $14063 \pm 53$ & $8.25 \pm 0.02$ & $-3.33 \pm 0.01$\\[0.2cm]
Spec (BOSS) & $-$ & $-$                              & $13958 \pm 170$ & $8.38 \pm 0.08$ & $-3.28 \pm 0.05$ & $14117 \pm 182$ & $8.36 \pm 0.07$ & $-3.29 \pm 0.04$\\[0.2cm]
Phot (PS1) & $14174^{+1047}_{-829}$ & $7.87 \pm 0.13$ & $13262_{-841}^{+1145}$ & $7.73 \pm 0.10$ & $-3.30$ & $13193_{-698}^{+730}$ & $7.70 \pm 0.18$ & $-3.33$ \\[0.2cm]
Phot (Gaia) & $12238^{+930}_{-735}$ & $7.63 \pm 0.16$ & $11220_{-658}^{+852}$ & $7.43 \pm 0.21$ & $-3.30$  & $11881_{-580}^{+730} $ & $7.54 \pm 0.16$ & $-3.33$ \\[0.2cm]
Phot (SDSS) & $13470^{+307}_{-271}$ & $7.77 \pm 0.10$ & $12530 \pm 320$       & $7.63 \pm 0.15$ & $-3.30$  & $12681_{-291}^{+323}$ & $7.64 \pm 0.15$ & $-3.33$ \\[0.2cm]
\hline
\end{tabular}
\end{center}
\label{tab:2324-0018}
\end{table*}




\bsp	
\label{lastpage}
\end{document}